\documentclass[12pt]{iopart}

\usepackage{graphicx}
\usepackage{color}
\usepackage[dvipdfmx,bookmarks=true,colorlinks,%
            citecolor=blue,linkcolor=blue,anchorcolor=blue,filecolor=blue,urlcolor=blue,%
           ]{hyperref}
\usepackage{amsfonts}
\usepackage{amssymb}

\begin{document}
\bibliographystyle{iopart-num}


\newcommand{\lc}{\left\langle}
\newcommand{\rc}{\right\rangle}
\newcommand{\lr}{\left|}
\newcommand{\rl}{\right|}
\newcommand{\lb}{\left(}
\newcommand{\rb}{\right)}
\newcommand{\ls}{\left[}
\newcommand{\rs}{\right]}
\newcommand{\Lb}{\left\{}
\newcommand{\Rb}{\right\}}
\newcommand{\lrl}[1]{\left\vert#1\right\vert}
\newcommand{\lrc}[1]{\left<#1\right>}
\newcommand{\lrb}[1]{\left(#1\right)}
\newcommand{\lrs}[1]{\left[#1\right]}
\newcommand{\Lrb}[1]{\left\{#1\right\}}

\newcommand{\bea}{\begin{array}}
\newcommand{\eea}{\end{array}}
\newcommand{\bm}[1]{\mbox{\boldmath{$#1$}}}
\newcommand{\dfrac}{\displaystyle\frac}
\newcommand{\hatr}{\hat{\bm r}}
\newcommand{\beq}{\begin{equation}}
\newcommand{\eeq}{\end{equation}}
\newcommand{\beqn}{\begin{eqnarray}}
\newcommand{\eeqn}{\end{eqnarray}}
\newcommand{\bsub}{\begin{subequations}}
\newcommand{\esub}{\end{subequations}}
\newcommand{\benu}{\begin{enumerate}}
\newcommand{\eenu}{\end{enumerate}}
\newcommand{\beit}{\begin{itemize}}
\newcommand{\enit}{\end{itemize}}

\newcommand{\ff}[1]{\frac{1}{#1}}
\newcommand{\no}{\nonumber\\}
\newcommand{\re}{\nonumber\\}
\newcommand{\ivec}{\vec}
\newcommand{\svec}[1]{{\mbox{\boldmath$#1$}}}
\newcommand{\cals}[1]{{\cal #1}}
\newcommand{\scr}[1]{{\cal #1}}
\newcommand{\ket}[1]{{\left| #1 \right\rangle }}
\newcommand{\bra}[1]{{\left\langle #1 \right|}}
\newcommand{\ovl}{\overline}
\newcommand{\unl}{\underline}

\newcommand{\bJ}{\mbox{\boldmath$J$}}
\newcommand{\bp}{\mbox{\boldmath$p$}}
\newcommand{\rto}{\rightarrow}
\newcommand{\bo}{\boldsymbol}
\newcommand{\ee}[1]{(-)^{#1}}
\newcommand{\mean}[3]{\langle #1|#2|#3\rangle}
\newcommand{\redu}[3]{\langle#1||#2||#3\rangle}

\newcommand{\alp}{\alpha}
\newcommand{\bet}{\beta}
\newcommand{\gam}{\gamma}
\newcommand{\del}{\delta}
\newcommand{\eps}{\epsilon}
\newcommand{\vep}{\varepsilon}
\newcommand{\zet}{\zeta}
\newcommand{\tht}{\theta}
\newcommand{\vth}{\vartheta}
\newcommand{\iot}{\iota}
\newcommand{\kap}{\kappa}
\newcommand{\lam}{\lambda}
\newcommand{\vrh}{\varrho}
\newcommand{\sig}{\sigma}
\newcommand{\vsi}{\varsigma}
\newcommand{\ups}{\upsilon}
\newcommand{\vph}{\varphi}
\newcommand{\ome}{\omega}
\newcommand{\dig}{\digamma}
\newcommand{\gim}{\gimel}
\newcommand{\dal}{\daleth}
\newcommand{\nab}{\nabla}
\newcommand{\Gam}{\Gamma}
\newcommand{\Del}{\Delta}
\newcommand{\The}{\Theta}
\newcommand{\Lam}{\Lambda}
\newcommand{\Sig}{\Sigma}
\newcommand{\Ups}{\Upsilon}
\newcommand{\Ome}{\Omega}

\newcommand{\half}{\frac{1}{2}}
\newcommand{\quat}{\frac{1}{4}}

\newcommand{\balp}{\mbox{\boldmath$\alpha$}}
\newcommand{\bbet}{\mbox{\boldmath$\beta$}}
\newcommand{\bgam}{\mbox{\boldmath$\gamma$}}
\newcommand{\bdel}{\mbox{\boldmath$\delta$}}
\newcommand{\beps}{\mbox{\boldmath$\epsilon$}}
\newcommand{\bvep}{\mbox{\boldmath$\varepsilon$}}
\newcommand{\bzet}{\mbox{\boldmath$\zeta$}}
\newcommand{\bmeta}{\mbox{\boldmath$\eta$}}
\newcommand{\bthe}{\mbox{\boldmath$\theta$}}
\newcommand{\bvth}{\mbox{\boldmath$\vartheta$}}
\newcommand{\biot}{\mbox{\boldmath$\iota$}}
\newcommand{\bkap}{\mbox{\boldmath$\kappa$}}
\newcommand{\blam}{\mbox{\boldmath$\lambda$}}
\newcommand{\bmu}{\mbox{\boldmath$\mu$}}
\newcommand{\bnu}{\mbox{\boldmath$\nu$}}
\newcommand{\bxi}{\mbox{\boldmath$\xi$}}
\newcommand{\bpi}{\mbox{\boldmath$\pi$}}
\newcommand{\bvpi}{\mbox{\boldmath$\varpi$}}
\newcommand{\brho}{\mbox{\boldmath$\rho$}}
\newcommand{\bvrh}{\mbox{\boldmath$\varrho$}}
\newcommand{\bsig}{\mbox{\boldmath$\sigma$}}
\newcommand{\bvsi}{\mbox{\boldmath$\varsigma$}}
\newcommand{\btau}{\mbox{\boldmath$\tau$}}
\newcommand{\bups}{\mbox{\boldmath$\upsilon$}}
\newcommand{\bphi}{\mbox{\boldmath$\phi$}}
\newcommand{\bvph}{\mbox{\boldmath$\varphi$}}
\newcommand{\bchi}{\mbox{\boldmath$\chi$}}
\newcommand{\bome}{\mbox{\boldmath$\omega$}}
\newcommand{\bdig}{\mbox{\boldmath$\digamma$}}
\newcommand{\bgim}{\mbox{\boldmath$\gimel$}}
\newcommand{\bdal}{\mbox{\boldmath$\daleth$}}
\newcommand{\bGam}{\mbox{\boldmath$\Gamma$}}
\newcommand{\bDel}{\mbox{\boldmath$\Delta$}}
\newcommand{\bThe}{\mbox{\boldmath$\Theta$}}
\newcommand{\bLam}{\mbox{\boldmath$\Lambda$}}
\newcommand{\bXi}{\mbox{\boldmath$\Xi$}}
\newcommand{\bPi}{\mbox{\boldmath$\Pi$}}
\newcommand{\bSig}{\mbox{\boldmath$\Sigma$}}
\newcommand{\bUps}{\mbox{\boldmath$\Upsilon$}}
\newcommand{\bPhi}{\mbox{\boldmath$\Phi$}}
\newcommand{\bPsi}{\mbox{\boldmath$\Psi$}}
\newcommand{\bOme}{\mbox{\boldmath$\Omega$}}
\newcommand{\bnab}{\mbox{\boldmath$\nabla$}}

\newcommand{\RD}{{\rm D}}
\newcommand{\RE}{{\rm E}}
\newcommand{\RV}{{\rm V}}
\newcommand{\RS}{{\rm S}}
\newcommand{\RVT}{{\rm VT}}
\newcommand{\RTV}{{\rm TV}}
\newcommand{\RPV}{{\rm PV}}
\newcommand{\RT}{{\rm T}}
\newcommand{\mev}{{\rm MeV}}
\newcommand{\kev}{{\rm keV}}

\newcommand{\red}[1]{\textcolor{red}{#1}}
\newcommand{\blue}[1]{\textcolor{blue}{#1}}

\title[Halos and deformed halos from CDFT]
      {Halos in medium-heavy and heavy nuclei with covariant density functional
       theory in continuum}

\author{J Meng$^{1,2,3}$ and S G Zhou$^{4,5,6}$}

\address{$^{1}$ State Key Laboratory of Nuclear Physics and Technology,
                School of Physics, Peking University, Beijing 100871, China}
\address{$^{2}$ School of Physics and Nuclear Energy Engineering,
                Beihang University, Beijing 100191, China}
\address{$^{3}$ Department of Physics, University of Stellenbosch, Stellenbosch 7602, South Africa}
\address{$^{4}$ State Key Laboratory of Theoretical Physics, Institute of Theoretical Physics,
                Chinese Academy of Sciences, Beijing 100190, China }
\address{$^{5}$ Center of Theoretical Nuclear Physics, National Laboratory
                of Heavy Ion Accelerator, Lanzhou 730000, China }
\address{$^{6}$ Center for Nuclear Matter Science, Central China Normal University,
                Wuhan 430079, China}

\ead{\mailto{mengj@pku.edu.cn} (Jie Meng), \mailto{sgzhou@itp.ac.cn} (Shan-Gui Zhou)}

\begin{abstract}
The covariant density functional theory with a few number of parameters
has been widely used to describe the ground-state and excited-state properties for the nuclei all over the nuclear chart. In order to describe exotic properties of unstable nuclei,
the contribution of the continuum and its coupling with bound states
should be treated properly.
In this Topical Review, the development of the covariant density functional theory in continuum will be introduced, including the relativistic continuum Hartree-Bogoliubov theory,
the relativistic Hartree-Fock-Bogoliubov theory in continuum, and
the deformed relativistic Hartree-Bogoliubov theory in continuum.
Then the descriptions and predictions of the neutron halo phenomena in both spherical and deformed nuclei will be reviewed. The diffuseness of the nuclear potentials, nuclear shapes and density distributions, and the impact of the pairing correlations on nuclear size will be discussed.
\end{abstract}

\submitto{Topical Review \JPG}

\maketitle

\tableofcontents

\section{Introduction}

The building blocks of nucleus are protons and neutrons.
An arbitrary combination of protons and neutrons doesn't necessarily make a nucleus.
Among thousands of nuclides expected to exist,
only 288 of them are stable or at least with half-lives longer than
the expected life of the Solar System,
which form the valley of stability in the nuclear chart.
These nuclides are sandwiched by about 3,000 nuclides produced in
laboratories~\cite{Thoennessen2013_RPP76-056301}.
The nuclear density functional theory (DFT) has predicted that
about 7,000 nuclides are bound with respect to neutron or proton emission
\cite{Erler2012_Nature486-509, Afanasjev2013_PLB726-680,
Agbemava2014_PRC89-054320}.
If the continuum contribution is included, as demonstrated in Ref.~\cite{Qu2013_SciChinaPMA56-2031} from O to Ti isotopes, this number may exceed
10,000.
In addition, there are also nuclides (resonances) which, though beyond the drip lines,
can have lifetimes long enough to be measured and investigated.

The nuclides lying above the valley of stability in the nuclear chart are
proton-rich and the proton drip line has been identified
experimentally up to protactinium ($Z=91$).
However, for those lying below the valley of stability, which are neutron-rich,
the border of the nuclear territory is known only up to oxygen ($Z=8$).
Experimental exploration of very neutron-rich nuclei is extremely challenging
because of very low production rates in studies involving the fragmentation
of stable nuclei, as well as the separation and identification of the products.

In nuclear physics, the study of the properties of exotic nuclei ---
nuclei with extreme numbers of proton or neutron --- is one of
the top priorities as it can lead to new insights on the origin of
chemical elements in stars and star explosions.
Although the radioactive ion beams (RIB) have extended our knowledge of nuclear physics
from stable nuclei to exotic ones far away from the valley of stability,
it is still a dream to reach the neutron drip line up to
mass number $A\sim 100$ with the new generation RIB facilities
developed around the world, including the RIKEN Radioactive Ion Beam Factory (RIBF) in Japan~\cite{Motobayashi2010_NPA834-707c},
Facility for Antiproton and Ion Research (FAIR) in Germany~\cite{Sturm2010_NPA834-682c},
Second Generation System On-Line Production of Radioactive Ions (SPIRAL2) at {GANIL} in France~\cite{Gales2010_NPA834-717c},
Facility for Rare Isotope Beams (FRIB) in USA~\cite{Thoennessen2010_NPA834-688c},
Rare Isotope Science Project (RISP) in Korea~\cite{Choi2010_ISNPA},
Cooler Storage Ring (CSR) at Heavy Ion Research Facility in Lanzhou (HIRFL) in
China~\cite{Xia2002_NIMA488-11, Zhan2010_NPA834-694c}, etc.

The development of RIB facilities around
the world significantly stimulates the study of exotic nuclei
far from the valley of stability~\cite{Mueller1993_ARNPS43-529, Tanihata1995_PPNP35-505, Hansen1995_ARNPS45-591,
Casten2000_PPNP45-S171, Bertulani2001_PhysRIB, Johnson2004_PR389-1,
Jensen2004_RMP76-215, Meng2006_PPNP57-470, Ershov2010_JPG37-064026,
Frederico2012_PPNP67-939, Riisager2013_PST152-014001}.
New and exotic phenomena were observed in nuclei close to drip lines
such as neutron or proton halos~\cite{Tanihata1985_PRL55-2676},
changes of nuclear magic numbers~\cite{Ozawa2000_PRL84-5493},
the island of inversion~\cite{Warburton1990_PRC41-1147},
pygmy resonances~\cite{Adrich2005_PRL95-132501}, etc.
More exotic nuclear phenomena have also been predicted, e.g.,
giant halos~\cite{Meng1998_PRL80-460, Meng2002_PRC65-041302R},
shape decoupling between core and halo
\cite{Zhou2010_PRC82-011301R, Li2012_PRC85-024312},
etc.

In halo nuclei, the extremely weak binding of valence nucleons leads to many
new features, such as the coupling between bound states and
the continuum due to pairing correlations and the very extended
spatial density distributions. Therefore, one must consider
properly the asymptotic behavior of nuclear densities at a large
distance from the center and treat in a self-consistent way the
discrete bound states, the continuum, and the coupling between
them in order to give a proper theoretical description of the
halo phenomenon~\cite{Meng2006_PPNP57-470, Dobaczewski2007_PPNP59-432}.
There are mainly two key points to achieve a self-consistent description
of halo nuclei with density functional theories:
the pairing correlations and the suitable basis.

In the mean field descriptions of nuclear halos, nuclear pairing plays an
important role~\cite{Bertsch1991_APNP209-63, Meng1996_PRL77-3963,
Meng1998_PRC57-1229, Meng1998_PRL80-460, Meng2002_PRC65-041302R}.
The conventional BCS method for the pairing is not appropriate for exotic nuclei.
To include properly the contribution of continuum states,
the Bogoliubov transformation has been justified to be very useful
\cite{Bulgac1980_nucl-th9907088, Dobaczewski1984_NPA422-103}.
By solving the nonrelativistic Hartree-Fock-Bogoliubov
(HFB)~\cite{Bulgac1980_nucl-th9907088,Dobaczewski1984_NPA422-103} or
the relativistic Hartree-Bogoliubov (RHB)~\cite{Meng1996_PRL77-3963,
Poschl1997_PRL79-3841, Meng1998_NPA635-3} equations
in coordinate ($r$) space, one can fully take into account
the coupling to the continuum.
As an alternative, one may first locate resonant states in
the continuum and compute the resonance parameters with various methods
\cite{Yang2001_CPL18-196, Cao2002_PRC66-024311,
Zhang2004_PRC70-034308, Zhang2008_PRC77-014312, Zhou2009_JPB42-245001,
Li2010_PRC81-034311, Guo2010_PRC82-034318},
then by using the resonant-BCS (rBCS) approach~\cite{Sandulescu2000_PRC61-061301R},
include the contribution of these resonant states and
study halo phenomena~\cite{Sandulescu2003_PRC68-054323,
Zhang2013_EPJA49-77, Zhang2014_PLB730-30}.
The Green's function method~\cite{Belyaev1987_SovJNP45-783}, which can
take into account properly the asymptotic behavior of continuum
states, has also been adopted
for solving the HFB~\cite{Shlomo1975_NPA243-18, Matsuo2001_NPA696-371,
Oba2009_PRC80-024301}, the self-consistent
Skyrme-HFB \cite{Zhang2011_PRC83-054301, Zhang2012_PRC86-054318,Zhang2014_PRC90-034313} 
or RMF \cite{Sun2014_PRC90-054321}
equations in the coordinate space.

The solution of the coupled differential equations of HFB
and RHB theories is relatively easier for spherical systems
with local potentials, where one-dimensional Numerov or
Runge-Kutta methods can be applied. This is also true
even for nonlocal problems where the finite element method
(FEM) can be used~\cite{Meng1998_NPA635-3}.
A more widely used method for solving such equations
is to expand the single particle wave functions in a basis and
this is more convenient particularly for deformed systems.
In the past decades, the harmonic oscillator (HO) basis has been used with
a great success for stable nuclei in both nonrelativistic and
relativistic mean field approximations such as
the Skyrme Hartree-Fock, Hartree-Fock-Bogoliubov, relativistic Hartree,
and relativistic Hartree-Bogoliubov theories for deformed or nonlocal systems
\cite{Vautherin1972_PRC5-626, Decharge1980_PRC21-1568,
Gambhir1990_APNY198-132}.
However, due to the incorrect asymptotic behavior of the
HO wave functions, the expansion in a conventional HO basis is
not proper for the description of exotic nuclei~\cite{Zhou2000_CPL17-717}.
Considerable efforts have been made to develop mean field models
in a basis with an improved asymptotic behavior at large
distances, e.g., a transformed HO basis~\cite{Stoitsov2000_PRC61-034311,
Stoitsov2003_PRC68-054312}
or a Woods-Saxon (WS) basis~\cite{Zhou2003_PRC68-034323}.

The DFT with a small number of parameters is
widely used to study the ground-state and
excited-state properties of the nuclei all over the nuclear chart.
In particular, the covariant density functional theory (CDFT),
incorporating the Lorentz symmetry in a self-consistent way,
has received wide attention
\cite{Serot1986_ANP16-1, Ring1996_PPNP37-193,
Vretenar2005_PR409-101, Meng2006_PPNP57-470,
Niksic2011_PPNP66-519, Meng2013FrontiersofPhysics55}.

There exist a number of attractive features to describe atomic
nuclei in the relativistic framework.
The most obvious one is the natural inclusion of the nucleon spin degree
of freedom, resulting in the nuclear spin-orbit potential that emerges
automatically with the empirical strength in a covariant way.
Moreover, it provides a new saturation mechanism for nuclear matter~\cite{Brockmann1990_PRC42-1965},
reproduces well the measured isotopic shifts in the Pb region~\cite{Sharma1993_PLB317-3},
reveals more naturally the origin of the
pseudospin symmetry~\cite{Arima1969_PLB30-517, Hecht1969_NPA137-129}
as a relativistic symmetry~\cite{Ginocchio1997_PRL78-436,
Meng1998_PRC58-R628, Meng1999_PRC59-154, Long2006_PLB639-242,
Liang2011_PRC83-041301R, Lu2012_PRL109-072501, Lu2013_PRC88-024323,Liang2015PhysicsReports1},
and predicts the spin symmetry in the anti-nucleon spectrum
\cite{Zhou2003_PRL91-262501, Liang2010_EPJA44-119}.
It can also include the nuclear magnetism~\cite{Koepf1989_NPA493-61},
i.e., a consistent description of currents and time-odd fields,
which plays an important role in the nuclear magnetic moments
\cite{Yao2006_PRC74-024307, Arima2011_SciChinaPMA54-93,
Li2011_PTP125-1185,Wei2012_PTPSuppl196-400} and
nuclear rotations
\cite{Afanasjev2000_NPA676-196, Zhao2012_PRC85-054310, Zhao2011_PRL107-122501,
Zhao2011_PLB699-181, Meng2013FrontiersofPhysics55},
etc.

The CDFT is a reliable and useful model for nuclear structure study
in the whole nuclear chart,
and therefore, it is natural and necessary to describe the halo nuclei based on the CDFT.

Since the discovery of the neutron halo, i.e., the anomalously large
nuclear radius in $^{11}$Li~\cite{Tanihata1985_PRL55-2676}
which deviates from the conventional $A^{1/3}$ law on the mass dependence
of the nuclear radius,
considerable efforts have been undertaken to understand this interesting
phenomenon in exotic nuclei close to the neutron drip line.

The relativistic mean field models have been used to study nuclear
halo phenomema since early 1990s~\cite{Koepf1991_ZPA340-119,
Sharma1994_PRL72-1431, Zhu1994_PLB328-1,
Ren1995_PRC52-R20, Ren1995_PLB351-11, Ren1996_PRC53-R572, Ren1998_PRC57-2752}.
By incorporating the Bogoliubov transformation
in the relativistic Hartree theory, the relativistic continuum Hatree-Bogoliubov (RCHB)
theory has been developed and extensively used to study the halo phenomeon
in spherical nuclei
\cite{Meng1996_PRL77-3963, Poschl1997_PRL79-3841,
Meng1998_PRL80-460, Meng1998_PLB419-1, Meng1998_NPA635-3}.
The RCHB theory was also extended to including the hyperon and
used to study neutron halo in $\Lambda$-hypernuclei
\cite{Vretenar1998_PRC57-R1060, Lu2002_CPL19-1775, Lu2003_EPJA17-19,
Lu2003_HEPNP27-411, Lu2008_CPL25-3613}. 
The relativistic Hartree-Fock-Bogoliubov theory in continuum (RHFBc) 
theory has been developed and used to study the influence of exchange terms on the formation of neutron halos~\cite{Long2010_PRC81-031302R}.
Over the past years, lots of efforts have been made to develop
a deformed relativistic Hartree-Bogoliubov theory in continuum (DRHBc)
\cite{Zhou2008_ISPUN2007}.
Halo phenomena in deformed Ne and Mg isotopes have been investigated
with the DRHBc theory~\cite{Zhou2010_PRC82-011301R,Li2012_PRC85-024312, Chen2012_PRC85-067301}.

In this Topical Review, recent progresses on the application of
the CDFT for the neutron halo phenomena
in spherical and deformed nuclei will be presented. As an extensive review is available~\cite{Meng2006_PPNP57-470}, the RCHB theory and its application will be briefly mentioned here for completeness. In Section~\ref{sec:formalism}, we give
the formalism for CDFT in continuum,
including the general framework of CDFT,
the formalism for RHB theory, and the RCHB, RHFBc and DRHBc theories.
In Section~\ref{sec:spherical_halos},
the applications of the CDFT for halos and giant halos in spherical nuclei will be reviewd.
We will discuss the RCHB description of the neutron halo in $^{11}$Li,
the prediction of giant halos in Zr and Ca isotopes, the
continuum contribution and the extension of nuclear landscape,
halos from relativistic and non-relativistic approaches, and
halos with RHFBc theory.
The densities and potentials in exotic nuclei become more diffuse
and the surface diffuseness has important influences on
spin-orbit splittings, single particle level structure,
and nuclear masses. Recent progresses on these topics will be
presented in Section~\ref{sec:surface_diffuseness}.
In Section~\ref{sec:pairing_size}, we will discuss the influence
of pairing correlations on the nuclear size.
The necessity to carry out a self-consistent calculation will be shown in order to understand the impact of the pairing correlations on the nuclear size.
In Section~\ref{sec:deformed_halo},
the study of neutron halo in deformed nuclei will be reviewed, including the neutron separation energies,
quadrupole shapes, density distributions, and radii. The generic conditions for
the occurrence of halos in
deformed nuclei and the predicted shape decoupling between the
core and the halo will be discussed.
Finally we give a summary and discuss perspectives in
Section~\ref{sec:summary}.

\section{\label{sec:formalism}Covariant density functional theory in continuum}

\subsection{General framework of CDFT}

The covariant density functionals represent a number of
relativistic mean field (RMF) and relativistic Hartree-Fock (RHF) models
based on the framework of quantum hadrodynamics (QHD)~\cite{Serot1986_ANP16-1}.
The advantages in using covariant functionals include the natural inclusion
of the nucleon spin degree of freedom, the consistent treatment of
isoscalar Lorentz scalar and vector self-energies which provides a
unique parametrization of time-odd components of the nuclear mean field,
a natural explanation of the empirical pseudospin symmetry,
and a new saturation mechanism of nuclear matter with
a distinction between scalar and four-vector nucleon self-energies
\cite{Ring1996_PPNP37-193, Afanasjev1999_PR322-1,
Ginocchio2005_PR414-165, Vretenar2005_PR409-101, Meng2006_PPNP57-470,
Meng2013_FPC8-55,Liang2015PhysicsReports1}.

The conventional CDFTs are based on the finite-range meson-exchange
representation, in which a nucleus is described as a self-bound system of
Dirac nucleons interacting with each other via the exchange of mesons
through an effective Lagrangian density, which is associated with the nucleon
($\psi$), the $\sigma$-, $\omega$-, $\rho$-, and $\pi$-meson fields,
and the photon field ($A$),
\begin{equation}
   {\cal L}={\cal L}_{N}+{\cal L}_{\sigma}+{\cal L}_{\omega }+{\cal L}_{\rho}+{\cal L}_{\pi}+{\cal
      L}_{A}+{\cal L}_{I}, \label{Eq:Lagrangian}
\end{equation}
containing the Lagrangian densities for the nucleon $\scr L_N$,
\begin{eqnarray}\label{Eq:Lagrangian1}
    {\cal L_N }       =  & \bar\psi\lrb{i\gamma^\mu\partial_\mu - M}\psi,
\end{eqnarray}
for the $\sigma$-, $\omega$-, $\rho$-, and $\pi$-mesons, and the photon $A$,
\begin{eqnarray}\label{Lagrangian1}
    {\cal L}_{\sigma} =  & + \frac{1}{2}\partial^{\mu}\sigma\partial_{\mu}\sigma-\frac{1}{2}m_{\sigma}^{2}\sigma^{2},
                         \nonumber \\
    {\cal L}_{\omega}  =  & - \frac{1}{4}\Omega^{\mu\nu}\Omega_{\mu\nu}+\frac
    {1}{2}m_{\omega}^{2}\omega_{\mu}\omega^{\mu},
                        \nonumber \\
    {\cal L}_{\rho}   =  & - \frac{1}{4} \vec{R}_{\mu\nu}  \cdot
     \vec{R}^{\mu \nu}   +   \frac{1}{2} m_{\rho}^{2}\vec{\rho}^{\mu} \cdot \vec{\rho}_{\mu},
                         \nonumber \\
    {\cal L}_{\pi}=  & +\frac{1}{2}\partial_{\mu}\vec{\pi}\cdot
                         \partial^{\mu}\vec{\pi}-\frac{1}{2}m_{\pi}^{2}\vec{\pi}\cdot \vec{\pi},
                         \nonumber \\
    {\cal L}_{A}=  &  -\frac{1}{4}F^{\mu\nu}F_{\mu\nu}, \nonumber
\end{eqnarray}
and for the interactions between nucleon and meson fields,
\begin{eqnarray}
    \scr L_I =& \bar\psi \left[ -g_\sigma\sigma -
    g_\omega\gamma^\mu\omega_\mu -
    g_\rho\gamma^\mu\ivec\tau \cdot \ivec\rho_\mu -
      \frac{f_{\rho}} {2M} \sigma^{\mu\nu} \partial_{\nu} \ivec\rho_\mu \cdot \vec{\tau} \right.\nonumber\\
       &\left.\qquad -\frac{f_\pi}{m_\pi}\gamma_5\gamma^\mu\partial_\mu
    \ivec\pi \cdot \ivec\tau
    - e\gamma^\mu\frac{1-\tau_3}{2}
    A_\mu \right] \psi.
\end{eqnarray}

In above Lagrangian densities, $M$ denotes the mass of nucleon, and $m_{\sigma}$ ($g_{\sigma}$), $m_{\omega}$ ($g_{\omega}$), $m_{\rho}$ ($g_{\rho},f_{\rho}$), and $m_{\pi}$
($f_{\pi}$) are respectively the masses (coupling constants) for $\sigma$-, $\omega$-, $\rho$-, and $\pi$-mesons. The $\rho$-tensor coupling term, which is practically negligible at Hartree approximation, can significantly improves the descriptions of nuclear shell structures
at Hartree-Fock approximation~\cite{Long2007_PRC76-034314}.

The field tensors of the vector mesons $\Omega$ and $\vec{R}$ and the electromagnetic
field $F$ are defined as,
\begin{eqnarray}
   \Omega^{\mu\nu}  &=& \partial^\mu\omega^\nu
                   - \partial^\nu\omega^\mu, \nonumber \\
   \vec R^{\mu\nu} &=& \partial^\mu\vec\rho^\nu
                   - \partial^\nu\vec\rho^\mu, \\
   F^{\mu\nu}    &=& \partial^\mu A^\nu
                   - \partial^\nu A^\mu. \nonumber
\end{eqnarray}

The isoscalar-scalar $\sigma$-meson provides the
long-range attractive part of the nuclear interaction
whereas the short-range repulsion is governed by the
isoscalar-vector $\omega$-meson. The photon field $A^\mu(x)$
accounts for the Coulomb interaction and the isospin dependence
of the nuclear force is described by the isovector-vector
$\rho$-meson.
The $\pi$-meson field and the rho-tensor coupling contribute only if the Fock terms are included.

In the following, the vectors in the isospin space are denoted by arrows and
the space vectors by bold type. Greek indices $\mu$ and $\nu$ run over
the Minkowski indices $0$, $1$, $2$, $3$ or $t$, $x$, $y$, $z$,
while Roman indices $i$, $j$, etc. denote the spatial components.

The variation of the Lagrangian ${L}=\int d^{3}x{\cal L}(x)$
with respect to nucleon ($\psi$) and meson fields ($\phi_{i}$)
leads to the Dirac equation of nucleon,
 \begin{equation}
\left(  -i\gamma^{\mu}\partial_{\mu}+M+\Sigma\right)  \psi(x)=0, \label{Dirac}
 \end{equation}
with the self-energy $\Sigma$ to be self-consistently determined,
and the Klein-Gordon equations for the meson fields
($\phi=\sigma,\omega_{\mu}  ,\vec{\rho}_{\mu}$ and $\vec{\pi}$),
 \begin{equation}
     \left( \partial^{\mu} \partial_{\mu}  + m_{\phi}^{2}\right)  \phi=S_{\phi},
 \end{equation}
with the source terms
\begin{equation}
S_{\phi}=\left\{
\begin{array}[c]{ll}
  -g_{\sigma}\bar{\psi} \psi,           & \phi=\sigma,\\[0.5em]
   g_{\omega}\bar{\psi}\gamma_{\mu}\psi,& \phi=\omega_{\mu},\\[0.5em]
   g_{\rho}\bar{\psi}\gamma_{\mu}\vec{\tau}\psi + \partial^{\nu}\frac{f_{\rho}}
           {2M}\bar{\psi}\sigma_{\mu\nu}\vec{\tau}\psi,~~~~
                                        & \phi=\vec{\rho}_{\mu },\\[0.5em]
   \partial^{\nu}\frac{f_{\pi}}{m_{\pi}}\bar{\psi}\gamma_{5}\gamma_{\nu}\vec {\tau}\psi,
                                        & \phi=\vec{\pi},
\end{array} \right.
 \end{equation}
and the Proca equation for the photon field ($A$),
 \begin{equation}
\partial^{\nu}F_{\nu\mu}=e\bar{\psi}\frac{1-\tau_{3}}{2}\gamma_{\mu}\psi.
 \end{equation}

Under the no-sea and mean field approximations, the Dirac Hartree-Fock equation
can be derived as
 \beq\label{Spherical}
\int d\svec r' h(\svec r, \svec r') \psi(\svec r')  = \varepsilon\psi(\svec r),
 \eeq
where $\varepsilon$ is the single particle energy ({including the rest mass}) and
the single particle Dirac Hamiltonian $ h(\svec r, \svec r')$ contains
the kinetic energy $h^{\rm{kin}}$,
the direct {local potential $h^{\rm{D}}$, and the
exchange non-local potential $h^{\rm{E}}$,}
\begin{eqnarray} 
&h^{\rm{kin}}(\svec r, \svec r') = \lrs{\svec\alpha\cdot\svec p +
              \beta M }\delta(\svec r - \svec r'),\\
&h^{\rm{D}}(\svec r, \svec r') = \lrs{\Sigma_T(\svec r)\gamma_5 + \Sigma_0(\svec r) +
              \beta\Sigma_S(\svec r)}\delta(\svec r - \svec r'),\\
&h^{\rm{E}}(\svec r, \svec r') =  \lrb{\bea{cc}Y_G(\svec r, \svec r') &Y_F(\svec r, \svec r') \\[0.5em]
                                         X_G(\svec r, \svec r')&X_F(\svec r, \svec r')\eea}.
 \end{eqnarray} 
In the above expressions, the local scalar $\Sigma_{S}$, vector
$\Sigma_{0}$, and tensor $\Sigma_{T}$ self-energies contain the contributions from the direct {(Hartree)} terms
and the rearrangement terms, while the nonlocal self-energies
$X_G$, $X_F$, $Y_G$, and $Y_F$ come from the exchange (Fock)
terms. More details can be found in Ref.~\cite{Long2010_PRC81-024308}.

If one sticks to the Hartree level (RMF), for a nucleus
with time reversal symmetry, the Dirac equation (\ref{Dirac}) is reduced as
\begin{equation}
 \hat{h}\psi(\bm{r}) = \epsilon \psi_{k}(\bm{r}) ,
 \label{eq:Diracequation}
\end{equation}
where
\begin{equation}
 \hat{h} = \bm{\alpha} \cdot \bm{p}
         + \beta \left[ M+ \Sigma_S(\bm{r}) \right]
         + \Sigma_0(\bm{r}),
 \label{eq:Dirac}
\end{equation}
is the single particle Hamiltonian.

More recently, the RMF framework has been reinterpreted by
the relativistic Kohn-Sham density functional theory,
and the functionals have been developed based on
the zero-range point-coupling interaction
\cite{Nikolaus1992_PRC46-1757, Burvenich2002_PRC65-044308,
Niksic2008_PRC78-034318, Zhao2010_PRC82-054319}. In fact, by solving formally the Klein-Gordon equations for mesons and taking the large meson mass limit, one can derive
point-coupling functionals
in which the meson in each channel (scalar-isoscalar,
vector-isoscalar, scalar-isovector, and vector-isovector)
is replaced by the corresponding local four-point
(contact) interaction between nucleons.

The basic building blocks of CDFT with point-coupling vertices are
\begin{equation}
 (\bar\psi{\cal O}\Gamma\psi),~~~~~{\cal O}\in\{1,\vec{\tau}\},
 ~~~~~\Gamma\in\{1,\gamma_\mu,\gamma_5,    \gamma_5\gamma_\mu,\sigma_{\mu\nu}\},
\end{equation}
where $\psi$ is the Dirac spinor field of nucleon, $\vec{\tau}$ is
the isospin Pauli matrix, and $\Gamma$ generally denotes the $4\times4$ Dirac matrices.
There are ten of such building blocks characterized by their transformation
characteristics in isospin and Minkowski space.
A general effective Lagrangian can be written as a power series in
$\bar\psi{\cal O}\Gamma\psi$ and their derivatives,
with higher-order terms representing in-medium many-body correlations.
For details, see Refs.~\cite{Meng2013_FPC8-55, Nikolaus1992_PRC46-1757,
Burvenich2002_PRC65-044308, Niksic2008_PRC78-034318,
Zhao2010_PRC82-054319}.

The point-coupling model has attracted more and more attention
owing to the following advantages.
First, it avoids the possible physical constrains introduced
by the explicit usage of the Klein-Gordon equation to describe the mean meson fields,
especially the fictitious $\sigma$ meson.
Second, it is possible to study the role of naturalness
\cite{Friar1996_PRC53-87, Manohar1984_NPB234-12}
in effective theories for nuclear structure related problems.
Third, it is relatively easy to include the Fock terms~\cite{Liang2012_PRC86-021302R},
and provides more opportunities to investigate its relationship to
the nonrelativistic approaches~\cite{Sulaksono2003_APNP308-70}.

\subsection{Relativistic Hartree-Fock-Bogoliubov theory}

Pairing correlations are crucial in the description of
open-shell nuclei. For exotic nuclei, the conventional BCS
approach turns out to be only a poor approximation
\cite{Dobaczewski1984_NPA422-103,Meng2006_PPNP57-470}.
Starting from the RMF Lagrangian density, a relativistic theory
of pairing correlations in nuclei has been developed by
Kucharek and Ring~\cite{Kucharek1991_ZPA339-23,Ring1996_PPNP37-193},
\beq\label{RHFBEQ}
  \int d\svec r' H (\svec r,\svec r')
     \lrb{\bea{c}\psi_U(\svec r') \\[0.5em] \psi_V(\svec r')\eea}
  = E \lrb{\bea{c} \psi_U(\svec r)\\[0.5em] \psi_V(\svec r)\eea},
\eeq
with
\beq\label{RHFBHam}
H (\svec r,\svec r') =
     \lrb{ \bea{cc}
         h(\svec r,\svec r') - \lambda\delta(\svec r-\svec r') & \Delta(\svec r,\svec r') \\[0.5em]
         \Delta(\svec r,\svec r')   & -h(\svec r,\svec r') +\lambda\delta(\svec r-\svec r')
           \eea},
\eeq
where the chemical potential $\lambda$ $\ $is introduced to preserve
the particle number on the average.

The single particle Hamiltonian $h(\svec r,\svec r')$ is derived from
Eq.~(\ref{Dirac}) with the retardation effects neglected, e.g.,
the Hamiltonian given in Eq.~(\ref{eq:Diracequation})
under the Hartree approximation.
The pairing potential can be written as,
\begin{equation}
\Delta_{\alpha}({\mbox{\boldmath${ r}$}},{\mbox{\boldmath${ r}$}}^{\prime
})=-\frac{1}{2}\sum_{\beta}V_{\alpha\beta}^{pp}\left( {\mbox{\boldmath${ r}$}},{\mbox{\boldmath${
r}$}}^{\prime}\right) \kappa_{\beta}({\mbox{\boldmath${ r}$}},{\mbox{\boldmath${ r}$}}^{\prime}),
\label{pairingp}
\end{equation}
where the pairing tensor $\kappa$ is
\begin{equation}
\kappa_{\alpha}({\mbox{\boldmath${ r}$}},{\mbox{\boldmath${ r}$}}^{\prime
})=\psi_{V_{\alpha}}({\mbox{\boldmath${ r}$}})^{\ast}\psi_{U_{\alpha}%
}({\mbox{\boldmath${ r}$}}^{\prime}).
\end{equation}
The pairing force is either taken as a density-dependent two-body force in a zero range limit,
\begin{equation}
V({\mbox{\boldmath${ r}$}},{\mbox{\boldmath${ r}$}}^{\prime})=V_{0} \delta({\mbox{\boldmath${
r}$}}-{\mbox{\boldmath${ r}$}}^{\prime})\frac{1}{4}\left(  1-{\mbox{\boldmath${ \sigma
}$}}\cdot{\mbox{\boldmath${ \sigma}$}}^{\prime}\right) \left(  1-\frac {\rho(r)}{\rho_{0}}\right) ,
\label{Delta}
\end{equation}
with {an adjusted} strength $V_{0}$, or
as the pairing part of the Gogny force~\cite{Berger1984_NPA428-23},
 \beq
\label{Gogny}
V(\svec r, \svec r') = \sum_{i = 1, 2}e^{\lrb{\lrb{r-r'}/\mu_i}^2}
\lrb{W_i +
B_iP^\sigma - H_i P^\tau - M_iP^\sigma P^\tau},
\eeq
with the parameters $\mu_{i}$, $W_{i}$, $B_{i}$, $H_{i}$ and $M_{i}$.
It has been shown that after proper renormalization,
the above density-dependent force of zero range and
the finite range Gogny force
produce essentially the same results~\cite{Meng1998_PRC57-1229}.

\subsection{Relativistic continuum Hartree-Bogoliubov theory}

If the Fock terms are neglected,
as it is usually done in the covariant density functional theory,
the Dirac Hartree-Bogoliubov equation for the nucleons
still takes the same form as that given in Eq.~(\ref{RHFBEQ}).
For spherical nuclei, the quasiparticle wave functions read,
\begin{eqnarray}
   \psi_U^k & = & \frac 1 r
      \left( \mbox{i} {  G_U^{k}(r)}  {\mathcal{Y}}^l_{jm} (\theta,\phi)
             \atop
                      { -F_U^{k}(r)}  {\mathcal{Y}}^{l'}_{jm} (\theta,\phi)
       \right),\ \
 \label{eq:RHB_U}
 \\
   \psi_V^k & = & \frac 1 r
       \left( \mbox{i}   {G_V^{k}(r)}  {\mathcal{Y}}^l_{jm} (\theta,\phi)
               \atop
                        -{F_V^{k}(r)}  {\mathcal{Y}}^{l'}_{jm} (\theta,\phi)
       \right).
 \label{eq:RHB_V}
\end{eqnarray}

The angular part of the RHB equation can be integrated out and
the radial RHB equations are derived as
\begin{eqnarray}
 \hspace*{-2.5cm}
\left\{
   \begin{array}{lll}
      \displaystyle
      \frac {d G_U(r)} {dr} + \frac {\kappa} r G_U(r) -
       ( E + \lambda-\Sigma_0(r) + \Sigma_S(r) ) F_U(r) +
         r \int r'dr' \Delta_F(r,r')  F_V(r') &=& 0, \\
      \displaystyle
      \frac {d F_U(r)} {dr} - \frac {\kappa} r F_U(r) +
       ( E + \lambda-\Sigma_0(r)-\Sigma_S(r) ) G_U(r) +
         r \int r'dr' \Delta_G(r,r') G_V(r') &=& 0,\\
      \displaystyle
      \frac {d G_V(r)} {dr} + \frac {\kappa} r G_V(r) +
       ( E - \lambda+\Sigma_0(r)-\Sigma_S(r) ) F_V(r) +
         r \int r'dr' \Delta_F(r,r') F_U(r') &=& 0,\\
      \displaystyle
      \frac {d F_V(r)} {dr} - \frac {\kappa} r F_V(r) -
       ( E - \lambda+\Sigma_0(r)+\Sigma_S(r) ) G_V(r) +
         r \int r'dr' \Delta_G(r,r')  G_U(r') &=& 0. \\
\end{array}
\right.
\label{CoupEq}
\end{eqnarray}
The pairing potentials are calculated as
\begin{eqnarray}
    \Delta_{G} ( r, r')
    &=& \frac 1 4 \sum_{\kappa'}
    V_{\kappa \kappa '}^{J=0} ( r, r' )  g_{\kappa'}
    \sum_{E_k > 0} [ G_U^k(r) G_V^k (r') + G_U^k(r') G_V^k (r) ] , \\
    \Delta_{F} ( r, r')
    &=& \frac 1 4 \sum_{\kappa'}
    V_{\kappa \kappa '}^{J=0} ( r, r' )  g_{\kappa'}
    \sum_{E_k > 0} [ F_U^k(r) F_V^k (r') + F_U^k(r') F_V^k (r) ] .
\label{gap1}
\end{eqnarray}
If the zero-range pairing force is used, the above coupled integro-differential
equations are reduced to differential ones, which can be directly solved in coordinate space~\cite{Meng1998Nucl.Phys.A3}.

\subsection{Relativistic Hartree-Fock-Bogoliubov theory in continuum}

For spherical systems, the solution of the RHFB equations,
i.e., the Dirac spinor $\psi_{U_\alpha}$ and $\psi_{V_\alpha}$,
can be written {similarly to} Eqs.~(\ref{eq:RHB_U}) and (\ref{eq:RHB_V}),
\begin{eqnarray}
   \psi_{U_{\alpha}}({\mbox{\boldmath${ r}$}})=  &  \frac{1}{r}
   \left(
       \begin{array} [c]{c}%
           iG_{U_{a}}(r){\mathcal{Y}}_{j_{a}m_{a}}^{l_{a}}(\hat{{\mbox{\boldmath${ r}$}}%
           }) \\[0.5em]
           -F_{U_{a}}(r){\mathcal{Y}}_{j_{a}m_{a}}^{l_{a}^{\prime}}(\hat {{\mbox{\boldmath${ r}$}}})
       \end{array}
   \right)  ,\label{SpinorUV} \\
   \psi_{V_{\alpha}}({\mbox{\boldmath${ r}$}})=  &  \frac{1}{r}
   \left(
       \begin{array} [c]{c}
          iG_{V_{a}}(r){\mathcal{Y}}_{j_{a}m_{a}}^{l_{a}}(\hat{{\mbox{\boldmath${ r}$}}%
          }) \\[0.5em]
          -F_{V_{a}}(r){\mathcal{Y}}_{j_{a}m_{a}}^{l_{a}^{\prime}}(\hat {{\mbox{\boldmath${ r}$}}})
       \end{array}
   \right).
\end{eqnarray}
The RHFB equations (\ref{RHFBEQ}) are then reduced to the coupled integro-differential equations,
\begin{eqnarray}
\label{RHFB-R}
   &\lrs{\frac{d}{dr} + \frac{\kappa_a}{r} + \Sigma_T} G_{U_a}(r) - \lrb{E_a + \lambda - \Sigma_-}
     F_{U_a}(r)\re&~~~~~~~~~~ + X_{U_a} (r) + r\int r'dr' \Delta_a(r, r') F_{V_a}(r') = 0,
     \\
   &\lrs{\frac{d}{dr} - \frac{\kappa_a}{r} - \Sigma_T} F_{U_a}(r) + \lrb{E_a + \lambda - \Sigma_+}
     G_{U_a}(r)\re&~~~~~~~~~~ - Y_{U_a} (r) + r\int r'dr' \Delta_a(r, r') G_{V_a}(r') = 0,
     \nonumber \\
   &\lrs{\frac{d}{dr} + \frac{\kappa_a}{r} + \Sigma_T} G_{V_a}(r) + \lrb{E_a - \lambda + \Sigma_-}
     F_{V_a}(r)\re&~~~~~~~~~~ + X_{V_a} (r) + r\int r'dr' \Delta_a(r, r') F_{U_a}(r') = 0,
     \nonumber \\
   &\lrs{\frac{d}{dr} - \frac{\kappa_a}{r} - \Sigma_T} F_{V_a}(r) - \lrb{E_a - \lambda + \Sigma_+}
     G_{V_a}(r)\re&~~~~~~~~~~ - Y_{V_a} (r) + r\int r'dr' \Delta_a(r, r') G_{U_a}(r') = 0,
     \nonumber
 \end{eqnarray}
where $E_{a}$ are the quasi-particle energies (without the rest mass),
and the local self-energies $\Sigma_{+}$ and $\Sigma_{-}$ are
 \begin{eqnarray} 
\Sigma_{+}\equiv&\Sigma_{0}+\Sigma_{S},\\
\Sigma_{-}\equiv&\Sigma_{0}-\Sigma _{S}-2M.
 \end{eqnarray} 
More details can be found in Ref.~\cite{Long2010_PRC81-024308}.

The pairing potentials $\Delta_a(r,r^{\prime})$ in Eq. (\ref{RHFB-R}) can be expressed as
\begin{equation}
\Delta_{a}(r,r^{\prime})=-\sum_{b}V_{ab}^{pp}(r,r^{\prime})\kappa
_{b}(r,r^{\prime}),
\end{equation}
where the pairing tensor $\kappa(r,r^{\prime})$ reads
\beq
\hspace*{-2.5cm}
\kappa_b(r,r') = \ff2\hat j_b^2\lrs{G_{U_b}(r) G_{V_b}(r') + F_{U_b}(r) F_{V_b}(r')}
 + \ff2\hat
j_b^2\lrs{G_{V_b}(r) G_{U_b}(r') + F_{V_b}(r) F_{U_b}(r')},
\eeq
where $\hat j^2=2j+1$.
Details of the pairing interaction matrix element $V_{ab}^{pp}$ can be found
in Ref.~\cite{Meng1998_NPA635-3}.

In contrast to the RHB approach with $\delta$-force in the pairing channel
where the radial equations (\ref{RHFB-R}) {become} differential equations,
in RHFB theory the radial equations are {fully} integro-differential.
The integral terms may arise from the Fock terms or
the pairing channel if the finite-range pairing force is used.
In coordinate space, it is difficult to solve such equations
by the localization procedure adopted in
Refs.~\cite{Bouyssy1987_PRC36-380, Long2006_PLB640-150} when solving the relativistic Hartree-Fock equations.
It is more convenient to solve them by an expansion of
the Dirac-Bogoliubov spinors in an appropriate basis.
This has been done by using the Dirac Woods-Saxon (DWS) basis
\cite{Zhou2003_PRC68-034323}.
This basis has been constructed for the investigation of weakly-bound nuclei.
The set of DWS basis functions
\begin{equation}
\left\{  \left[  \varepsilon_{b},g_{\beta}({\mbox{\boldmath${ r}$}},\tau)\right]  ;\varepsilon_{b}\gtrless0\right\}  ,
\label{WSbase}%
\end{equation}
are eigenfunctions (with eigenvalues $\varepsilon_{b}$) of a Dirac equation with Woods-Saxon-like potentials for $\Sigma_{0}(r)\pm\Sigma_{S}(r)$. They are determined by the shooting method in {coordinate} space within a spherical box of size $R_{\rm{max}}$~\cite{Koepf1991_ZPA339-81}.

The $U$ and $V$ components of the Dirac-Bogoliubov spinors (\ref{SpinorUV}) can be expanded as,
\begin{eqnarray} 
\label{WS-Expand}
\psi_{U}=&\sum_{p=1}^{N_{F}}U_{p}g_{p}+\sum_{d=1}^{N_{D}}U_{d}g_{d},\\
\psi_{V}=&\sum_{p=1}^{N_{F}}V_{p}g_{p}+\sum_{d=1}^{N_{D}}V_{d}g_{d},  \nonumber
 \end{eqnarray} 
where $N_{F}$ and $N_{D}$ respectively correspond to the numbers of positive ($\varepsilon_{p}>0$) and negative ($\varepsilon_{d}<0$) energy states in the DWS basis. Obviously, because of spherical symmetry the quantum number $\kappa$ is preserved, i.e., the RHFB equations can be solved for each value of $\kappa$ and the sums in the expansion (\ref{WS-Expand}) run only over states with the same $\kappa$.  For a fixed value of $\kappa$ we have the radial basis spinors
\begin{eqnarray} 
g_{p}({r})=&\left(\begin{array}[c]{c}G_{p}(r)\\[0.5em]
F_{p}(r)\end{array} \right),
\\
g_{d}({r})=&\left(\begin{array}[c]{c}G_{d}(r)\\[0.5em]
F_{d}(r)\end{array} \right),%
\label{DWS-basis}
\end{eqnarray} 
where the sub-indices $p$ and $d$ correspond to the number of nodes of the basis functions $G_{p}$ for positive energy and $F_{d}$ for negative energy.

In the DWS basis (\ref{WS-Expand}) the radial RHFB equations (\ref{RHFB-R}) are transformed to a matrix eigenvalue problem,
 \beq\label{eigenUV}
\lrb{\bea{cc}H-\lambda & \Delta \\[0.5em] \Delta & -H+\lambda\eea}\lrb{\bea{c}U\\[0.5em] V\eea}
= E\lrb{\bea{c}U\\[0.5em] V\eea},
 \eeq
where $H$ and $\Delta$ are $\left(N_{F}+N_{D}\right)\times\left( N_{F}+N_{D}\right)$-dimensional matrices, $U$ and $V$ are the column vectors with $N_{F}+N_{D}$ elements. From the expressions of the single particle Hamiltonian $h$ and pairing potential $\Delta$ given in the previous part we obtain the matrix elements of $H$ and $\Delta$ as
\begin{eqnarray} 
 H_{nn'}^{\rm{kin}}&=&\int dr G_{n}(r) \lrb{-\frac{d}{dr}+\frac{\kappa}{r}} F_{n'}(r) +
                      \int dr F_{n}(r) \lrb{ \frac{d}{dr}+\frac{\kappa}{r}} G_{n'}(r), \\
 H_{nn'}^{\rm{D}}  &=&\int dr \lrs{ G_{n}(r)G_{n'}(r) \Sigma_+(r) +
                                    F_{n}(r)F_{n'}(r) \Sigma_-(r) } + \nonumber \\
                   & &
                      \int dr{\lrs{ G_{n}(r)F_{n'}(r) + G_{n}(r)F_{n'}(r)} \Sigma_T(r)},\\
 H^{\rm{E}}_{nn'}  &=&\int dr \int dr'\lrb{\bea{cc}G_{n}& F_{n}\eea}_r \lrb{\bea{cc}
Y_{G} &Y_{F}\\[0.5em]X_{G}&X_{F}\eea}_{(r,r')}\lrb{\bea{c}G_{n'}\\[0.5em] F_{n'}\eea}_{r'},
\label{eq:HEnnprime} \\
\Delta_{nn'}&=&\int dr  \int dr'\Delta_\kappa(r,r')  \lrs{ G_{n}(r)G_{n'}(r') +
F_{n}(r)F_{n'}(r')},
 \end{eqnarray} 
where $n$ and $n^{\prime}$ run over the radial quantum numbers of
the DWS basis states in Eq.~(\ref{DWS-basis}) with both positive energies ($n, n'=p$)
and negative energies ($n, n'=d$).

\subsection{Deformed relativistic Hartree-Bogoliubov theory in continuum}

For deformed nuclei the solution of HFB or RHB equations in
$r$ space is numerically a very demanding task. Considerable efforts have been made to develop mean field models
either in $r$ space or in a basis with an improved asymptotic
behavior at large
distances~\cite{Terasaki1996_NPA600-371, Stoitsov2000_PRC61-034311,
Teran2003_PRC67-064314, Zhou2003_PRC68-034323,
Tajima2004_PRC69-034305, Stoitsov2008_PRC77-054301, Nakada2008_NPA808-47}.
In particular, an expansion in a Woods-Saxon (WS) basis was proposed
as a reconciler between the HO basis and the integration in coordinate space
\cite{Zhou2003_PRC68-034323}.
Woods-Saxon wave functions have a more realistic asymptotic behavior at large $r$ than
the harmonic oscillator wave functions.
A discrete set of Woods-Saxon wave functions is obtained by using box boundary conditions
to discretize the continuum.
It has been shown in Ref.~\cite{Zhou2003_PRC68-034323} for spherical systems that the solution
of the relativistic Hartree equations in a Woods-Saxon basis
is equivalent to the solution in coordinate space.
The Woods-Saxon basis has been used recently for both
nonrelativistic~\cite{Schunck2008_PRC78-064305}
and relativistic Hartree-Fock-Bogoliubov theories~\cite{Long2010_PRC81-024308}
with a finite range pairing force.

Over the past years, lots of efforts have been made to develop
a deformed relativistic Hartree (RH) theory~\cite{Zhou2006_AIPCP865-90} and
a deformed relativistic Hartree-Bogoliubov theory in continuum (DRHBc theory)~\cite{Zhou2008_ISPUN2007}.
As a first application, halo phenomena in deformed nuclei
have been investigated within the DRHBc theory~\cite{Zhou2010_PRC82-011301R}.
The detailed theoretical framework is
available in Ref.~\cite{Li2012_PRC85-024312}.

For axially deformed nuclei with spatial reflection symmetry, the potentials $\Sigma_S(\bm{r})$ and $\Sigma_0(\bm{r})$ in Eq.~(\ref{eq:Dirac})
and various densities can be expanded in terms of the Legendre
polynomials~\cite{Price1987_PRC36-354},
\begin{equation}
 f(\bm{r})   = \sum_\lambda f_\lambda({r}) P_\lambda(\cos\theta),\
 \lambda = 0,2,4,\cdots
 ,
 \label{eq:expansion}
\end{equation}
with
\begin{equation}
 f_{\lambda}(r) = \frac{2\lambda+1}{2}\int_{-1}^{1}d(\cos\theta)f(\bm r)P_{\lambda}(\cos\theta).
\end{equation}
The quasiparticle wave functions $\psi_{U_k}$ and $\psi_{V_k}$ are expanded in terms of
spherical Dirac spinors $\varphi_{n\kappa m}(\bm r{s} p)$
with the eigenvalues $\epsilon_{n\kappa}$ obtained from the solution of a Dirac equation
$h^{(0)}_D$ containing spherical potentials $S^{(0)}(r)$ and $V^{(0)}(r)$ of
Woods-Saxon shape~\cite{Zhou2003_PRC68-034323, Koepf1991_ZPA339-81}:
\begin{eqnarray}
 \psi_{U_k} (\bm r{s} p)
 & = & \displaystyle
 \sum_{n\kappa} u^{(m)}_{k,(n\kappa)}     \varphi_{n\kappa m}(\bm r{s} p),
 \label{eq:Uexpansion0} \\
 \psi_{V_k} (\bm r{s} p)
 & = & \displaystyle
 \sum_{n\kappa} v^{(m)}_{k,(n\kappa)} \bar\varphi_{n\kappa m}(\bm r{s} p).
\label{eq:Vexpansion0}
\end{eqnarray}
The basis wave function reads
\begin{equation}
 \varphi_{n\kappa m}(\bm{r}{s}) =
   \frac{1}{r}
   \left(
     \begin{array}{c}
       i G_{n\kappa}(r) {\mathcal{Y}}^l _{jm} (\Omega{s})
       \\
       - F_{n\kappa}(r) {\mathcal{Y}}^{\tilde l}_{jm}(\Omega{s})
     \end{array}
   \right) ,
\label{eq:SRHspinor}
\end{equation}
where $G_{n\kappa}(r) / r$ and $F_{n\kappa}(r) / r$ respectively the radial wave
functions for the upper and lower components. The state
$\bar\varphi_{n\kappa m}(\bm r{s} p)$ is the time reversal state of $\varphi_{n\kappa
m}(\bm r{s} p)$.
The spherical spinor $\varphi_{n\kappa m}$ is characterized
by the radial quantum number $n$, angular momentum $j$, and the parity $\pi$;
the latter two are combined to the relativistic quantum number
$\kappa=\pi(-1)^{j+1/2} (j+1/2)$ which runs over positive and negative
integers $\kappa=\pm 1,\pm 2,\cdots$. For the spinor spherical harmonics
$\mathcal{Y}^l _{jm}$ and $\mathcal{Y}^{\tilde{l}}_{jm}$, $l = j + \frac{1}{2}{\rm sign}(\kappa)$ and
$\tilde l = j - \frac{1}{2}{\rm sign}(\kappa)$.
These states form a complete spherical and discrete basis in Dirac space
(see Ref.~\cite{Zhou2003_PRC68-034323} for details).
Because of the axial symmetry the $z$-component $m$ of
the angular momentum $j$ is a conserved quantum number and the RHB Hamiltonian can
be decomposed into blocks characterized by $m$ and parity $\pi$. For
each $(m\pi)$-block, solving the RHB equation is
equivalent to the diagonalization of the matrix
\begin{equation}
 \left( \begin{array}{cc}
  {\cal A}-\lambda & {\cal B} \\
  {\cal B^\dag} & -{\cal A}^\ast+\lambda \\
 \end{array} \right)
 \left(
  { {\cal U}_k
    \atop
    {\cal V}_k
  }
 \right)
 = E_k
 \left(
  { {\cal U}_k
    \atop
    {\cal V}_k
  }
 \right),
 \label{eq:RHB1}
\end{equation}
where
\begin{equation}
 {\cal U}_k = \left(u^{(m\pi)}_{k,(n\kappa)}\right),\
 {\cal V}_k = \left(v^{(m\pi)}_{k,(n\kappa)}\right),
\end{equation}
and
\begin{eqnarray}
 {\cal A}
 & = &
 \left( h^{(m\pi)}_{D(n\kappa)(n'\kappa')} \right)
 =
 \left( \langle n\kappa m\pi|h_D|n'\kappa'm\pi\rangle \right) ,
 \\
 {\cal B}
 & = &
 \left( \Delta^{(m\pi)}_{(n\kappa)(n'\kappa)} \right)~
 =
 \left( \langle n\kappa m\pi |\Delta| \overline{n'\kappa'm\pi} \rangle \right).
\label{eq:pairing_matrix}
\end{eqnarray}
Further details are given in Ref.~\cite{Li2012_PRC85-024312}.

After solving the RHB or RHFB equations self-consistently, one gets
the quasiparticle wave functions and meson fields of a nucleus.
Then various physical quantities can be calculated.
Here we take the deformed RHBc as an example and show how the total energy, the radius, and the deformation
parameters are calculated~\cite{Li2012_PRC85-024312}.

The total energy of a nucleus is
\begin{eqnarray}
 E
 & = &
  E_\mathrm{nucleon} + E_{\sigma} + E_{\omega} + E_{\rho}
  + E_c + E_\mathrm{c.m.}
 \nonumber  \\
 & = &
  \sum_{k,(m\pi)} 2(\lambda - E_{k}) v^2_{k,(m\pi)} - E_\mathrm{pair}
 \nonumber  \\
 &   &
 -\dfrac{1}{2} \int d^3\bm r \left[ g_\sigma\sigma(\bm r)\rho_s(\bm r) + U(\sigma)
                           \right]
 \nonumber  \\
 &   &
 -\dfrac{1}{2} \int d^3\bm r g_\omega \omega(\bm r) \rho_v(\bm r)
 \nonumber  \\
 &   &
 -\dfrac{1}{2} \int d^3\bm r g_\rho \rho(\bm r)
                \left[ \rho_v^Z(\bm r)-\rho_v^N(\bm r) \right]
 \nonumber  \\
 &   &
 -\dfrac{1}{2} \int d^3\bm r   A_0 \rho_v^Z(\bm r)
 + E_{\rm c.m.},
\end{eqnarray}
where
\begin{equation}
 v^2_{k,(m\pi)} = \int d^3r V_k^\dagger({\bm r})V_k^{}({\bm r})
       = \sum_{n\kappa} \left( v^{(m\pi)}_{k,(n\kappa)} \right)^2,
\end{equation}
and $U(\sigma)$ is the nonlinear self-coupling term of $\sigma$ meson.

For a zero-range force the pairing field $\Delta({\bm r})$ is local, and
the pairing energy is calculated as
\begin{equation}
 E_\mathrm{pair} = -\frac{1}{2}\int d^3r \kappa({\bm r})\Delta({\bm r}).
\end{equation}
Note that in the DRHBc theory, a smooth cutoff with two parameters,
similar to the soft cutoff proposed in Ref.~\cite{Bonche1985_NPA443-39},
has been introduced to regularize the zero-range pairing force;
for the details, the readers are referred to
Refs.~\cite{Zhou2010_PRC82-011301R, Li2012_PRC85-024312}.
The center of mass correction energy
\begin{equation}
 E_{\rm c.m.} = - \frac{1}{2Am} \langle \hat{\mathbf{P}}^2 \rangle,
\label{Ecm}
\end{equation}
is calculated after variation with the wave functions of the
self-consistent solution~\cite{Long2004_PRC69-034319, Zhao2009_CPL26-112102}
or in the oscillator approximation
\begin{eqnarray}
 E_{\rm c.m.} = -\dfrac{3}{4} \times 41 \times A^{1/3}\ \mathrm{MeV},
 \label{eq:ECM}
\end{eqnarray}
The root mean square (rms) radius is calculated as
\begin{eqnarray}
 R_{\tau,\mathrm{rms}} \equiv \langle r^2 \rangle^{1/2}
 & = &
  \left( \int d^3\bm r \left[ r^2 \rho_\tau(\bm r) \right] \right)^{1/2}
 \nonumber \\
 & = &
  \left( \int dr \left[ r^4 \rho^\tau_{v,\lambda=0}(r)\right] \right)^{1/2}
 ,
\end{eqnarray}
where $\tau$ represents the proton, the neutron, or the nucleon.
The rms charge radius is calculated simply as $r_{\rm ch}^2 =
r_{\mathrm p}^2 + 0.64$ fm$^2$. The intrinsic multipole moment is
calculated by
\begin{eqnarray}
 Q_{\tau,\lambda}
 & = &
  \sqrt{\frac{16\pi}{2\lambda+1}} \langle r^2 Y_{\lambda0}(\theta,\phi) \rangle
  = 2 \langle r^2 P_{\lambda}(\theta) \rangle
 \nonumber \\
 & = &
  \frac{8\pi}{2\lambda+1} \int dr  \left[ r^4 \rho^\tau_{v,\lambda}(r) \right]
 .
\end{eqnarray}
The quadrupole deformation parameter is obtained from the quadrupole
moment by
\begin{equation}
 \beta_{\tau,2} = \frac{ \sqrt{5 \pi}  Q_{\tau,2} } {3 N_\tau \langle r_\tau^2\rangle}
 \ ,
\end{equation}
where $N_\tau$ refers to the number of neutron, proton, or nucleon.

\section{\label{sec:spherical_halos}Neutron halos and giant halos}

Halo is one of the most interesting exotic nuclear phenomena.
There are many new features in the halo nuclei because of
their large spatial distribution.
To give a proper theoretical description of the halo phenomena,
it is crucial to treat properly the coupling between
bound states and the continuum due to pairing correlations
by either the finite range or the zero range pairing
forces and the very extended spatial density distributions.
By assuming spherical symmetry, the RCHB theory~\cite{Meng1998_NPA635-3} provides
a proper treatment of pairing correlations in the presence of
the continuum and an excellent description for exotic nuclei
\cite{Meng1996_PRL77-3963, Meng1998_PRL80-460,
Meng2003_NPA722-C366, Meng2002_PRC65-041302R,
Meng1997_ZPA358-123, Poschl1997_PRL79-3841,
Meng1998_PLB419-1, Meng1998_PRC57-1229, Meng1999_NPA650-176,
Meng2002_PLB532-209, Zhang2002_CPL19-312, Zhang2003_SciChinaG46-632}.
The first microscopic self-consistent description of halo in $^{11}$Li
has been achieved by using the RCHB theory~\cite{Meng1996_PRL77-3963}.
Later, the giant halos in light and medium-heavy nuclei have been predicted
\cite{Meng1998_PRL80-460, Meng1998_NPA635-3,Meng2002_PRC65-041302R,
Zhang2002_CPL19-312,Zhang2003_SciChinaG46-632}.
The RCHB theory has been generalized to treat the odd particle system
\cite{Meng1998_PLB419-1} and extended to including hyperons~\cite{Lu2003_EPJA17-19}.
Recently, the relativistic Hartree-Fock-Bogoliubov (RHFB)
theory in continuum has also been developed and used to study
neutron halos~\cite{Long2010_PRC81-031302R, Long2010_PRC81-024308}.

In this Section, progresses on the description of halos
in spherical nuclei will be reviewed, including
the self-consistent description of the halo in $^{11}$Li,
the giant halo and the extension of nuclear landscape,
halo and giant halo from the non-relativistic and relativistic approaches, and
halos from the RHFBc, etc.

\subsection{Neutron halo in $^{11}$Li}

The ground state properties of Li isotopes have been
investigated by using the RCHB theory~\cite{Meng1996_PRL77-3963}.
A satisfactory agreement with experimental values is found for
the binding energies and the radii of $^{6-11}$Li.
The matter radius, showing a considerable increase from the nucleus $^9$Li to $^{11}$Li,
has been reproduced by the RCHB theory.
In contrast to the earlier mean field calculations,
e.g., Refs.~\cite{Bertsch1989_PRC39-1154, Sagawa1992_PLB286-7, Zhu1994_PLB328-1},
the results given in Ref.~\cite{Meng1996_PRL77-3963}
were obtained without any artificial modifications of the potential.

\begin{figure}
\centering
\includegraphics[width=6.0cm,angle=270]{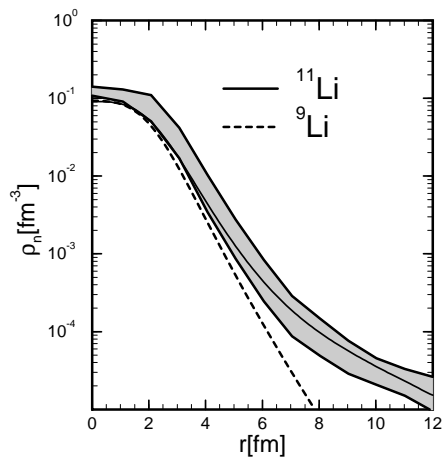}
\caption{Calculated and experimental density distributions in
$^{11}$Li and $^9$Li. The solid line shows the result of $^{11}$Li
while the dashed line corresponds to the calculation for $^9$Li. The
shaded area gives the experimental results with error bars. Taken
from Ref.~\cite{Meng1996_PRL77-3963}.} \label{Fig01}
\end{figure}

In Fig.~\ref{Fig01} the neutron density distributions of
$^9$Li and $^{11}$Li are shown.
It is clearly seen that the increase of the matter radius is caused
by the large neutron density distribution in $^{11}$Li.
Its density distribution is fully coincident with the experimental
density shown with its error bars by the shaded area.

By examining the mean fields for protons and neutrons and the
single particle levels in the canonical basis, it has been
shown that the halo in $^{11}$Li is formed by Cooper-pairs
scattered in the two levels $1p_{1/2}$ and $2s_{1/2}$ with the
former below the Fermi surface and the latter in the continuum;
the pairing interaction couples those levels below
the Fermi surface with the continuum.
The RCHB theory provides a much more general mechanism for halo:
a halo appears if the Fermi level is close to the continuum and around the Fermi level there are single particle levels with low orbital angular momenta and correspondingly
low centrifugal barriers.

\subsection{Giant halos and the extension of nuclear landscape }

The halo nuclei observed so far are composed of one or two halo nucleons only.
The RCHB theory has been used to study the influence of correlations
and many-body effects in nuclei with a larger number of neutrons
distributed in the halo channel.
A giant neutron halo has been predicted in Zr isotopes
close to the neutron drip line~\cite{Meng1998_PRL80-460}.
It is formed by up to six neutrons scattered to weakly-bound or
continuum states beyond the $^{122}$Zr core with the magic neutron number $N=82$.

Before we proceed, it is worthwhile to discuss more about the characterization
of a halo in medium-heavy nuclei which is one of the most relevant topics
in the study of the halo phenomenon.  The critical conditions for the appearance of a halo require the existence of two groups of orbitals with significantly different asymptotic slopes corresponding to the core and halo nucleons. The one-nucleon or multi-nucleon separation energies are close to zero, and there is a significant energy gap for the core and valent nucleons in the spectrum.

Although the theoretical description of light halo nuclei is well under control, as discussed in
Refs.~\cite{Rotival2009_PRC79-054308, Rotival2009_PRC79-054309}, existing definitions and tools are often
too qualitative and the associated observables are incomplete for heavier ones.
There have been lots of efforts in quantifying halos by examining
the density profiles or the particles in the classical forbidden area obtained from mean field calculations
\cite{Meng1998Phys.Lett.B1,Im2000_PRC61-047302, Mizutori2000_PRC61-044326}.
In Ref.~\cite{Rotival2009_PRC79-054308},
a quantitative  analysis method is proposed in which
the halo part of the density is identified as the region beyond
a radius $r_0$ where the core density is one order of magnitude smaller
than the halo one. The method does not require an a priori separation of the density into a halo and a core part. Once $r_0$ is fixed, two criteria are introduced to characterize halo systems:
(i) the average number of fermions participating in the halo and
(ii) the influence of the latter on the system extension \cite{Rotival2009_PRC79-054308}.
This method has been used to characterize halo features
not only in nuclei with the spherical Hartree-Fock-Bogoliubov model
but also in other finite many-fermion systems like
atom-positron/ion-positronium complexes \cite{Rotival2009_PRC79-054309}.
Similar attempts are highly demanded for systematic studies of nuclear halos
within the frame of both relativistic and non-relativistic DFT.
In this Review, we mainly analyze the halo properties based on
a combined analysis of the following features:
i) the deviation of the rms radius from the $r_0 N^{1/3}$ law;
ii) the large spatial extension of the density profile; and
iii) the occupation of weakly bound and/or continuum orbitals in the canonical basis.
For those weakly bound or continuum orbitals,
we also discuss in detail their contribution to the halo.

\begin{figure}
\centering
\includegraphics[width=8.0cm]{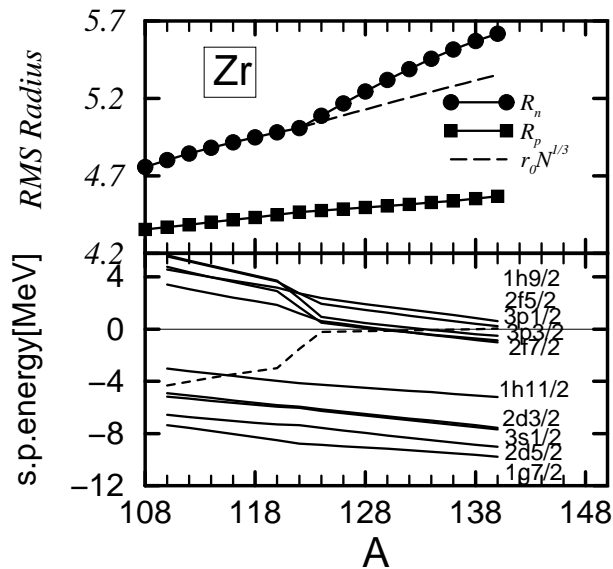}
\caption{Upper part: Root mean square radii for neutrons and protons in Zr
isotopes close to the neutron drip line as a function of the mass
number $A$. Lower part: single particle energies for neutrons in
the canonical basis. The dashed line indicates the chemical
potential. Taken from Ref.~\cite{Meng1998_PRL80-460}.} \label{Fig02}
\end{figure}

In the upper panel of Fig.~\ref{Fig02}, the rms
radii of the protons and neutrons are presented for the Zr isotopes.
There is a clear kink for the neutron rms radius at the magic neutron
number $N=82$. This kink can be understood by
examining the single particle levels in the canonical basis
which is shown in the lower panel of Fig.~\ref{Fig02}.
Going from $N=70$ to $N=100$, there is a big gap above the 1$h_{11/2}$ orbital.
For $N>82$, the neutrons are filled to the levels in the continuum
or weakly bound states in the order of $3p_{3/2}$, $2f_{7/2}$,
$3p_{1/2}$, $2f_{5/2}$ and $1h_{9/2}$.
The neutron chemical potential is given by a dashed
line. It approaches rapidly the continuum threshold already shortly after
the magic neutron number $N=82$ and crosses the continuum at
$N=100$ for the nucleus to $^{140}$Zr. In this region the chemical
potential keeps a very small but negative value for a large $A$ range. This means that the additional neutrons are added with a
very small, nearly vanishing binding energy at the edge of the
continuum provided by pairing correlations. The total binding energies $E$ for the isotopes above
$^{122}$Zr are therefore almost identical. This has been
recognized in Ref.~\cite{Sharma1994_PRL72-1431}.
However, the RMF calculations with the BCS approximation expanded
on an oscillator basis is definitely not appropriate for
exotic nuclei close to the continuum limit.

From the occupation probabilities in the canonical basis, one
can get the number of halo neutrons.
It was found that there are 2 valence neutrons in $^{124}$Zr,
4 in $^{126}$Zr, 6 in $^{128}$Zr, ..., roughly 5 in $^{138}$Zr,
and roughly 6 in $^{140}$Zr where the neutron drip line is reached.
Note that this does not mean that there are around six neutrons
in the halo region of the nuclear density in $^{140}$Zr.
We emphasize that these figures, 2, 4, 5, or 6, are numbers of ``valence'' neutrons,
just as it is well recognized that there are two ``valence'' neutrons in $^{11}$Li.
With the very large neutron rms radii of these systems,
one can estimate the number of valence neutrons which fill in
the same volume outside the $^{122}$Zr core if packed
with normal neutron density.
This number is 24 for $^{134}$Zr and 34 for $^{140}$Zr.
This phenomenon is therefore clearly a neutron halo and
was called a Giant Halo because of the large
number of particles participating in the halo formation \cite{Meng1998_PRL80-460}.

It is very difficult for experimentalists to study giant halos
in Zr isotopes because they are too heavy to be synthesized
by the RIB facilities at present.
Thus it is useful to investigate the giant halo phenomena in lighter nuclei
which are easily accessible with available facilities.
In Refs.~\cite{Meng2002_PRC65-041302R, Zhang2003_SciChinaG46-632},
ground state properties of all the even-even O, Ca, Ni, Zr,
Sn, and Pb isotopes ranging from the proton drip line to the neutron
drip line were investigated by using the RCHB theory with the
effective interaction NLSH~\cite{Sharma1993_PLB312-377}.
Recently, a systematic calculation for nuclei above oxygen has also been
performed with the effective interaction PC-PK1~\cite{Zhao2010_PRC82-054319},
aiming to investigate the global impact of the continuum
for the nuclear boundary;
the results ranging from O to Ti isotopes have been published in
Ref.~\cite{Qu2013_SciChinaPMA56-2031}.
In such a systematic study, it is found that the appearance of giant halos,
the contribution of continuum and boundaries of nuclear chart
are tightly connected. We note that the functional PC-PK1 has proved to be very successful
in describing the isospin dependence of the nuclear masses
\cite{Zhao2012_PRC86-064324}, the Coulomb displacement energies
between mirror nuclei~\cite{Sun2011_SciChinaPMA54-214},
fission barriers~\cite{Lu2012_PRC85-011301R, Lu2014_PRC89-014323}
and nuclear rotations~\cite{Zhao2011_PRL107-122501,
Zhao2011_PLB699-181, Zhao2012_PRC85-054310, Meng2013_FPC8-55}, etc.

For nuclei far from the valley of stability and with small
nucleon separation energy, the Fermi
surface is very close to the continuum threshold and
the valence nucleons may extend over quite a wide space to
form low density nuclear matter.
Therefore the nucleon separation energies are sensitive quantities to
test theory and examine the halo phenomena.
The two-neutron separation energies for O, Ca, Ni, Zr, Sn, and Pb isotopes
have been investigated to examine their connection with systematic
behavior of the nuclear size~\cite{Meng2002_PRC65-041302R,
Zhang2002_CPL19-312, Zhang2003_SciChinaG46-632}.

\begin{figure}[h]
\begin{center}
\includegraphics[width=9cm]{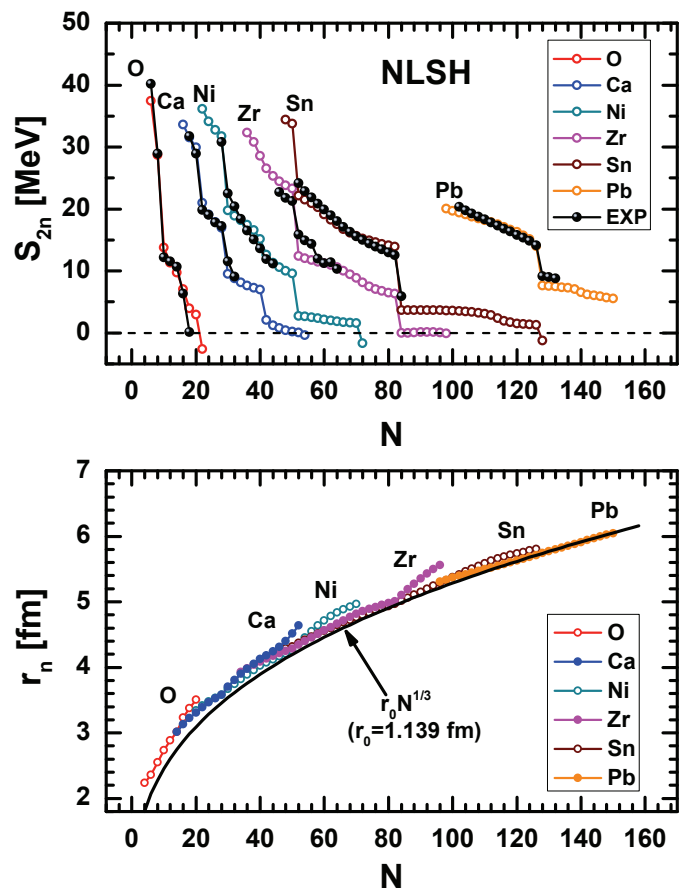}
\end{center}
\vspace{-0.22cm}
\caption{\label{Fig:Fig03}
(Color online) (a) The two-neutron separation energies  $S_{2{\rm n}}$ calculated by RCHB theory
\cite{Meng1998_NPA635-3} with NLSH~\cite{Sharma1993_PLB312-377} (Open
symbols) in comparison with data available (solid ones) for the even O,
Ca, Ni, Zr, Sn, and Pb isotopes against the neutron number $N$. (b)
The root mean square (rms) neutron radii $r_n$ for even O, Ca, Ni, Zr,
Sn, and Pb isotopes by RCHB calculations against the neutron number
$N$. The curve $r_0N^{1/3}$ with $r_0=1.139$ fm has been included to
guide the eye. Taken from Ref.~\cite{Meng2002_PRC65-041302R} with modified color.
}
\end{figure}

In Fig.~\ref{Fig:Fig03}, similar to Ref.~\cite{Meng2002_PRC65-041302R},
the two-neutron separation energies $S_{\rm 2n}$ and the neutron radii $r_n$
from the RCHB calculation for the even-even nuclei of the O, Ca, Ni, Zr, Sn, and
Pb isotope chains have been shown.
The predicted $r_n$ curve using the simple empirical equation
$r_n=r_0 N^{1/3}$ with $r_0=1.139$ fm
normalizing to $^{208}$Pb is also represented in the figure.
This simple formula for $r_n$ agrees with the calculated neutron radii
with exceptions in neutron rich Ca and Zr isotopes.

The comparison between experimental and calculated $S_{2{\rm n}}$ is shown
in Fig.~\ref{Fig:Fig03}.
The experimental magic or submagic numbers $N=20, 28$, and $40$ are
reproduced. The $S_{2{\rm n}}$ values for several nuclei in exotic Ca and Zr isotopes
are extremely close to zero. As discussed in Refs.~\cite{Meng1998_PRL80-460,
Meng2002_PRC65-041302R},
if taking $^{60}$Ca or $^{122}$Zr respectively as a core, the valence neutrons
will gradually occupy the loosely bound states and the continuum above
the sub-shell of $N=40$ in Ca or $N=82$ in Zr isotopes. Such typical behavior
of  $S_{2{\rm n}}$ can be taken as an evidence of the occurrence of giant halos
in Ca chain and Zr chain~\cite{Meng1998_PRL80-460}.

In Fig.~\ref{Fig:Fig03}, it is very interesting to see
that $r_n$ follows the $N^{1/3}$
systematics well for stable nuclei although their proton numbers are
quite different. Near the drip line, distinct abnormal behaviors appear at
$N=40$ in Ca isotopes and at $N=82$ in Zr isotopes. As pointed in
Ref.~\cite{Meng2002_PRC65-041302R}, nuclei having the abnormal
$r_n$ increase correspond to those having small $S_{2{\rm n}}$, which provides
another evidence to the emergence of giant halos.
The increase of $r_n$ in exotic Ni, Sn and Pb nuclei is not as rapid as those in Ca and Zr
chains. The regions of the abnormal increases of the neutron radii
are just the same as those for $S_{2n}$.  Both behaviors are
connected with the formation of giant halo~\cite{Meng1998_PRL80-460,
Meng2002_PRC65-041302R,Zhang2002_CPL19-312,
Zhang2003_SciChinaG46-632}.

\subsection{Halos from relativistic and non-relativistic approaches}

By using the RCHB theory, predictions of
halo in the Ne~\cite{Poschl1997_PRL79-3841, Lalazissis1998_NPA632-363},
Na~\cite{Meng1998_PLB419-1,Lalazissis1998_NPA632-363}, Ca
\cite{Meng2002_PRC65-041302R} and Zr~\cite{Meng1998_PRL80-460} isotopes near the neutron drip
line have been made. In particular, in Ca and Zr isotopes near the
neutron drip line, the halo phenomena were addressed as
giant halo~\cite{Meng1998_PRL80-460,Meng2002_PRC65-041302R}.
In the non-relativistic HFB approaches, Skyrme calculations with the
parameter set SLy4 predicted the halo in Sn and Ni~\cite{Mizutori2000_PRC61-044326}. It
was also claimed that the pairing gaps have an effect to reduce the
halo at the neutron drip line~\cite{Mizutori2000_PRC61-044326, Bennaceur2000_PLB496-154,
Dobaczewski2001_NPA693-361}.

In Refs.~\cite{Terasaki2006_PRC74-054318, Terasaki2006_IJMPE15-1833},
the phenomena of the giant halo and halo
of the neutron-rich even-Ca isotopes have been investigated
and compared between the frameworks of the RCHB and the
Skyrme HFB calculations.
With the two parameter sets for
each of the RCHB and the Skyrme HFB calculations, it has
been found that although the halo phenomena exist for Ca
isotopes near the neutron drip line in both calculations, the
halo from the Skyrme HFB calculations (SkM*) starts at a more
neutron-rich nucleus than that from the RCHB calculations, and
SLy4 does not predict a halo, because the drip line is much closer
to the stable region than the other parameter sets.

\begin{figure}
\begin{center}
\includegraphics[width=8cm]{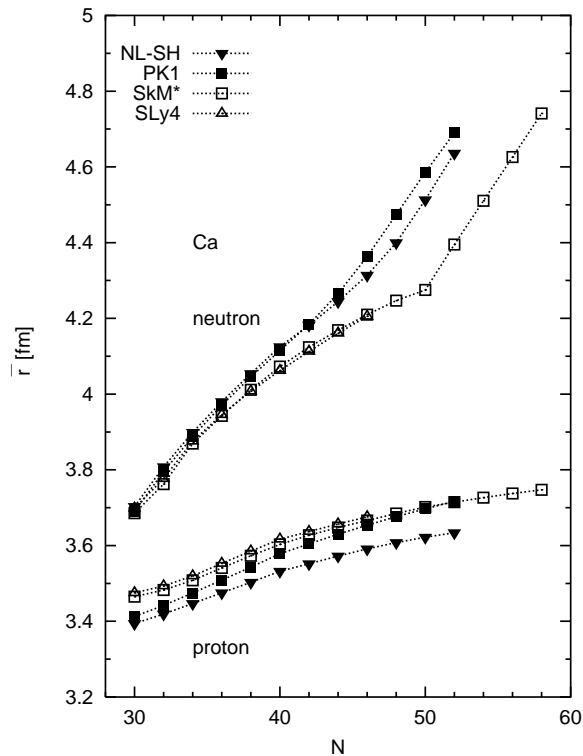}
\end{center}
\caption{\label{fig:Fig04}
\baselineskip=13pt
The neutron and proton rms radii
$\bar{r}_n$ and $\bar{r}_p$ of even-Ca isotopes in RCHB calculations
with NLSH and PK1 and the Skyrme HFB calculations with SkM$^\ast$
and SLy4 in the neutron-rich region.
Taken from Ref.~\cite{Terasaki2006_PRC74-054318}.}
\end{figure}

Figure \ref{fig:Fig04} shows systematics of the rms radii of
neutrons $\bar{r}_n$ and protons $\bar{r}_p$ of even-Ca from $N=30$
to the two-neutron drip lines. The increase in the curvature of
$\bar{r}_n$ indicates the halo.
For Ca, the number of nucleons in continuum, $N_h$, of
$^{62-72}$Ca is 0.6--2.2, of which the average is 1.7, in the NL-SH
calculation \cite{Meng2002_PRC65-041302R}; the corresponding value of the
SkM$^\ast$ calculation is always smaller than 0.5 (the average
0.27). As there are more than two halo neutrons, it was referred to as giant halo
\cite{Meng1998_PRL80-460,Meng2002_PRC65-041302R}.
The SkM$^\ast$ calculation predicts the halo
from $N=52$, and the RCHB calculations show the gradual occurrence
of the giant halo. Apparently, the starting nucleus of the halo of
SkM$^\ast$ corresponds to the extra lowering of the $S_{2n}$. The
SLy4 calculation does not predict the halo, because the particle-stable
region ends at $N=46$. The large difference between $\bar{r}_n$ and
$\bar{r}_p$ shown in Fig.~\ref{fig:Fig04} was identified as neutron skin in the region with no halo.

\begin{figure}
\begin{center}
\includegraphics[width=10cm]{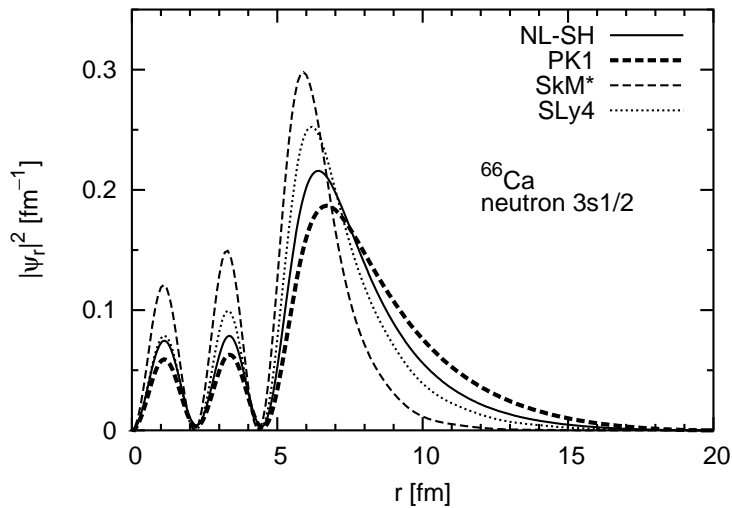}
\end{center}
\caption{\label{fig:Fig05}
\baselineskip=13pt
Radial wave function squared $|\psi_r|^2$
(canonical basis) of the neutron $3s_{1/2}$ of $^{66}$Ca. They are
normalized as $\int dr\,|\psi_r|^2=1$.
Taken from Ref.~\cite{Terasaki2006_PRC74-054318}.}
\end{figure}

The RCHB calculations have larger neutron rms radii systematically in
$N\ge40$ than the Skyrme HFB calculations. This difference
comes partially from the occupation of the neutron $3s_{1/2}$ orbital, which causes 50\%
of the maximum difference in the neutron rms radii among the
four calculations at $^{66}$Ca.
The difference in the neutron $3s_{1/2}$ orbital
is shown in Fig.~\ref{fig:Fig05}, which illustrates
the radial wave functions squared.
The tail of $3s_{1/2}$ from PK1 has appreciably extended distribution
than the others, and as a consequence, the amplitude of PK1 is smaller in the inner region.
The number and locations of the nodes are the same for the four curves.

As shown in the previous Subsection and in
Ref.~\cite{Meng1998_PRL80-460},
the occupation of $3p_{1/2}$ and $3p_{3/2}$ orbitals are responsible
for the formation of giant halos in Zr isotopes.
In Ref.~\cite{Grasso2006_PRC74-064317},
the Hartree-Fock-Bogoliubov (HFB) approach with Skyrme interactions
SLy4 and SkI4
has been used to study halos in Ca and Zr isotopes.
It was found that the appearance of giant halos depends
sensitively on the effective interaction adopted:
SkI4 predicts a neutron halo in the Zr chain with $A\ge 122$
due to the weakly bound orbitals $3p_{1/2}$ and $3p_{3/2}$.

In the above mentioned work, the box boundary condition is
adopted and the continuum is discretized.
Although the convergence of the halo properties with respect to the box size has
always been achieved, it is still desirable to examine
the influence of the asymptotic behavior of
the radial wave functions of those valence orbitals
on the formation of halo or giant halo.
The Skyrme HFB equations in $r$ space
have been solved for spherical systems
with correct asymptotic boundary conditions
for the continuous spectrum~\cite{Grasso2001_PRC64-064321}
or the Green's function technique~\cite{Zhang2011_PRC83-054301,
Zhang2014_PRC90-034313}.
In Ref.~\cite{Grasso2001_PRC64-064321},
it was shown that close to the drip line the amount of pairing correlations
depends on how the continuum coupling is treated.
In Ref.~\cite{Zhang2011_PRC83-054301},
it was found that, in even-even $N=86$ isotones
in the Mo-Sn region, the $l=1$ broad
quasiparticle resonances persist in feeling the pairing potential and
contribute to the pairing correlation even when the widths are
comparable with the resonance energy. In Ref.~\cite{Zhang2012_PRC86-054318}, the investigation of the pair correlation in the tail of the giant halo revealed that the asymptotic exponential behavior of the neutron pair condensate is dominated by nonresonant continuum states with low asymptotic kinetic energy.

\subsection{Halos with relativistic Hartree-Fock-Bogoliubov theory in continuum}

In spite of successes of RCHB theory, there are still a number of questions
needed to be answered in the conventional CDFT, e.g., the contributions due
to the exchange (Fock) terms and the pseudo-vector $\pi$-meson.
Earlier attempts to include the exchange terms with
relativistic Hartree-Fock (RHF) method led to underbound nuclei
due to the missing of the meson self-interactions 
\cite{Boguta1977Nucl.Phys.A413,Bouyssy1987_PRC36-380}.
For a long time, the RHF theory failed
in a quantitative description of nuclear systems
\cite{Brockmann1977_PLB69-167, Horowitz1984_PLB140-86,
Bielajew1984_APNY156-215, Bouyssy1985_PRL55-733,
Blunden1987_PLB196-295, Bouyssy1987_PRC36-380,
Bernardos1993_PRC48-2665, Marcos2004_JPG30-703}.
With the development of the computational facilities and numerical techniques,
the density dependent relativistic Hartree-Fock (DDRHF)
theory has been developed and shown
significant improvements in a quantitative description of nuclear phenomena
\cite{Long2006_PLB640-150, Long2006_PLB639-242, Long2007_PRC76-034314,
Long2008_EPL82-12001, Liang2008_PRL101-122502, Sun2008_PRC78-065805,
Liang2009_PRC79-064316, Liang2012_PRC85-064302}
with comparable accuracy.
With a number
of adjustable parameters comparable to those of RMF Lagrangians,
the DDRHF theory can give a equally good description of
nuclear systems without dropping
the Fock terms. Furthermore, important features like the behavior
of neutron and proton effective masses~\cite{Jaminon1989_PRC40-67}
can be interpreted.

In DDRHF theory~\cite{Long2006_PLB640-150}, the tensor forces due to
$\pi$ and $\rho$ meson exchanges can be naturally taken into account and
have brought significant improvements on the consistent description of
shell evolution~\cite{Long2008_EPL82-12001, Tarpanov2008_PRC77-054316, Long2009_PLB680-428, Moreno-Torres2010_PRC81-064327}. The relativistic Hartree-Fock-Bogoliubov (RHFB) theory in continuum with
density dependent meson-nucleon couplings
is developed in Refs.~\cite{Long2010_PRC81-031302R, Long2010_PRC81-024308}.
In the descriptions of nuclear halos, by using the finite range Gogny force D1S
in the pairing channel, systematic RHFBc calculations are performed both
for stable and weakly bound nuclei. It is demonstrated that
an appropriate description of both mean field and pairing effects
can be obtained within RHFBc theory
\cite{Long2010_PRC81-031302R, Long2010_PRC81-024308}.

\begin{figure*}[ptbh]
\begin{center}
\includegraphics[width = 0.35\textwidth]{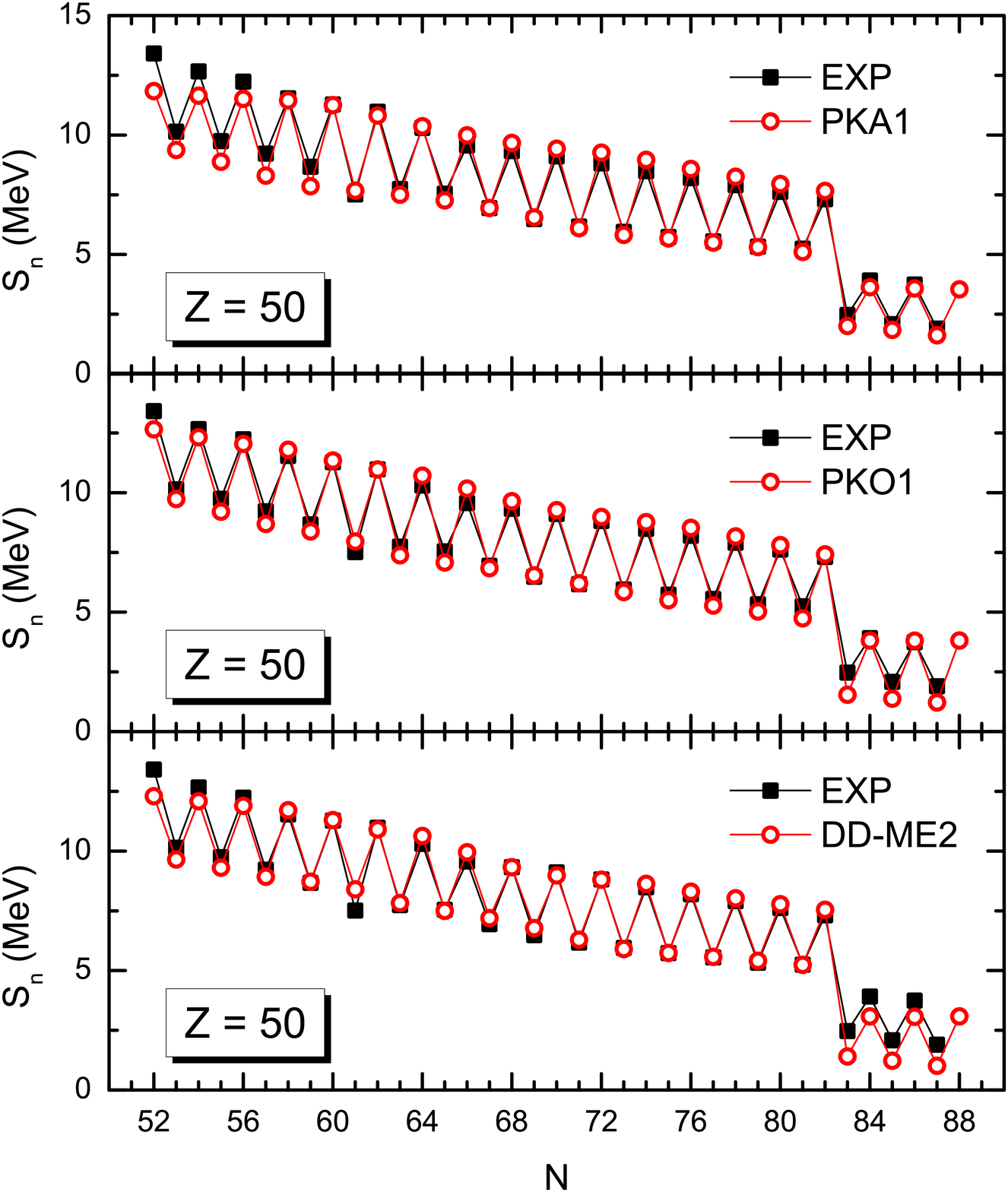}~~~
\includegraphics[width = 0.35\textwidth]{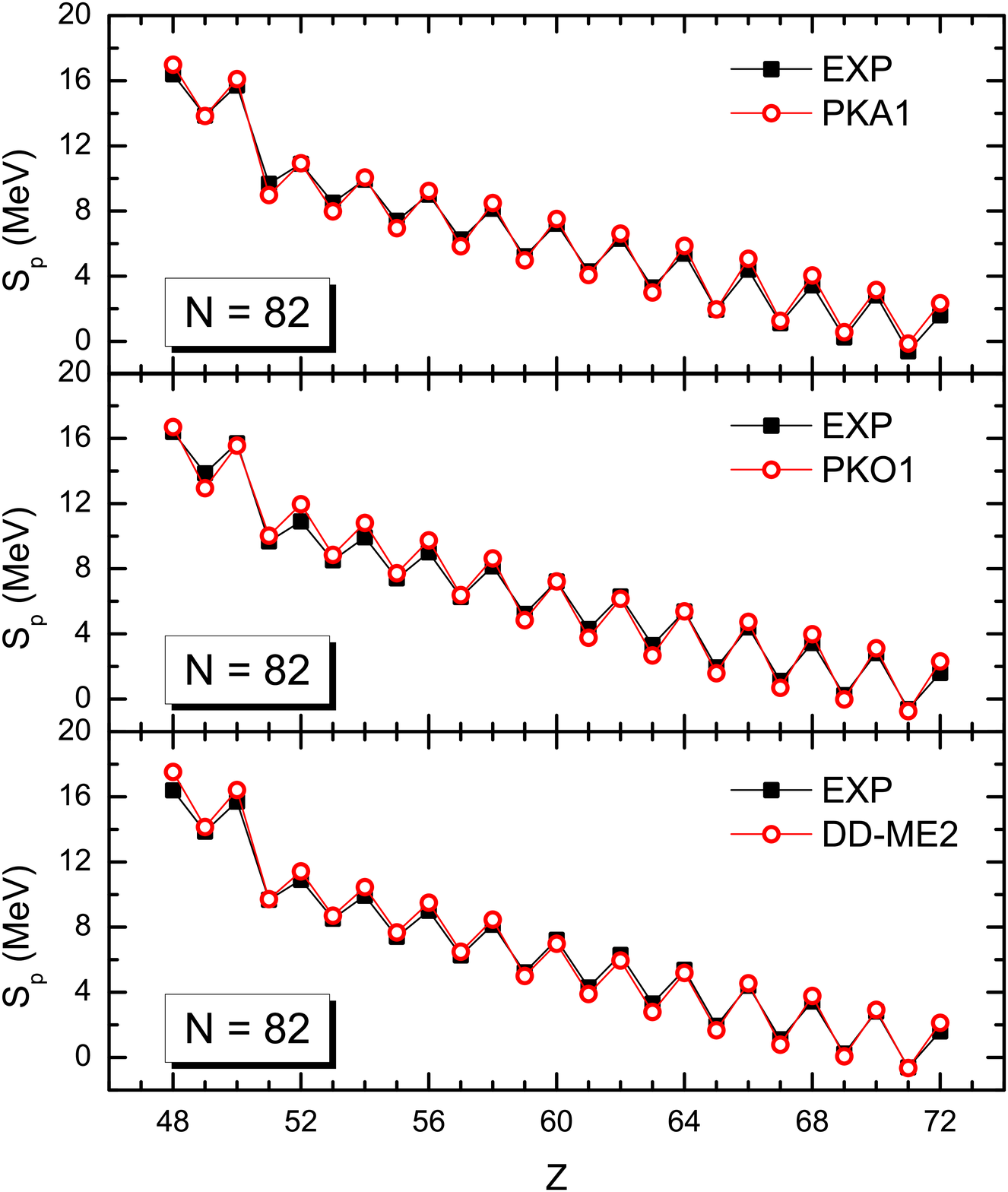}  
\end{center}
\caption{(Color online) Single-nucleon separation energies along Sn
isotopic ($S_{n}$: left
panels) and $N=82$ isotonic ($S_{p}$: right panels) chains.
The results are calculated by RHFB with
PKA1~\cite{Long2007_PRC76-034314}, PKO1~\cite{Long2006_PLB640-150},
and by RHB with DD-ME2~\cite{Lalazissis2005_PRC71-024312}, in
comparison to the experimental data~\cite{Audi2003_NPA729-337}.
Taken from Ref.~\cite{Long2010_PRC81-024308}.}\label{fig:Sn&FIG06B}
\end{figure*}

In Fig. \ref{fig:Sn&FIG06B}, the single-neutron separation energies $S_{n}$
of Sn isotopes from $^{101}$Sn to $^{138}$Sn (left panels) and the single-proton
separation energies $S_{p}$ of $N=82$ isotones from $^{130}$Cd to $^{153}$Lu
(right panels) are given.
The odd-even differences on the single-nucleon separation energies reflect
the effects of the pairing correlations. In Fig.~\ref{fig:Sn&FIG06B},
RHFBc with PKA1~\cite{Long2007_PRC76-034314} and PKO1~\cite{Long2006_PLB640-150}
as well as RHB with DD-ME2~\cite{Lalazissis2005_PRC71-024312} present
comparable and satisfied
quantitative agreements with the data for both isotopes and isotones.
From Fig. \ref{fig:Sn&FIG06B}, one can find some systematics in
the results. On the neutron rich side, i.e., after $^{132}$Sn for Sn
isotopes and before $_{~64}^{146}$Gd for $N=82$ isotones, PKA1 shows better
agreements than PKO1
and DD-ME2. This is mainly due to the improvements in the shell structures
with exchange terms included, especially the shell closures beyond 50 and 82, i.e.,
58 and 92, which may change
the level density and then influence the pairing effects. On the proton rich side,
these three approaches
provide comparable agreement with the data.

\begin{figure*}[ht]
\begin{center}
\includegraphics[width = 0.95\textwidth]{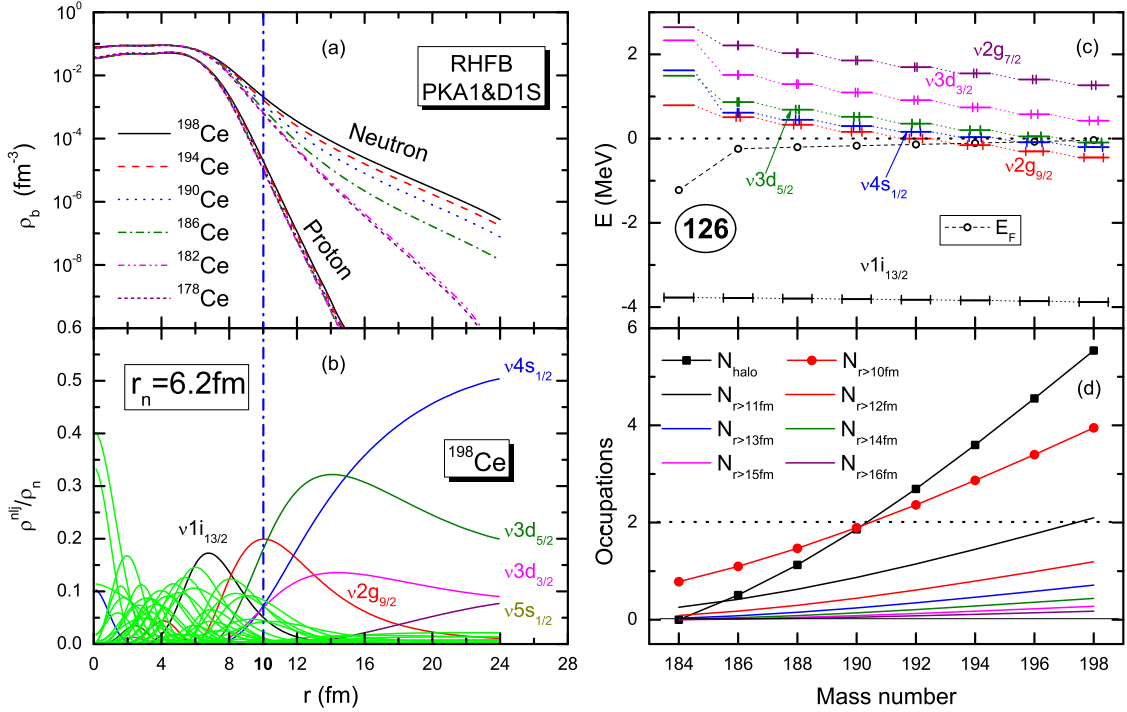}
\end{center}
\caption{(Color online) (a) Neutron and proton densities,
(b) the relative contributions of different orbitals to the full
neutron density in $^{198}$Ce, (c) neutron canonical single particle energies
occupation probability (in $x$-error bars) and Fermi energy $E_F$ (in open circles),
and (d) neutron numbers filling in the halo orbitals $4s_{1/2}$, $3d_{5/2}$,
and $3d_{3/2}$ ($N_{\rm halo}$), and the ones lying beyond the spheres
with the radii $r=10$, 11, 12, 13, 14, 15, 16 fm, respectively
$N_{r>10\ {\rm fm}}$, $N_{r>11\ {\rm fm}}$, $N_{r>12\ {\rm fm}}$, $N_{r>13\ {\rm fm}}$,
$N_{r>14\ {\rm fm}}$, $N_{r>15\ {\rm fm}}$ and $N_{r>16\ {\rm fm}}$.
The results are calculated by RHFB with PKA1~\cite{Long2007_PRC76-034314}
plus the Gogny pairing force D1S~\cite{Berger1984_NPA428-23}.
 The spherical box radius is adopted as $R  = 28$ fm.
Taken from Ref.~\cite{Long2010_PRC81-031302R}.
}
\label{fig:Density-SDG}
\end{figure*}

The Ce isotopes have been chosen to demonstrate the nuclear halo
phenomenon within the RHFBc theory~\cite{Long2010_PRC81-031302R, Long2010_PRC81-024308}.
In Fig.~\ref{fig:Density-SDG} the nuclear matter distributions (left panels)
and neutron canonical single particle configurations (right panels) for Ce isotopes
close to the drip line are shown.
The neutron drip line calculated here with PKA1~\cite{Long2007_PRC76-034314} is
$N=140$ whereas the calculations with PKO1~\cite{Long2006_PLB640-150} and
DD-ME2~\cite{Lalazissis2005_PRC71-024312} predict a shorter one as $N=126$.
As shown in Fig.~\ref{fig:Density-SDG}(a), the neutron densities
become more and more diffuse after the isotope $^{186}$Ce ($N=128$),
a direct and distinct evidence of halo occurrence.

From Fig.~\ref{fig:Density-SDG}(b) one can see that such extremely extensive
matter distribution, e.g. in $^{198}$Ce, is mainly due to the low-$l$ states,
namely the halo orbitals $\nu4s_{1/2}$, $\nu3d_{5/2}$ and $\nu3d_{3/2}$.
From the occupations of the halo orbitals $N_{\rm halo}$ in
Fig.~\ref{fig:Density-SDG}(d), the halos were predicted in $^{186}$Ce, $^{188}$Ce and $^{190}$Ce
and giant halos may appear in $^{192}$Ce, $^{194}$Ce, $^{196}$Ce and $^{198}$Ce
because more than two neutrons are occupying the halo orbitals.
Similar conclusions can be obtained from the neutron numbers lying beyond
the sphere with the radius $r=10$ fm [$N_{r>10~\rm{fm}}$ in Fig.~\ref{fig:Density-SDG}(d)],
which is large enough (the neutron matter radius $r_n = 6.2$ fm in $^{198}$Ce) for halos.
Even extending to $r=16$ fm, which is sometime taken as the radial cut-off in
the calculations of stable nuclei, there are still some amount of neutrons
lying beyond this sphere in the isotopes from $^{192}$Ce to $^{198}$Ce.

In Fig.~\ref{fig:Density-SDG}(c), one finds that the halo orbitals ($\nu4s_{1/2}$, $\nu3d_{5/2}$
and $\nu3d_{3/2}$) are located around the particle continuum threshold where
they are gradually occupied.
For the isotopes beyond $^{184}$Ce ($N=126$), the Fermi levels (in open circles)
approach the continuum threshold rather closely such that the stability of
these halo isotopes becomes sensitive to pairing effects. Nearby the low-$l$ states,
there are the high-$l$ states $\nu2g_{9/2}$ and $\nu2g_{7/2}$.
Because of the relatively large centrifugal barrier for $g$-orbitals
they do not contribute much to the diffuse neutron distributions.
Nevertheless, the existence of the high-$l$ states nearby halo orbitals is
still particulary significant, because it leads to a rather high level density
around the Fermi surface, and evidently the pairing effects are enhanced to
stabilize the halo isotopes.

\begin{figure}[htbp]
\begin{center}
   \includegraphics[width = 0.45\textwidth]{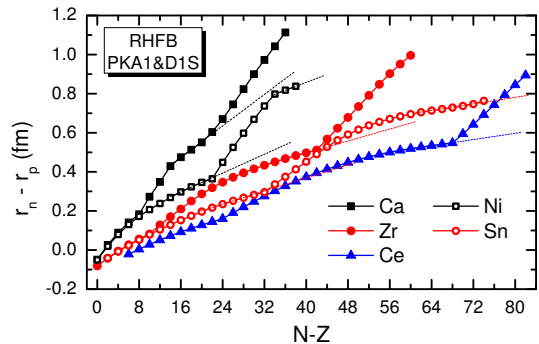}
\end{center}
\caption{(Color online) Neutron skin thickness $r_{n}-r_{p}$
for Ca, Zr, Ni, Sn, and Ce as a function of the isospin ($N-Z$).
The $r_n$ and $r_p$ are respectively the neutron and proton root mean square radii.
The results are calculated by RHFB with PKA1~\cite{Long2007_PRC76-034314}
plus the paring force D1S~\cite{Berger1984_NPA428-23}.
Taken from Ref.~\cite{Long2010_PRC81-031302R}.
} \label{fig:FIG08}
\end{figure}

Evidence for the existence of a halo can also be investigated by
the systematic behavior of nuclear bulk properties such as radii.
In Fig.~\ref{fig:FIG08}, the isospin dependence of
the neutron skin thickness ($r_{n}-r_{p}$) calculated in density dependent RHFB theory
in continuum using PKA1 for Ca, Ni, Zr, Sn, and Ce isotopes are given
\cite{Long2010_PRC81-031302R}. Taking the stable nuclei as references
(shown as the dashed lines), continuously growing deviations are found in
the Ca, Zr, and Ce chains until the neutron drip line.
This can be considered as evidence of a halo.
Despite the deviations in the mid-region, both Ni and Sn nuclei show
a similar isospin dependence in both stable and exotic nuclei regions,
which indicate a neutron skin since the growth of a halo is interrupted.
Compared with Ni, the Sn isotopes show even a much weaker skin effect.

The systematics for the Ca and Zr chains from the RHFBc calculations are similar to those from the RCHB
calculations~\cite{Meng2002_PRC65-041302R}. For the Ce isotopes,
RHFBc calculations with PKA1 show clear evidence for the existence
of halo structures.
The calculations of RHFBc with PKO1~\cite{Long2006_PLB640-150}
and RCHB with DD-ME2~\cite{Lalazissis2005_PRC71-024312}
predict that the isotopic chain ends at $N=126$, before
the halo occurrence as predicted by RHFB with PKA1.
This deviation is interpreted by the shell structure evolution
in Fig.~\ref{fig:Density-SDG}(c), where much weaker shell are predicted
with PKA1 in comparison with PKO1 and DD-ME2.
As shown in Fig.~\ref{fig:Density-SDG}(c), the neutron shell gap ($N=126$)
between $\nu1i_{13/2}$ and $\nu2g_{9/2}$ states is close to
the particle continuum threshold, which might essentially influence
the stability of the drip line isotopes.

From Ref.~\cite{Long2010_PRC81-031302R}, one may conclude that
the Fock term itself has little direct impact on the formation of a nuclear halo.
However, the Fock term may influence the shell structure,
particularly those orbitals close to the Fermi surface in drip-line nuclei,
thus affecting indirectly the formation of a potential halo.

\subsection{Halos in exotic hyper nuclei}

Motivated by knowledge of $\Lambda$-N interaction and
understanding on giant
halo~\cite{Meng1998_PRL80-460,Meng2002_PRC65-041302R, Zhang2003_SciChinaG46-632},
it is interesting to study the possible appearances of halos in exotic hyper nuclei.

Two neutron separation energies $S_{2n}$ for
normal nuclei, single-$\Lambda$ hyper nuclei and
double-$\Lambda$ hyper nuclei of Ca isotopes, labelled by $\Lambda
= 0$, $\Lambda =1$ and $\Lambda =2$, respectively, from the proton drip line
to neutron drip line were investigated~\cite{Lu2003_EPJA17-19}.
It was found that one or two
$\Lambda$ hyperons lower the Fermi level due to the attractive
$\Lambda$-N potential but keep the
neutron shell structure unchanged. Therefore, the neutron drip
line is pushed outside from $N=52$ in Ca isotope chain to
$N=54$ in hyper isotope chain. Meanwhile, giant halos due to
pairing correlation and the contribution from the continuum still
exist in Ca hyper nuclei similar to that in Ca
isotopes~\cite{Meng2002_PRC65-041302R}. This is a slight but rewarding step for
exploring the limit of drip line nuclei on the basis of the giant
halo. For more details, see Ref.~\cite{Lu2003_EPJA17-19}.

Apart from neutron halos in hyper nuclei, as $\Lambda$ hyperon is
less bound than the corresponding nucleon in nuclei, it is worth
investigating the existence of hyperon halos.

In Ref.~\cite{Rufa1990_PRC42-2469,
Rufa1991_PRC43-2020},
a $\Lambda$ halo was tentatively suggested in a RMF plus BCS model.
However, since the normal BCS method will have unphysical solution by
involving baryon gas~\cite{Meng1998_NPA635-3}, such prediction needs
further study. In Ref.~\cite{Lu2002_CPL19-1775}, hyper carbon isotopes are studied by the RCHB theory. By investigating the baryon density
distributions in hyper nuclei $^{13}_{\Lambda}$C, $^{14}_{2\Lambda}$C, and
$^{15}_{3\Lambda}$C, it was found that with up to two $\Lambda$ hyperons added to the core
$^{12}$C, the nucleon density distributions remain the same and
hyperon density distributions at the tail are comparable with
those of the nucleons. An intriguing phenomenon appears in
$^{15}_{3\Lambda}$C where the hyperon density distribution has a
long tail extended far outside of its core $^{14}_{2\Lambda}$C. This is a signature of hyperon halo due
to the weakly bound state $1p_{3/2}^\Lambda$ in
$^{15}_{3\Lambda}$C which has a small hyperon separation energy
and a density distribution with long tail~\cite{Lu2002_CPL19-1775}.

\section{\label{sec:surface_diffuseness}Surface diffuseness}

From the mean field point of view,
the properties of nucleons in an atomic nucleus are determined by
the mean potential which stems from their interaction with the other nucleons
and is determined by nucleon densities in the self-consistent models.
Therefore, the study of the isospin dependence of the mean potential
and the density distribution,
both become highly diffuse near the particle drip line,
is crucial to understand exotic nuclei.
For finite nuclei, the diffuseness of the nuclear surface
provides a measure of the thickness of the surface region.
The surface diffuseness is intimately related to
the spin-orbit splitting~\cite{Meng1999_NPA650-176},
the evolution of single particle shell structure~\cite{Meng1998_PRL80-460},
the pseudospin symmetry~\cite{Meng1998_PRC58-R628, Meng1999_PRC59-154, Liang2015_PR570-1} and
spin symmetry~\cite{Zhou2003_PRL91-262501, Liang2015_PR570-1},
and
the nuclear surface energy and thus nuclear masses~\cite{Wang2014_PLB734-215}.
In this Section, we will review recent progresses on
nuclear surface diffuseness and the corresponding impact on nuclear shell structure and size.

\subsection{Spin-orbit splitting and shell structure}

\begin{figure}
\centering
\includegraphics[width=8.0cm,angle=270]{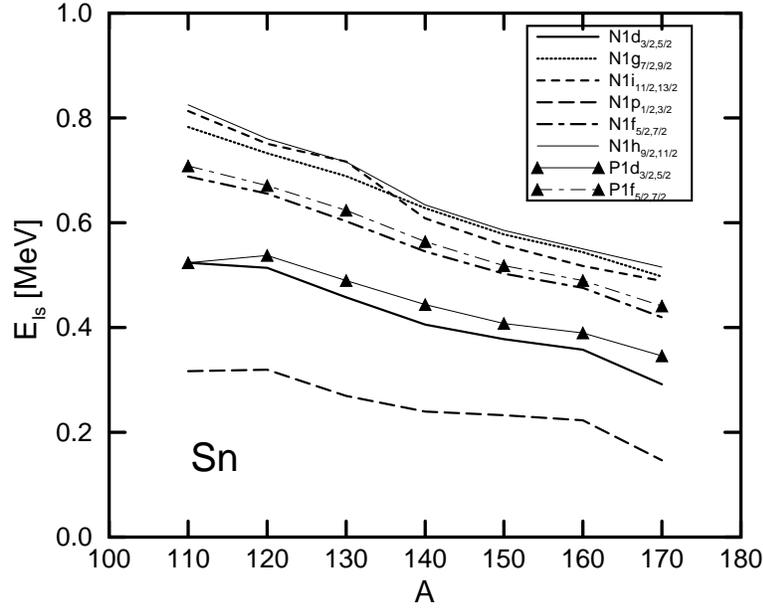}
\caption{The neutron spin-orbit splitting $E_{ls} = 
(E_{lj=l-1/2}-E_{lj=l+1/2}) / (2l+1)$ versus the mass number
$A$ in Sn isotopes for neutron ($1d_{3/2}, 1d_{5/2}$),
($1g_{7/2}, 1g_{9/2}$), ($1i_{11/2}, 1i_{13/2}$), ($1p_{1/2},
1p_{3/2}$), ($1f_{5/2}, 1f_{7/2}$) and ($1h_{9/2}, 1h_{11/2}$)
orbitals and proton ($1d_{3/2}, 1d_{5/2}$) and ($1f_{5/2},
1f_{7/2}$) orbitals, respectively. Taken from
Ref.~\cite{Meng1999_NPA650-176}.} \label{Fig09}
\end{figure}

The RCHB theory has been used to systematically study the isospin dependence of the mean potential
for the Sn isotopes from proton drip line to neutron drip
line and the spin-orbit splitting for the whole isotope chain was examined in details~\cite{Meng1999_NPA650-176}. As an example, the spin-orbit splitting
\begin{equation}
   E_{ls} = \displaystyle \frac {E_{lj=l-1/2}-E_{lj=l+1/2}} {2l+1}
\label{E-ls}
\end{equation}
versus the binding energy:
\begin{equation}
   E = \displaystyle \frac { ( l+1 ) E_{lj=l-1/2}+ l E_{lj=l+1/2}} {2l+1}
\label{abe}
\end{equation}
in $^A$Sn with $A=110, 120, \cdots, 170$ are given in Fig.~\ref{Fig09} for the
neutron spin-orbit partners ($1d_{3/2}, 1d_{5/2}$), ($1g_{7/2},
1g_{9/2}$), ($1i_{11/2}, 1i_{13/2}$), ($1p_{1/2}, 1p_{3/2}$),
($1f_{5/2}, 1f_{7/2}$) and ($1h_{9/2}, 1h_{11/2}$), and the
proton spin-orbit partners ($1d_{3/2}, 1d_{5/2}$) and
($1f_{5/2}, 1f_{7/2}$).

It is very interesting to see that the
spin-orbit splittings for the neutron and proton are very close to
each other, at least for ($1d_{3/2}, 1d_{5/2}$) and  ($1f_{5/2},
1f_{7/2}$) cases. The splitting decreases monotonically from the
proton drip line to the neutron drip line. To understand its mechanism, it is very helpful to
examine the origin of the spin-orbit splitting in the Dirac
equation.

For the Dirac nucleon moving in scalar and vector
potentials, its equation of motion could be de-coupled and reduced
for the upper component and the lower component, respectively.
For the spin-orbit splitting, the Dirac
equation can be reduced for the upper component as following:
\begin{eqnarray}
   & &  [ \frac {d^2} {dr^2}  - \frac 1  {E + 2M - \Sigma_0  + \Sigma_S}
        \frac {d(2M-\Sigma_0+\Sigma_S)} {dr} \frac {d} {dr} ] G^{lj}_i(r) \nonumber \\
   &-&  [  \frac { \kappa ( 1 + \kappa ) }  {r^2}
        - \frac 1 {E + 2M - \Sigma_0 + \Sigma_S} \frac {\kappa} r
        \frac {d(2M - \Sigma_0 + \Sigma_S)} {dr}  ] G^{lj}_i (r) \nonumber \\
 = &-&  (E + 2M - \Sigma_0 + \Sigma_S ) ( E - \Sigma_0 - \Sigma_S ) G^{lj}_i (r).
\label{larspinor4}
\end{eqnarray}

The spin-orbit splitting is due to the corresponding
spin-orbit potential
\begin{equation}
   \displaystyle \frac 1 {E + 2M- \Sigma_0 + \Sigma_S} \frac {\kappa} r
        \frac {d(2M - \Sigma_0 + \Sigma_S)} {dr}
\label{spp}
\end{equation}
with some proper normalization factor. The
spin-orbit splitting is energy dependent and depends
on the derivative of the potential $2M- \Sigma_0 + \Sigma_S$. Therefore the so-called
spin-orbit potential,
 $V_{ls} = \displaystyle \frac {\kappa} r \frac {d(2M - \Sigma_0 + \Sigma_S)} {dr}$,
is introduced.

\begin{figure}
\centering
\includegraphics[width=6.0cm,angle=270]{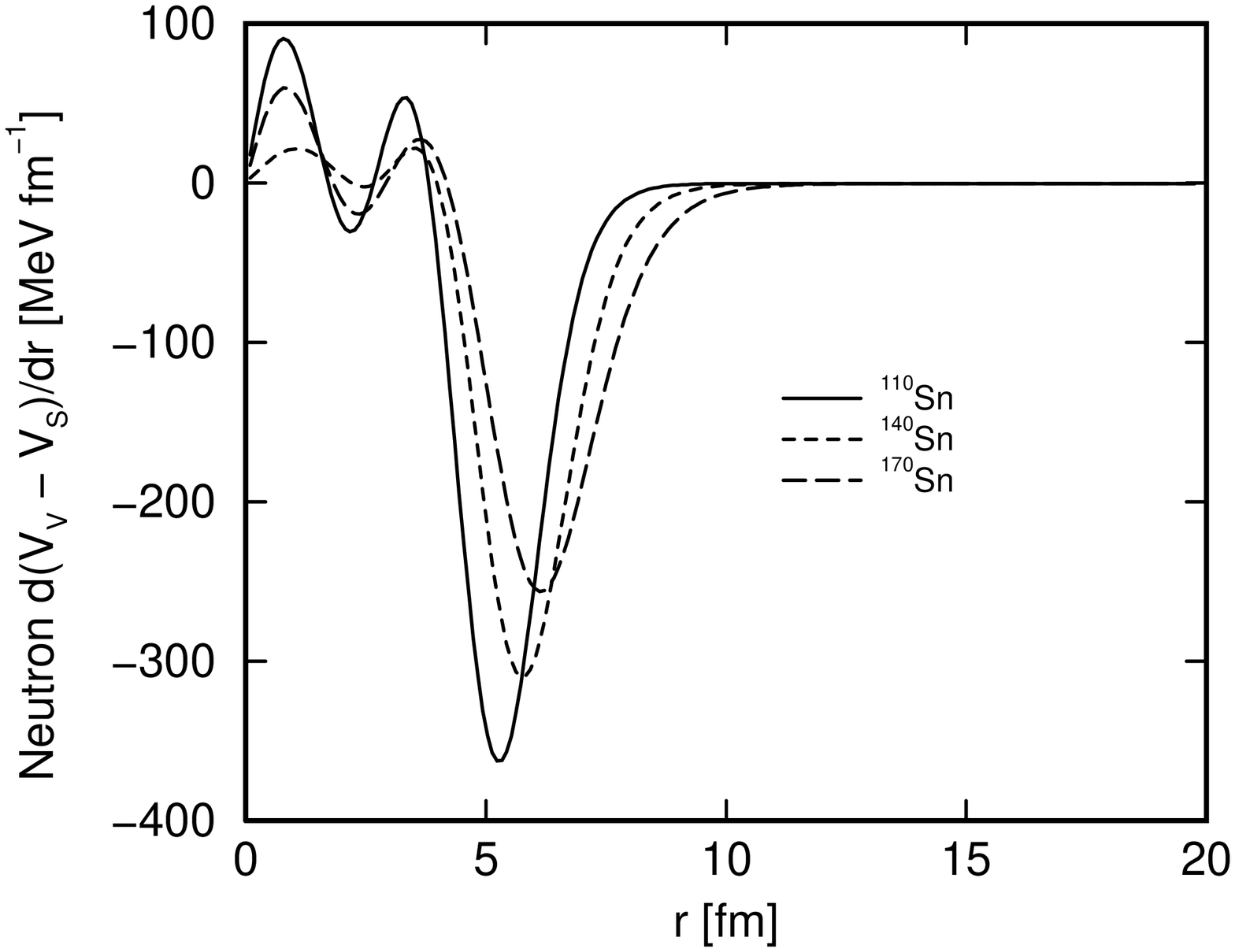}
\includegraphics[width=6.0cm,angle=270]{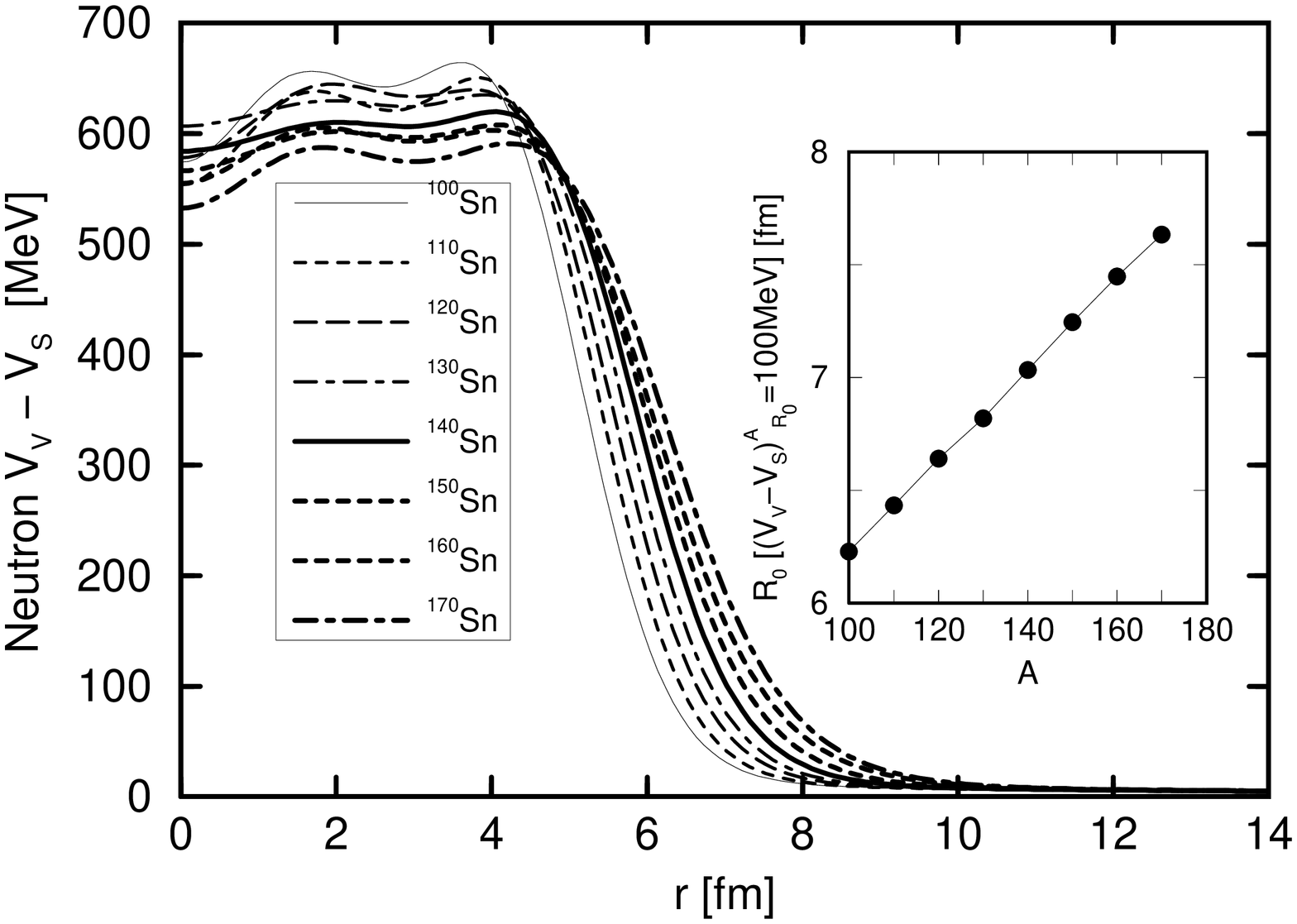}
\caption{
  Upper: The derivative of the neutron potentials $\Sigma_0( r )
- \Sigma_S( r )$ for $^A$Sn with $A=110, 120, \cdots, 170$;
  Lower: The neutron potentials $\Sigma_0( r ) - \Sigma_S( r )$ for $^A$Sn with $A=110, 120, \cdots, 170$.
In order to examine the surface
diffuseness more clearly, the radii $R_0$ at which $\Sigma_0(R_0) -
\Sigma_S(R_0) = 100$ MeV has been given as an inserted figure. Taken
from Ref.~\cite{Meng1999_NPA650-176}.} \label{Fig10}
\end{figure}

The derivatives of the neutron potentials $\Sigma_0( r )
- \Sigma_S( r )$ for $^{110}$Sn, $^{140}$Sn, and $^{170}$Sn, are given in the upper panel of Fig.~\ref{Fig10}. For both proton and neutron, the derivatives of the potentials $\Sigma_0( r )
- \Sigma_S( r )$ are almost the same. As the potential $\Sigma_0-\Sigma_S$ is a big quantity ($\sim 700$ MeV), the isospin dependence in the spin-orbit
potential could be neglected. That is the reason why
the spin-orbit splitting for the neutron and proton is very
close to each other in Fig.~\ref{Fig09}. From $^{110}$Sn to
$^{170}$Sn, the amplitude of the derivative for $\Sigma_0-\Sigma_S$ decreases monotonically due
to the surface diffuseness.

\begin{figure}
\centering
\includegraphics[width=6.0cm,angle=270]{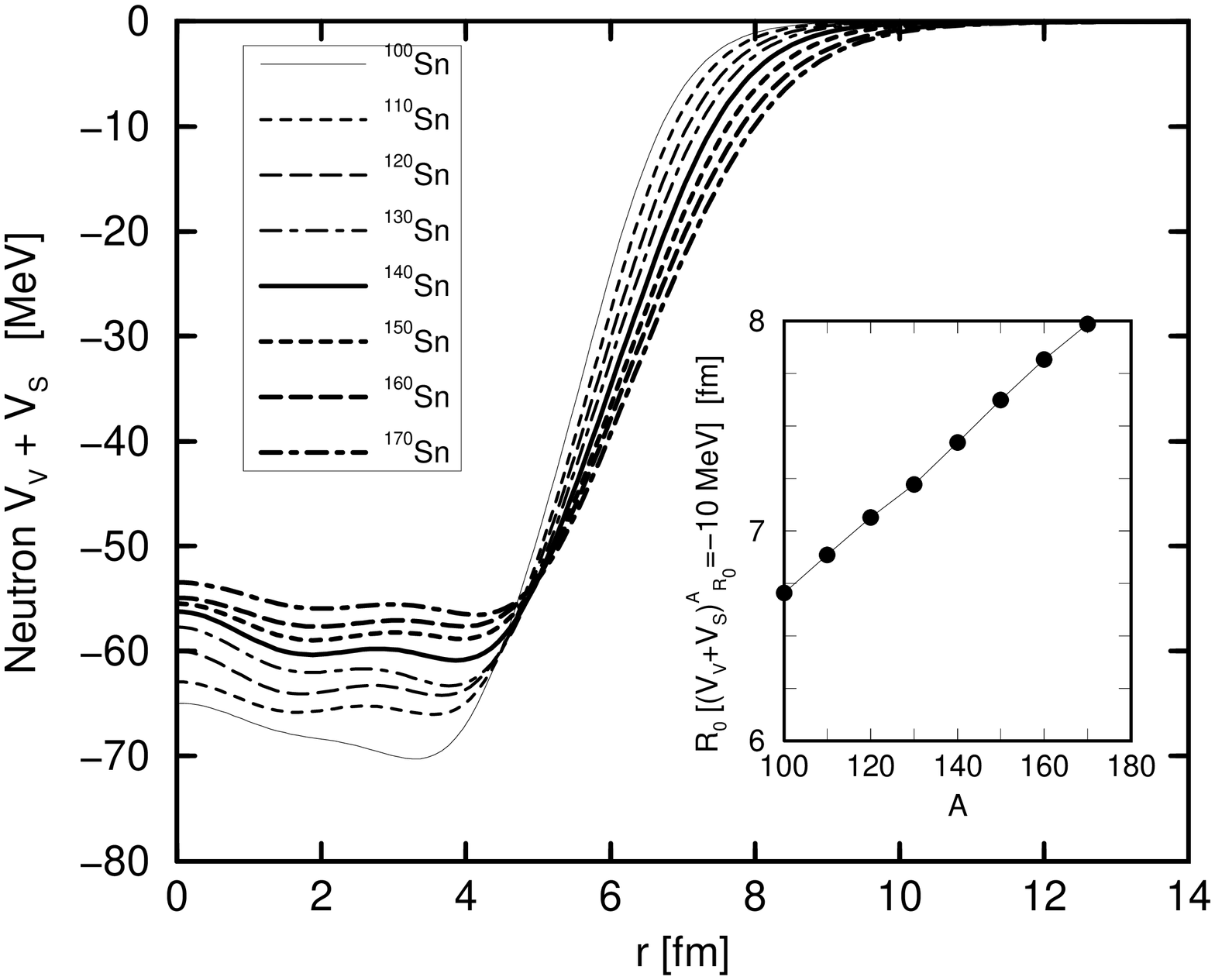}
\includegraphics[width=6.0cm,angle=270]{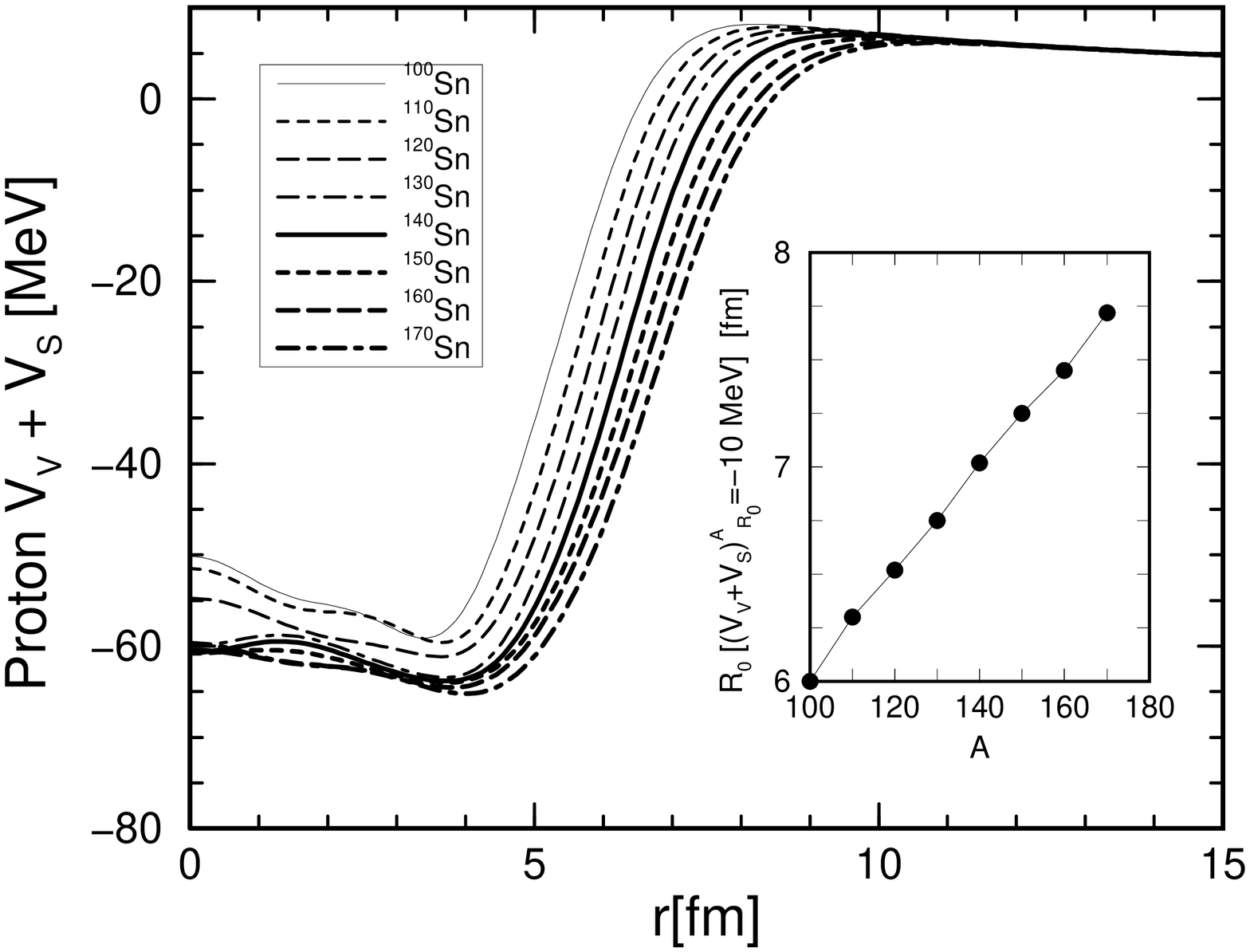}
\caption{
  Left: The neutron potentials $\Sigma_0( r ) + \Sigma_S( r )$ for
$^A$Sn with $A=110, 120, \cdots, 170$. In order to
examine the surface diffuseness more clearly, the radii $R_0$ at
which $\Sigma_0 (R_0) + \Sigma_S(R_0) = -10$ MeV has been given as an
inserted figure;
  Right: The proton potentials $\Sigma_0( r ) + \Sigma_S( r
)$ for $^A$Sn with $A=110, 120, \cdots, 170$.  In order to examine the surface
diffuseness more clearly, the radii $R_0$ at which $\Sigma_0 (R_0) +
\Sigma_S(R_0) = -10$ MeV has been given as an inserted figure. Taken
from Ref.~\cite{Meng1999_NPA650-176}.} \label{Fig11}
\end{figure}

The decline of the spin-orbit splitting is from the
diffuseness of the potential or the outwards tendency of the
potential. The neutron and proton potentials $\Sigma_0 + \Sigma_S $
are given in Fig.~\ref{Fig11} for $^A$Sn with $A=110, 120, \cdots, 170$. It
is seen that the depth of the neutron potential decreases monotonically
from the proton drip line to the neutron drip line and the surface
moves outwards. The inserted figures give the
radii $R_0$ at which $\Sigma_0 + \Sigma_S= -10$ MeV as functions of the
mass number. With the increase of the neutron number, the depth of the proton potential increases monotonically but is pushed towards outside as well due to the proton-neutron
interaction.

Similar neutron potentials $\Sigma_0 - \Sigma_S$ for $^A$Sn with $A=110, 120, \cdots, 170$ are given in the lower panel of
Fig.~\ref{Fig10} and its inserted figure gives the radii $R_0$
at which $\Sigma_0 - \Sigma_S=100$ MeV as a function of the mass number. As
seen above, the spin-orbit splitting is related
with the derivative of the potential $\Sigma_0 - \Sigma_S$. The surface
diffuseness happens for both the vector and scalar potential as shown in both $\Sigma_0 - \Sigma_S$ and $\Sigma_0 + \Sigma_S$.

\begin{figure}
\centering
\includegraphics[width=8.0cm,angle=270]{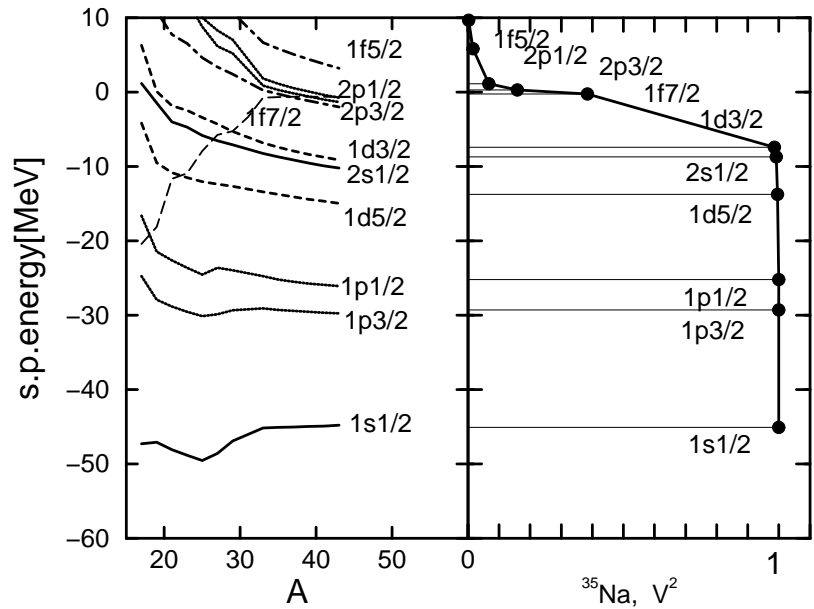}
\caption{Left part: Single particle energies for neutrons in the
canonical basis as a function of the mass number for Na isotopes. The dashed line
indicates the chemical potential. Right part: The occupation
probabilities in the canonical basis for $^{35}$Na. Taken from
Ref.~\cite{Meng1998_PLB419-1}.} \label{Fig12}
\end{figure}

Similar RCHB calculations have been carried out for
all the nuclei in Na isotopes with mass number $A$ ranging from 17
to 45 with effective interaction NLSH~\cite{Meng1998_PRL80-460}. Apart from the diffuseness of the potential, the shell evolution with neutron number has been examined there as well.
In Fig.~\ref{Fig12}, the microscopic structure of the single
particle energies in the canonical basis~\cite{Ring1980,Meng1998_NPA635-3}
is given. In the left panel of Fig.~\ref{Fig12}, the single
particle levels in the canonical basis for the Na isotopes with an
even neutron number are shown. The level density becomes dense with the neutron number due to the decrease of the spin-orbit potential or diffuseness of the potential. Going from $A=19$ to $A=45$, it can be observed a big gap above the $ N=8 $, $ N=20 $ major shell, and $
N=14 $ sub-shell. The $N=28$ shell for stable nuclei fails to
appear, as the $2p_{3/2}$ and $2p_{1/2}$ come so close to
$1f_{7/2}$. When $N \ge 20 $, the neutrons are filled to the
levels in the continuum or weakly bound states in the order of
$1f_{7/2}$, $2p_{3/2}$, $2p_{1/2}$, and $1f_{5/2}$. In the right
part, the occupation probabilities in the canonical basis of all
the neutron levels below  $E = 10$ MeV have been given for
$^{35}$Na to show how the levels are filled in nuclei near the
drip line. The importance of careful treatment of the pairing
correlation and the scattering of particle pairs
to higher lying levels are noted in the figure.

\subsection{Interaction cross section}

The neutron and proton distributions and radii are fundamental quantities to define the halo phenomena.
However, the radius for unstable nucleus is normally extracted from
the interaction cross sections by assuming an empirical form (e.g., a Gaussian function)
for the density distribution \cite{Tanihata1985_PRL55-2676}.
It would be better to use the density distribution from
the self-consistent models and calculate the interaction cross sections directly from Glauber model.

\begin{figure}
\centering
\includegraphics[width=8.0cm]{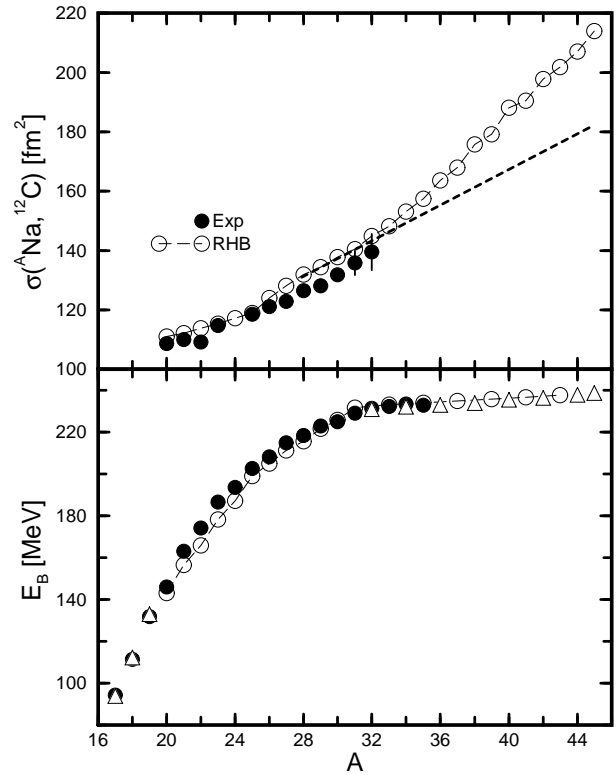}
\caption{Upper part: The interaction cross sections $\sigma_I$  of
$^A$Na isotopes on a carbon target at $950 A$ MeV: the open
circles are the result of RCHB calculation and the available experimental data
($A =$20--23, 25--32) are given by solid circles with their
error bars. The dashed line is a simple extrapolation based on the
RCHB calculation for $^{28-31}$Na. Lower part: Binding energies for
Na isotopes, the convention is the same as the upper part, but the
RCHB result for particle unstable isotopes are indicated by
triangle. Taken from Ref.~\cite{Meng1998_PLB419-1}.} \label{Fig13}
\end{figure}

Such systematic calculations have been carried out for all the nuclei in
Na isotopes with mass number $A$ ranging from 17 to 45 with the RCHB theory
and the effective interaction NLSH \cite{Meng1998_PLB419-1}. The calculated
binding energies $E$ and the interaction cross sections with the
Glauber model are shown in Fig.~\ref{Fig13}. The calculated
binding energies $E$ are compared with the empirical
values~\cite{Audi1995_NPA595-409}.
The proton (neutron) drip line nucleus has been predicted to be $^{19}$Na ($^{45}$Na).
The difference between
the calculations and the empirical values for the stable isotopes
is from the deformation, which has been neglected in Ref.~\cite{Meng1998_PLB419-1}.

As the interaction cross sections $\sigma_I$ is well reproduced,
the density distributions of the whole isotopes from the RCHB theory
have been examined and the relation between the development of halo and
shell effect has been studied in detail \cite{Meng1998_PLB419-1}.
It was found that the neutron $N=28$ shell closure fades due to
the lowering of the $2p$ orbitals.
The tail in the nuclear density profile depends strongly on the shell structure
and the development of a halo was connected with the changes in the occupation
in the next shell or sub-shell close to the continuum limit \cite{Meng1998_PLB419-1}.

Similar calculations have been performed for C, N, O and F isotopes
up to the neutron drip line by the self-consistent RCHB theory~\cite{Meng2002_PLB532-209}.
The agreement between the calculated one neutron separation energies $S_n$ and the available experimental values are similar as Na isotopes.
A Glauber model calculation for the total charge-changing cross section
has been carried out with the density distribution obtained from the RCHB
theory. The measured
cross sections for studied nuclei with $^{12}$C as a
target are nicely reproduced. An important conclusion was found that, contrary
to the usual impression, the proton density distribution
is less sensitive to the neutron number along
the isotope chain. Instead it is almost unchanged from
stability to the neutron drip line.
Such calculations are quite useful in extracting both the proton
and neutron distributions inside the nucleus and in defining the neutron/proton skin or halo.

\subsection{Surface diffuseness correction in global mass formula}

For finite nuclei, the diffuseness of nuclear surface is an important
degree-of-freedom in the calculations of nuclear masses~\cite{Wang2014_PLB734-215}.
The density distributions of most stable nuclei are of
the ``neutron skin-type'',  with a typical value around $0.5$ fm
for the surface diffuseness.
For nuclei near the neutron drip line, halos and giant halos may develop and the neutrons may distribute more extensive spatially than the protons, which implies the enhanced neutron surface diffuseness
for these extremely neutron-rich nuclei.
However, in global mass calculations,
the surface diffuseness of exotic nuclei near the drip lines
has not been properly considered. Recently, the surface diffuseness from
the RCHB theory and its impact  on the symmetry energy and shell correction
have been studied in Ref.~\cite{Wang2014_PLB734-215} and
the accuracy of the global mass formula has been considerably improved.

Inspired by the Skyrme energy-density functional,
a macroscopic-microscopic mass formula, the Weizs\"acker-Skyrme (WS) formula~\cite{Wang2010_PRC81-044322, Wang2010_PRC82-044304, Liu2011_PRC84-014333}, was proposed with a rms deviation around 336 keV
with respect to the 2149 measured masses~\cite{Audi2003_NPA729-337}
in 2003 Atomic Mass Evaluation (AME).
The WS models provide the best accuracy for nuclear masses
in several mass regions~\cite{Sobiczewski2014_PRC89-024311, Sobiczewski2014_PRC90-017302}
and has been used in r-process simulations~\cite{LiZhu2012ActaPhysicaSinica72601,Mendoza-Temis2014_arXiv1409.6135}.

In the WS formula, an axially deformed Woods-Saxon potential,
with a contant surface diffuseness parameter $a$ for all nuclei,
was used to obtain the single particle levels of nuclei.
In Ref.~\cite{Wang2014_PLB734-215}, by taking into account
the surface diffuseness correction for unstable nuclei,
a new global mass formula, WS4, has been proposed.
The surface diffuseness $a$ of the Woods-Saxon potential was
given by $a=a_0 \left (1+ 2 \varepsilon \delta_q \right )$, with
$\varepsilon=(I-I_0)^2-I^4$ the correction factor for surface diffuseness, $a_0$ the diffuseness of the Woods-Saxon potential, $I_0=0.4A/(A+200)$ the isospin asymmetry of the nuclei
along the $\beta$-stability line described by Green's formula, and
$\delta_q=1$ for neutrons (protons) in the nuclei with $I>I_0$ ($I<I_0$)
and $\delta_q=0$ for other cases.
This means that the surface diffuseness of neutron distribution is larger than
that of protons at the neutron-rich side and smaller than that of protons
at the proton-rich side.
The rms deviation with the available mass data
\cite{Audi2012_ChinPhysC36-1287} falls to 298 keV,
crossing for the first time the 0.3 MeV accuracy threshold for
mass formulas (models) within the mean field framework.

\begin{figure}
\includegraphics[angle=-0,width=1.0\textwidth]{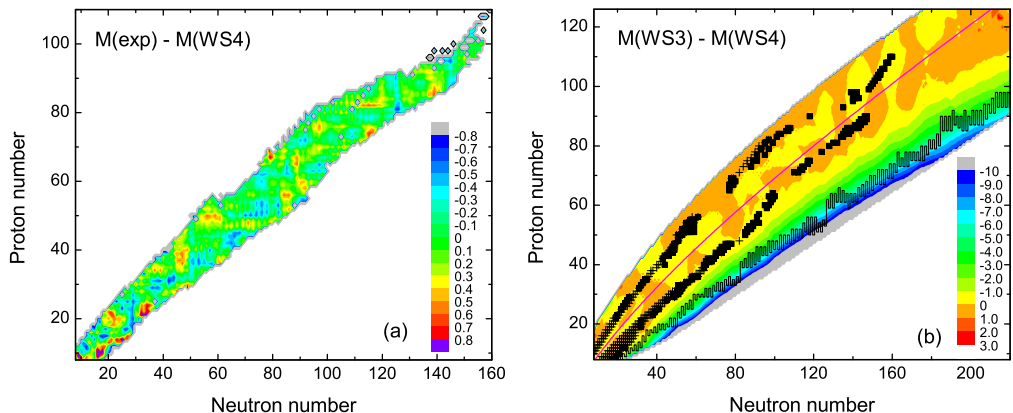}
\caption{(Color online)\label{fig:Wang2014_PLB734-215_Fig14}
(a) Difference between measured and the WS4 calculated masses.
(b) Difference between WS3 and WS4 calculated masses.
The squares and crosses denote the nuclei first appearing in AME2012 and
the nuclei with $|I-I_0|>0.1$, respectively.
The smooth and the zigzag curves denote the $\beta$-stability line from
the Green's formula and the neutron drip line from the WS4 formula, respectively.
Taken from Ref.~\cite{Wang2014_PLB734-215}.
}
\end{figure}

In Fig.~\ref{fig:Wang2014_PLB734-215_Fig14}(a), it was shown
the deviations of the calculated masses from the experimental values.
For all the 2353 nuclei with $Z$ and $N\ge 8$, the deviation is within 1.23 MeV.
In Fig.~\ref{fig:Wang2014_PLB734-215_Fig14}(b),
the difference between the WS3 (constant surface diffuseness $a_0$) and WS4 calculated masses.
For most nuclei, the results of WS3 and WS4 are consistent in general
(with deviations smaller than one MeV).
For nuclei near the neutron drip line, the masses given by WS4
are larger than the results of WS3 by several MeV.
This is due to the enhancement of the nuclear symmetry energy coming
from the surface diffuseness effect in the extremely neutron-rich nuclei.

It is worthwhile to note that with an accuracy of 258 keV for all
the available neutron separation energies and
of 237 keV for the alpha-decay $Q$-values of superheavy nuclei,
the proposed mass formula will be important not only for
the reliable description of the r-process of nucleosynthesis but also
for the study of the synthesis of superheavy nuclei.

\section{\label{sec:pairing_size}Pairing correlations and nuclear size}

In Ref. \cite{Bennaceur2000_PLB496-154}, it is shown that the neutron ground state Hartree-Fock (HF) asymptotic density equals
\begin{equation}
   \rho(r) \propto \exp{(-2 \mu r )/ r^2} ,
\label{densityHF}
\end{equation}
with $\mu = \sqrt{-2m \varepsilon_k} /\hbar$ and $\varepsilon_k$ the single particle energy
of the least bound $l=0$ neutron. The corresponding mean square radius deduced from the asymptotic solution is
\begin{equation}
   \langle r^2\rangle_{\mathrm{HF}} \propto \frac{\hbar^2}{2m|\varepsilon_k|},
\label{radiusHF}
\end{equation}
which diverges in the limit $\varepsilon_k \rightarrow 0$.

In the presence of pairing, the mean square radius deduced from the asymptotic Hartree-Fock-Bogoliubov (HFB) density is
\begin{equation}
   \langle r^2\rangle_{\mathrm{HFB}} \propto \frac{\hbar^2}{2m(E_k-\lambda)}
\label{radiusHFB}
\end{equation}
with the lowest discrete quasiparticle energy $E_k = \sqrt{(\varepsilon_k-\lambda)^2+\Delta_k^2}$. If the pairing gap $\Delta_k$ is finite, the radius will never diverge in the limit of small separation energy $\varepsilon_k\simeq \lambda \rightarrow 0$.

The asymptotic HF and HFB densities characterized by $l=0$ orbitals were compared in Ref.~\cite{Bennaceur2000_PLB496-154} and it has been emphasized that pairing
correlations reduce the nuclear size and an extreme halo with infinite
radius cannot be formed in superfluid nuclear systems. Here pairing correlations act against the formation of an infinite radius. Therefore this effect was called
``pairing anti-halo effect" in Ref.~\cite{Bennaceur2000_PLB496-154}.

In fact, the mechanism leading to a halo in Eq.~(\ref{radiusHF}), i.e., the valence nucleons occupy the weakly-bound orbital below and very close to the continuum threshold, has been used in early
interpretation for nuclear halo phenomena~\cite{Bertsch1989_PRC39-1154,
Sagawa1992_PLB286-7, Zhu1994_PLB328-1}.
The limiting condition $\varepsilon_k \rightarrow 0$ and the radius deduced from
this $l=0$ orbital alone correspond to an extremely ideal situation, which is
difficult to be found in real nuclei.

In Ref.~\cite{Chen2014_PRC89-014312}, the influences of pairing correlations
on the nuclear size and on the development of a nuclear halo were studied in details by the self-consistent RCHB theory~\cite{Meng1998_NPA635-3}. In order to simplify the problem,
neutron-rich nuclei with the neutron Fermi surface
below, between and above two weakly-bound $2p$ levels are investigated with a fixed Wood-Saxon potential, which is followed by the self-consistent calculation on the well
studied neutron halos in $^{11}$Li and $^{32}$Ne with low-$l$ orbitals in the
continuum.

\subsection{Fermi surface and nuclear size}

In order to investigate the impact of nuclear binding on nuclear size for varying pairing potential, a spherical Woods-Saxon shape mean field potential fitted to the corresponding neutron potentials resulting from a self-consistent calculation for $^{42}$Mg is adopted~\cite{Chen2014_PRC89-014312}.

The total neutron radius of the nucleus is given by
\begin{equation}
   R_N= \frac{1}{N}
        \sum_{nlj} (2j+1)\,\langle r\rangle_{nlj}\,v^2_{nlj},
   \label{eq:radius}
\end{equation}
which is determined by the rms radii of the orbitals $\langle r\rangle_{nlj}$ and
the corresponding occupation probabilities $v^2_{nlj}$.

\begin{figure*}[tbh]
\centering
\includegraphics[width=15.5cm]{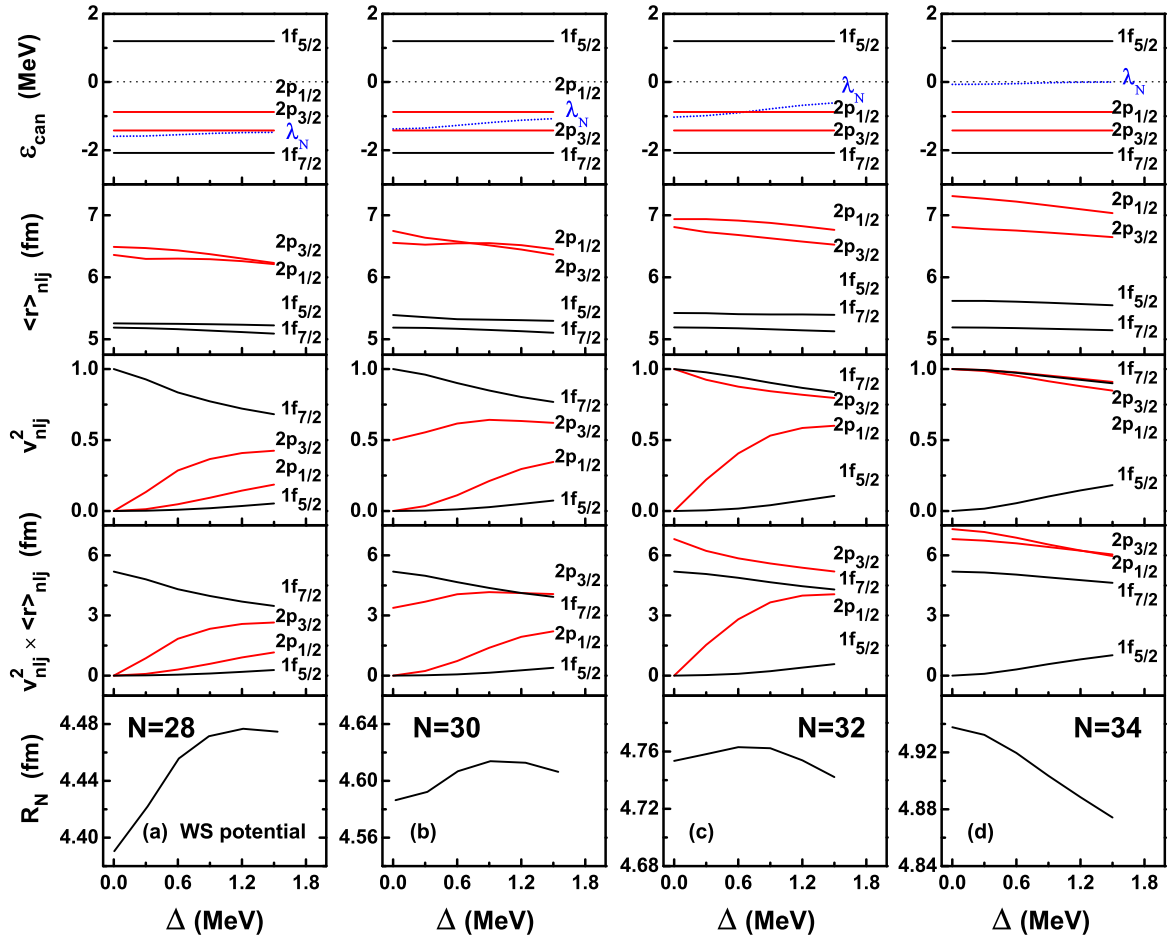}
\caption{(Color online) Single neutron levels $2p$ and $1f$ near the Fermi surface in the canonical
basis, the rms radii $\langle r\rangle_{nlj}$, occupation
probabilities $v^2_{nlj}$, contributions $v^2_{nlj}\times\langle
r\rangle_{nlj}$ to the neutron rms radii, and the total neutron rms radii $%
R_{N}$ for $N=28,~30,~32$, and $34$ as a function of average pairing gap $\Delta$.
The Fermi surfaces for the neutron numbers are plotted as dashed lines.
Taken from Ref.~\cite{Chen2014_PRC89-014312}.}
\label{Fig15}
\end{figure*}

In Fig.~\ref{Fig15}, the single neutron levels $\varepsilon_k$ in the canonical basis~\cite{Ring1980}, the rms radii $\langle r\rangle_{nlj}$, the occupation probabilities $v^2_{nlj}$, the contributions $v^2_{nlj}\times \langle r\rangle_{nlj}$ to the neutron rms radius for the orbitals near the Fermi surface ($2p$ and $1f$), and the total neutron rms radii $R_{N}$ for four neutron-rich Mg isotopes with $N=28,~30,~32$, and $34$ are plotted as functions of
the pairing gap
\begin{equation}
   \Delta = \frac{1}{N}\sum_{nlj} (2j+1) \Delta^{(lj)}_{n} v_{nlj}^2.
   \label{eq:pairing_gap}
\end{equation}
Here $\Delta^{(lj)}_{n}$ are the diagonal matrix elements of the pairing
field in the canonical basis.

The single neutron levels in canonical basis are obtained by solving the corresponding
RCHB equations for various pairing strength. In principle,
the energy levels $\varepsilon_{nlj}=h^{(lj)}_{nn}$ depend on $\Delta$.
However, since this solution was obtained for fixed
WS potentials, as shown in the upper panels of Fig.~\ref{Fig15}, these
states remain almost unchanged with increasing pairing correlations.

For Mg isotopes with $N=28,~30,~32$, and $34$, the neutron Fermi surface (shown as dashed line in upper
panels of Fig.~\ref{Fig15}) is located just below the two weakly-bound $2p$ levels,
between them, crossing the $2p_{1/2}$ level, and above the $2p$ levels.
As the pairing strength increases, more neutrons
are scattered from occupied levels in the Fermi sea to empty levels above
the Fermi level, and therefore the Fermi surface is raising.

In Fig.~\ref{Fig15}, the rms radii of the $2p$ orbitals are much larger than those of the $1f$ orbitals due to the lower centrifugal barrier. With increasing pairing correlations, the rms
radius decreases for $2p$ and $1f$ levels. This is the so-called ``pairing anti-halo
effect" discussed in Ref.~\cite{Bennaceur2000_PLB496-154}. However, this effect
concerns only the radii $\langle r\rangle_{nlj}$ of the individual orbitals.
The total neutron radius of the nucleus in Eq.~(\ref{eq:radius}) is determined by
the rms radii of the orbitals and the occupation probabilities,
which depend strongly on the pairing correlations.

As shown in Fig.~\ref{Fig15}, for $N=28$ the neutron Fermi surface is just
below the two weakly-bound $2p$ levels. Without pairing correlations, the
occupation probability is $1.0$ for the $1f_{7/2}$ orbital, and it vanishes
for the $2p$ and the $1f_{5/2}$ orbitals. As the pairing strength increases,
the neutrons on $1f_{7/2}$ orbital are scattered to $2p$ and $1f_{5/2}$ orbitals
which have much larger rms radii. Therefore, the contributions to the total
neutron rms radius from $2p$ and $1f_{5/2}$ states grow more than the
contribution from $1f_{7/2}$ state decreases. As a result, the total
neutron rms radius increases monotonically, where the contributions from $2p$
orbitals play a dominant role.

For $N=30$ in Fig.~\ref{Fig15}, the $2p_{1/2}$
and the $1f_{5/2}$ orbitals are empty for zero pairing, while the $1f_{7/2}$
orbital is fully occupied and the $2p_{3/2}$ orbital is half occupied. With
increasing pairing strength, the neutrons in the $1f_{7/2}$ orbital are scattered
to the $2p$ orbitals which contribute strongly to the neutron rms radius.
Therefore, the total neutron rms radius $R_N$ increases with the growing
pairing strength up to $\Delta=0.9$ MeV. Increasing the pairing strength
further, the neutrons are continuously to be scattered from the $1f_{7/2}$ to
the $2p_{1/2}$ state, while the neutron number in the $2p_{3/2}$ orbital
decreases slightly and the resonance state $1f_{5/2}$ begins to be occupied.
Together with the monotonic decreasing rms radii of the individual orbitals,
the increase resulting from the $2p_{1/2}$ and $1f_{5/2}$ states is smaller
than the decreases of the contributions from the $2p_{3/2}$ and $1f_{7/2}$
states. Therefore, for very strong pairing correlations, the total neutron
rms radius finally decreases. As it is clearly seen this decrease is not
caused by the pairing anti-halo effect, because the radius of the $1f_{7/2}$
state stays completely constant. It has its origin in the reoccupation
caused by pairing.

For $N=32$ in Fig.~\ref{Fig15}, the $1f_{7/2}$
and $2p_{3/2}$ orbitals are fully occupied without pairing correlations, while
the $2p_{1/2}$ and $1f_{5/2}$ orbitals are empty. As the pairing strength
increases, the neutrons in the $1f_{7/2}$ and $2p_{3/2}$ orbitals are
scattered to the $2p_{1/2}$ orbital, whose rms radius is larger than the ones
of $1f_{7/2}$ and $2p_{3/2}$ levels. The contribution of the $2p_{1/2}$
orbitals causes an increase in the total neutron rms radius up to a pairing
gap $\Delta =0.9$ MeV. As the pairing strength continues to grow beyond this
value, the neutrons in the $2p_{3/2}$ state begin to be scattered to the $%
1f_{5/2}$ state which provides a smaller contribution than the $2p_{3/2}$
state, while the occupation probability remains almost constant for the $%
1p_{1/2}$ state. Together with the decreasing rms radius of the individual
orbitals, the total neutron rms radius finally decreases for very large
pairing.

For $N=34$, the two weakly-bound $2p$ levels are fully occupied for
zero pairing, and the Fermi surface is just above the weakly-bound $2p$
levels. With increasing pairing strength, the neutrons in the $1f_{7/2}$ and
$2p$ orbitals are scattered to the $1f_{5/2}$ orbital with a much smaller rms
radius than the $2p$ orbitals. The contributions to the total neutron rms
radius from the $1f_{7/2}$ and $2p$ orbitals decrease more than the increase
of the contribution from $1f_{5/2}$ orbital, and therefore the total neutron
rms radius finally decreases monotonically.
Here the two effects, decreasing rms radius of the individual orbitals and
changes of the occupation probabilities, act in the same direction. The
decrease of the total neutron radius $R_N$ with increasing pairing
correlations is not only from the decreasing rms radii of the individual
orbitals but also from the change of the occupation probabilities.

It is clear that pairing correlations can change the rms radii of the
individual weakly-bound orbitals and simultaneously the corresponding
occupation probabilities. As a result, they have an influence on the nuclear
rms radius and the total nuclear size. Furthermore, the contributions of the
weakly-bound $2p$ orbitals to the nuclear rms radius play an important role.

\subsection{Crucial low $l$ orbitals}

Following the investigation on the impact of nuclear binding on nuclear size, the next question is the impact of orbital binding on the nuclear size as a function of the pairing potential. In Ref.~\cite{Chen2014_PRC89-014312}, this problem has been investigated for the orbitals $2p$ as well-bound, weakly-bound, around the
continuum threshold, and in the continuum cases.

\begin{figure}[tbh]
\begin{center}
\includegraphics[width=7cm]{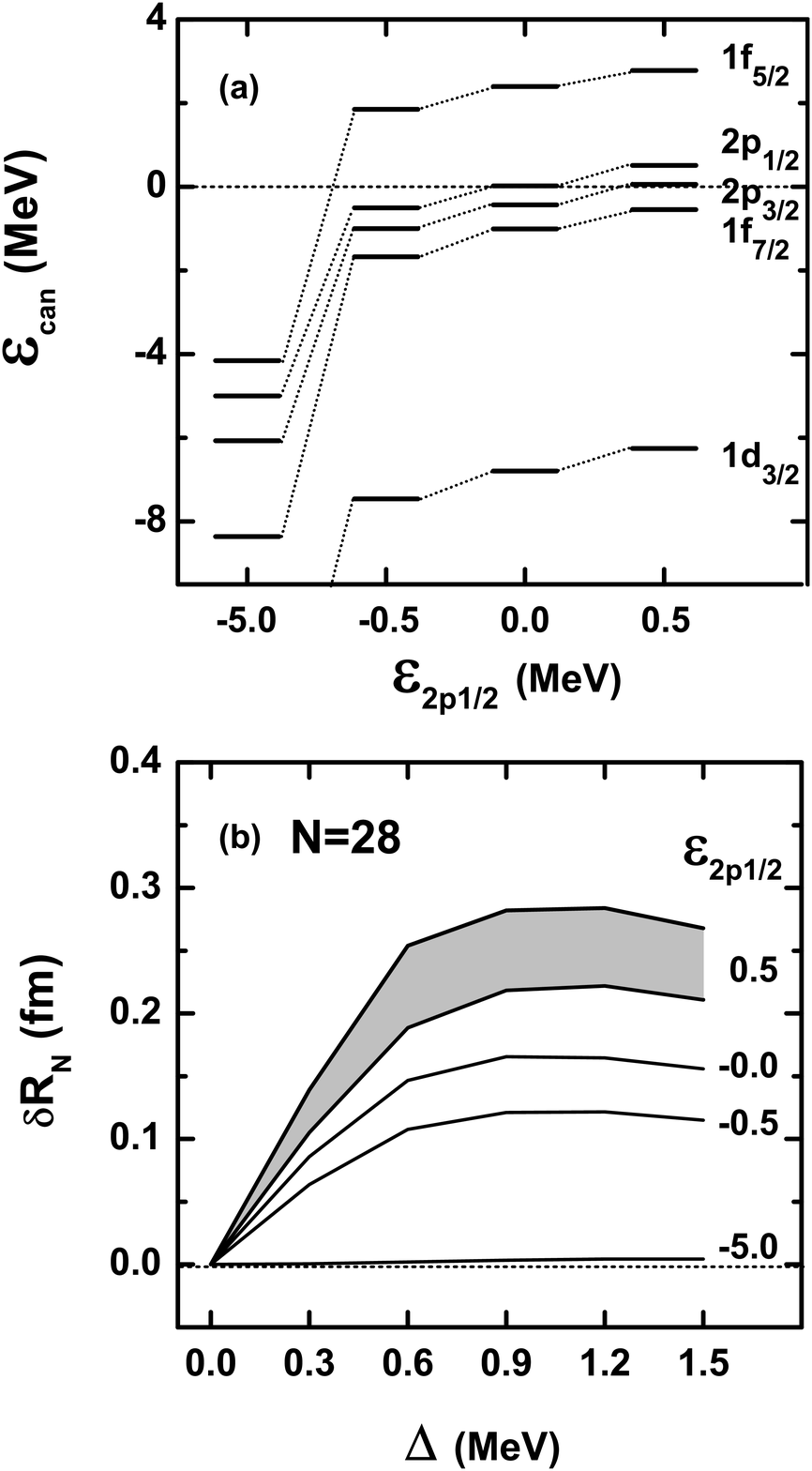}
\end{center}
\caption{(a) Single neutron levels $1d_{3/2}$, $1f_{7/2}$, $2p_{3/2}$, $%
2p_{1/2}$, and $1f_{5/2}$ in WS potentials of varying depth with $\protect%
\varepsilon_{2p_{1/2}}=-5.0,~-0.5,~0.0$, and $0.5$~MeV. (b) Changes of the total
neutron rms radii
$\protect\delta R_{N}=R_N(\Delta)-R_N(\Delta = 0)$
for the various cases of $\protect\varepsilon%
_{2p_{1/2}}$ as a function of average pairing gap $\Delta$. The shadow area
represents the results calculated with different box sizes in the WS
potential for $\protect\varepsilon_{2p_{1/2}}=0.5$ MeV.
Taken from Ref.~\cite{Chen2014_PRC89-014312}. }
\label{Fig16}
\end{figure}

In Fig.~\ref{Fig16}(a), the single neutron levels $%
1d_{3/2},~1f_{7/2},~2p_{3/2},~2p_{1/2}$, and $1f_{5/2}$ for $N=28$ in WS potentials with
varying depth such that $\varepsilon_{2p_{1/2}}=-5.0,~-0.5,~0.0$, and $+0.5$ MeV are shown.
In Fig.~\ref{Fig16}(b), for each $\varepsilon_{2p_{1/2}}$,
the change of the total neutron rms radius with respect to that of zero pairing,
$\protect\delta R_{N}=R_N(\Delta)-R_N(\Delta = 0)$, is plotted as a function of the pairing gap.

For $\varepsilon_{2p_{1/2}}=-5.0$ MeV which corresponds to a
well bound nucleus, the neutron rms radius $R_N(\Delta)$ increases slightly
with the pairing gap increasing from $0.0$ to $1.5$ MeV.
For $\varepsilon_{2p_{1/2}}=-0.5,~0.0,~+0.5$ MeV which
correspond to a weakly-bound nucleus, increases by more than $%
0.1$ fm are observed. It has to be noticed that for $\varepsilon_{2p_{1/2}}=+0.5$ MeV, both levels
$2p_{1/2}$ and $2p_{3/2}$ are in the continuum. If the pairing correlation is switched on, the system is basically not bound. The eigenvalue problem is solved in a finite box and the solution
depends on the box size. As an illustration, when the box size changes from $20$
to $25$ fm, the neutron rms radius calculated with a box
radius of $25$ fm is about $0.05$ fm larger than that obtained with with a
radius of $20$ fm box for $\Delta=1.5$ MeV as shown as shadowed region in Fig.~\ref{Fig16}(b).

For the discussion on the influence of pairing correlations on the nuclear size so far, the mean fields are fixed in the form of spherical WS potentials, and only the pairing strength and the Fermi level for $N=28,~30,~32,~34$ are considered. Here, with the changing pairing strength and the Fermi level, the single neutron levels stay almost constant except slight modification by the pairing field.
In Ref.~\cite{Chen2014_PRC89-014312}, fully self-consistent RCHB calculation
for neutron-rich nuclei $^{40,42,44,46}$Mg has been performed,
it should be emphasized that qualitatively the effect of pairing correlations on the
nuclear size agrees with that found for fixed WS potentials.

\begin{figure}[tbh]
\centering
\includegraphics[width=8.2cm]{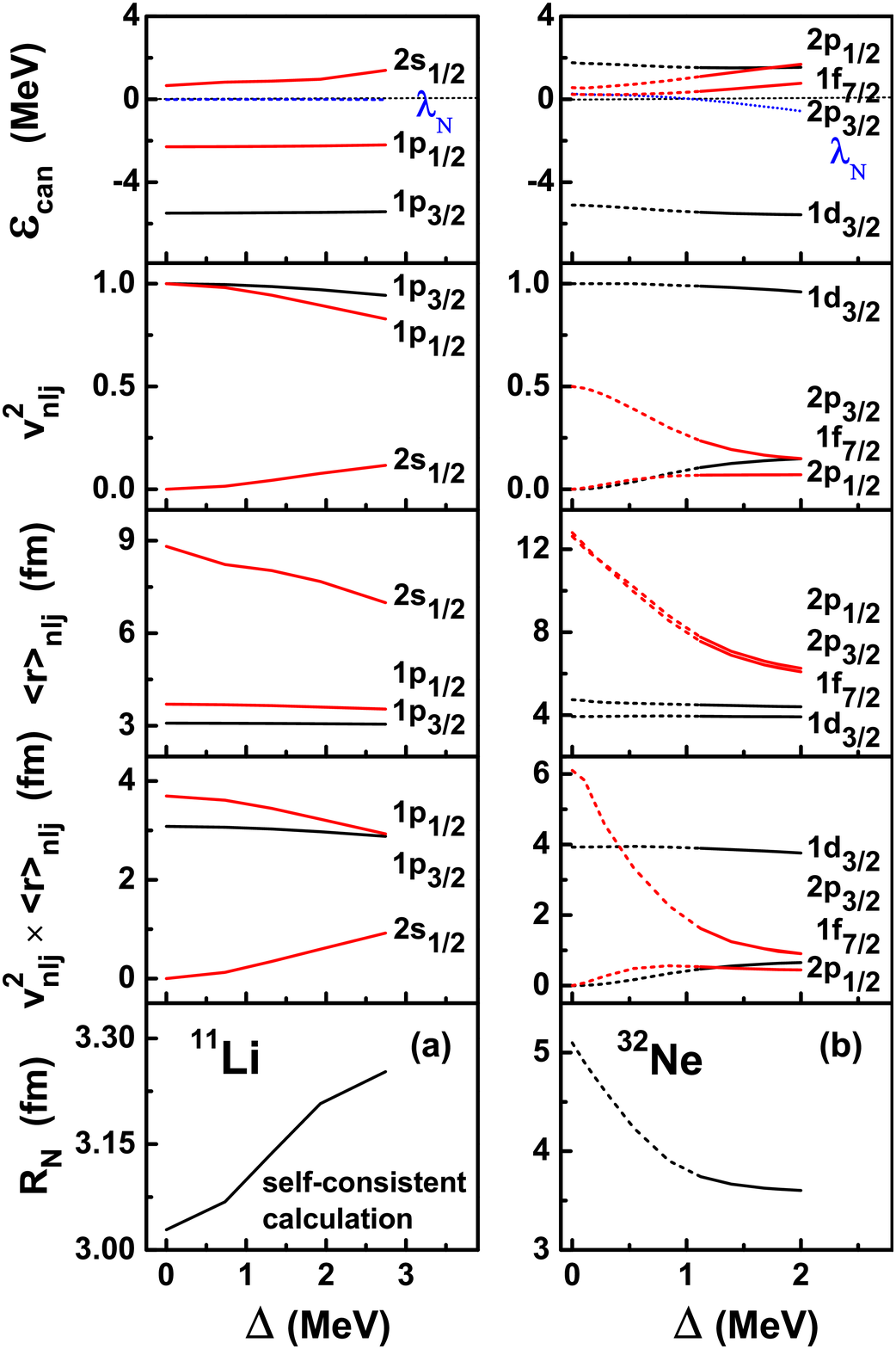}
\caption{(Color online) Same as Fig.~\protect\ref{Fig15}, but for $^{11}$Li
and $^{32}$Ne.
Taken from Ref.~\cite{Chen2014_PRC89-014312}.}
\label{Fig17}
\end{figure}

\subsection{Pairing correlations and halos}

In order to investigate the effect of pairing correlations on the
development of a nuclear halo, self-consistent calculations are performed with
different pairing strength for the well known neutron halo nuclei $^{11}$Li~%
\cite{Meng1996_PRL77-3963} and $^{32}$Ne~\cite{Poschl1997_PRL79-3841}%
, where the low-$l$ orbitals are in the continuum~\cite{Chen2014_PRC89-014312}.
Single neutron levels in the canonical basis, the rms radii $\langle r\rangle_{nlj}$,
the occupation probabilities $v^2_{nlj}$, the contributions to the neutron
rms radius $v^2_{nlj}\times\langle r\rangle_{nlj}$, and the total neutron
rms radii are plotted in Fig.~\ref{Fig17} as a function of the average
pairing gap for the nuclei $^{11}$Li (panel a) and $^{32}$Ne (panel b).

For $^{11}$Li in Fig.~\ref{Fig17}(a), a halo is developed by scattering the
neutrons to the $2s_{1/2}$ orbitals in continuum and
the increasing pairing correlations will promote the
development of this halo.

For $^{32}$Ne in Fig.~\ref{Fig17}(b), the $2p$ and $1f_{7/2}$ orbitals are in the
continuum. Without pairing correlations, the last two neutrons occupy the
$2p_{3/2}$ orbital and therefore this nucleus is unbound and the rms radius
depends on the box size. Increasing the pairing strength, the Fermi
surface comes down and becomes negative when the pairing gap gets larger
than $1.0$ MeV. It shows that for the halo nuclei such as $^{32}$Ne, where the
low-$l$ orbitals are in the continuum and unoccupied without pairing correlation,
increasing pairing correlations will decrease the size of the nuclear halo,
but nonetheless pairing plays an essential role: Without pairing the nucleus
would stay unbound.

It is concluded that the weakly-bounded orbitals with low orbital angular momenta $l$ in the
neighborhood of the threshold to the continuum play an important role~\cite{Chen2014_PRC89-014312}.
The pairing correlations have a two-fold influence on the total nuclear rms radius and the nuclear
size. First, they can change the rms radii of individual weakly bound orbitals
and, second, they can change the occupation probabilities of these orbitals.
With increasing pairing strength, the individual rms radii are reduced (pairing anti-halo
effect). Meanwhile pairing changes also the occupation probabilities. The
total nuclear size is determined by the competition between this two
effects.

In Ref.~\cite{Chen2014_PRC89-014312},
the strength of the pairing correlations was an external variable parameter.
In realistic nuclei, the strength
of pairing depends on the level density in the vicinity of the Fermi
surface. Therefore, even orbitals with high $l$-values and large centrifugal
barriers can contribute indirectly to changes of the nuclear size
\cite{Poschl1997_PRL79-3841}. If they come close to the Fermi surface, they
enhance pairing correlations and influence at the same time the occupation of
low-$l$ orbitals close to the continuum limit and therefore the nuclear size.

\section{\label{sec:deformed_halo}Deformed halo and shape decoupling}

Most open shell nuclei are deformed.
The interplay between deformation and weak binding raises interesting
questions, such as whether or not there exist halos in deformed
nuclei and, if yes, what are their new features.
Calculations in a deformed single particle model~\cite{Misu1997_NPA614-44}
have shown that the valence particles in specific orbitals with low
projection of the angular momentum on the symmetry axis
can give rise to halo structures in the limit of weak binding. The
deformation of the halo is in this case solely determined by the
intrinsic structure of the weakly bound orbitals. Indeed, halos in
deformed nuclei were investigated in several mean field calculations
in the past~\cite{Li1996_PRC54-1617,Pei2006_NPA765-29,Nakada2008_NPA808-47}. In Ref.~\cite{Hamamoto2004_PRC69-041306R}, it has been concluded that in
the neutron orbitals of an axially deformed Woods-Saxon potential
the lowest-$l$ component becomes dominant at large distances from
the origin and therefore all $\Omega^{\pi} = 1/2^+$ levels do not
contribute to deformation for binding energies close to zero. Such
arguments raise doubt about the existence of deformed halos. In
addition, a three-body model study~\cite{Nunes2005_NPA757-349} suggested that it
is unlikely to find halos in deformed drip line nuclei because the
correlations between the nucleons and those due to static or dynamic
deformations of the core inhibit the formation of halos.

Therefore a model which provides an adequate description of halos in
deformed nuclei must include in a self-consistent way the continuum,
deformation effects, large spatial distributions, and the coupling
among all these features. In addition it should be free of adjustable
parameters in order to make predictions reliable.

Over the past years, lots of efforts have been made to develop
a deformed relativistic Hartree (RH) theory~\cite{Zhou2006_AIPCP865-90} and
a deformed relativistic Hartree-Bogoliubov theory in continuum (DRHBc theory)~\cite{Zhou2008_ISPUN2007}.
As a first application, halo phenomena in deformed nuclei
have been investigated within the DRHBc theory~\cite{Zhou2010_PRC82-011301R}.
The detailed theoretical framework is
available in Ref.~\cite{Li2012_PRC85-024312}.
The DRHBc theory is useful to answer questions concerning exotic nuclear phenomena
in nuclei near the drip line
\cite{Li1996_PRC54-1617,
Misu1997_NPA614-44,Guo2003_CTP40-573, Meng2003_NPA722-C366, Nunes2005_NPA757-349,
Pei2006_NPA765-29}, such as whether there are deformed halos or not and
what new features can be expected in deformed exotic nuclei.
In Ref.~\cite{Li2012_CPL29-042101}, the DRHBc theory is extended to incorporate the blocking effect due to an odd nucleon.
In such a way, pairing correlations, continuum, deformation, blocking effects,
and extended spatial density distributions in exotic odd-$A$ or odd-odd nuclei can be
taken into account microscopically and self-consistently.
In recent years, the CDFT with the density-dependent meson-nucleon couplings
have attracted more and more attention owing to improved descriptions of
the equation of state at high density, the asymmetric nuclear matter and
the isovector properties of nuclei far from stability. Therefore, it's necessary to develop a density-dependent deformed relativistic Hartree-Bogoliubov (DDDRHB) theory in continuum. The challenge is the treatment on the density-dependent couplings, meson fields, and potentials in axially deformed system with partial wave method. Quite recently, the DDDRHB theory in continuum was developed in Ref.~\cite{Chen2012_PRC85-067301}.
In this Section, the halo in deformed exotic nucleus and its mechanism as well as possible shape decoupling will be briefly reviewed.

\subsection{Neutron separation energy and nuclear size }

\begin{figure}
\begin{center}
 \includegraphics[width=0.45\textwidth]{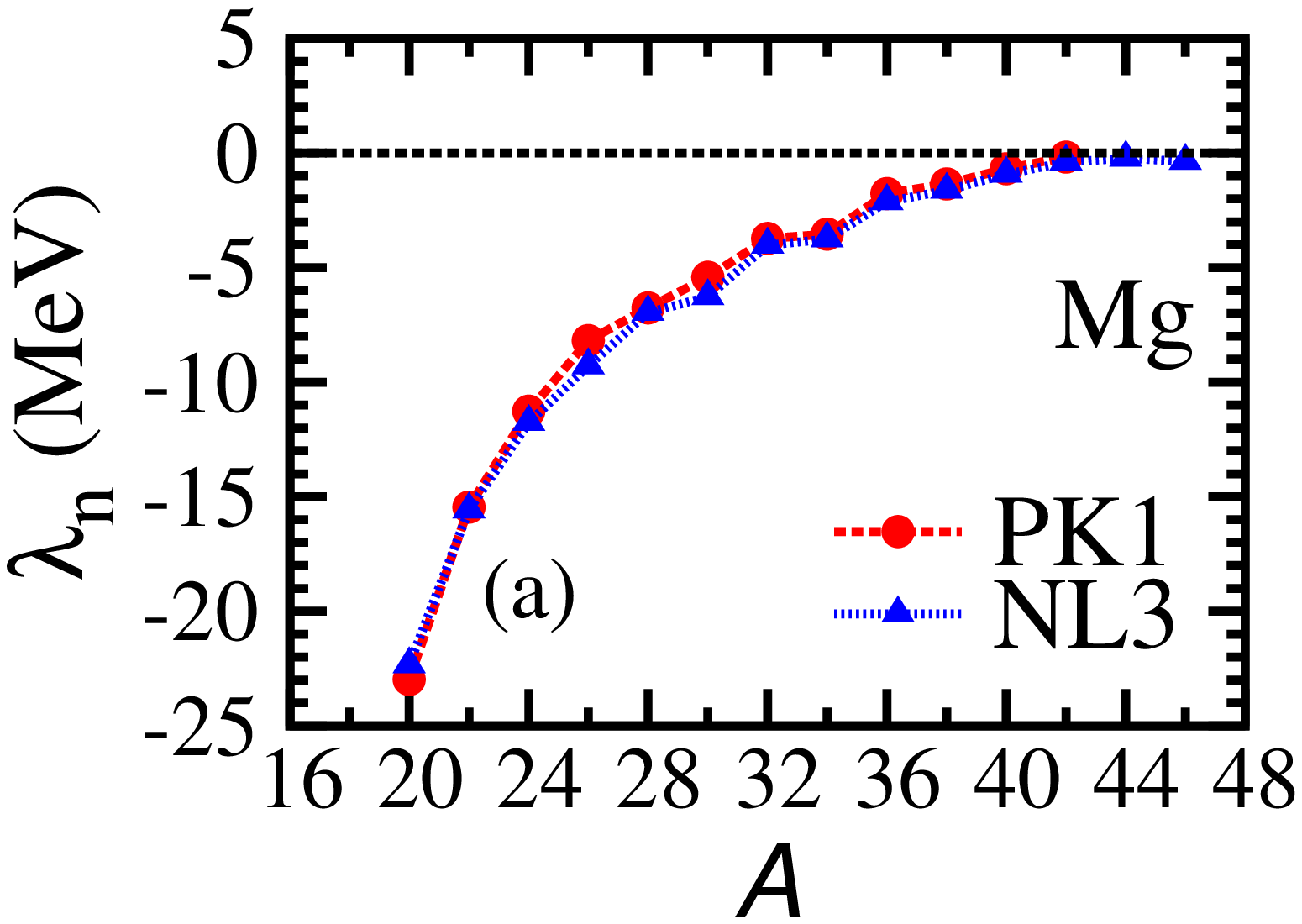}
~~~
 \includegraphics[width=0.45\textwidth]{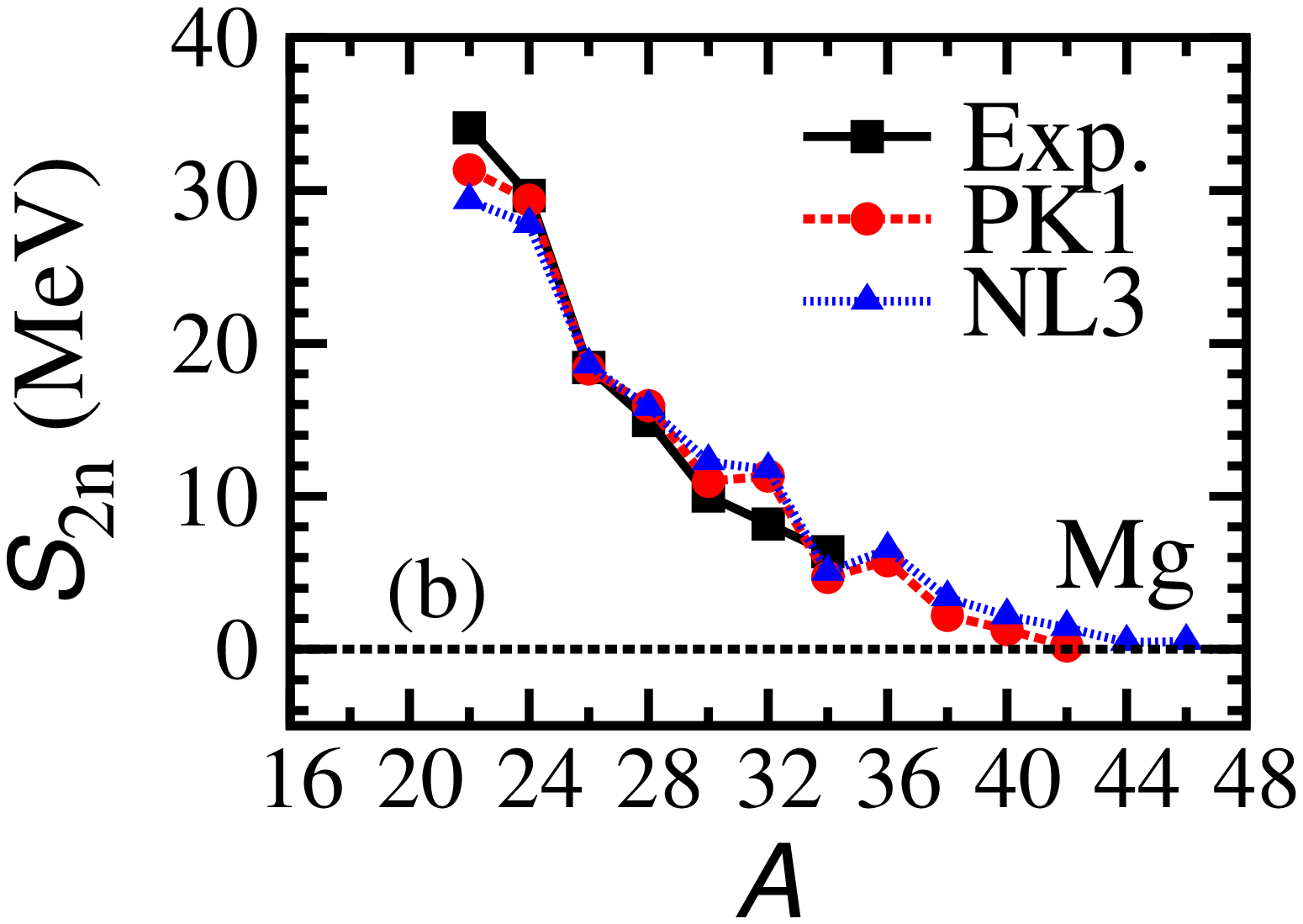}
\end{center}
\caption{(Color online) %
The neutron Fermi energy $\lambda_\mathrm{n}$ (left panel) and two neutron
separation energy $S_\mathrm{2n}$ (right panel) of Mg isotopes
calculated by DRHBc theory with the parameter sets NL3 and PK1.
The data of $S_\mathrm{2n}$ (labeled as ``Exp.'') are taken
from~\cite{Audi2003_NPA729-337}. Taken from Ref.~\cite{Li2012_PRC85-024312}.
}
\label{fig:Fig18}
\end{figure}

Figure~\ref{fig:Fig18} shows the neutron Fermi energy $\lambda_\mathrm{n}$
and two neutron separation
energy $S_\mathrm{2n}$ of Mg isotopes calculated in the DRHBc theory with the parameter
sets NL3 and PK1. Available data of $S_\mathrm{2n}$ taken from~\cite{Audi2003_NPA729-337}
are also included for comparison.
Except the different prediction of the drip line nucleus,
the neutron Fermi surfaces and two neutron separation
energies from both parameter sets are very similar.
The calculated two neutron separation energies $S_{2n}$ of Mg
isotopes agree reasonably well with the available experimental values
except for $^{32}$Mg. The large discrepancy in $^{32}$Mg is connected
to the shape and the shell structure at $N=20$. For details, see Ref.~\cite{Li2012_PRC85-024312}.

Experimentally $^{40}$Mg has been observed~\cite{Baumann2007_Nature449-1022}.
Theoretically there are different predictions on the last bound nucleus in Mg isotopes,
e.g., $^{44}$Mg from the finite range droplet model~\cite{Moller1995_ADNDT59-185},
$^{40}$Mg from a macroscopic-microscopic model~\cite{Zhi2006_PLB638-166},
a RMF model with the parameter
set NLSH~\cite{Ren1996_PLB380-241}, and a Skyrme HFB model solved in
a 3-dimensional Cartesian mesh~\cite{Terasaki1997_NPA621-706},
and $^{42}$Mg from a Skyrme HFB model solved in a transformed harmonic
oscillator basis~\cite{Stoitsov2003_PRC68-054312} and the HFB21 mass table
\cite{Goriely2010_PRC82-035804}.
Therefore the prediction of the neutron drip line nucleus in Mg isotopes
is model and parametrization dependent.
In the DRHBc calculations with the parameter set NL3, $^{46}$Mg is
the last nucleus of which the neutron Fermi surface is negative and
the two neutron separation energy is positive.
However, with the parameter set PK1, $^{42}$Mg is predicted to be
the last nucleus within the neutron drip line.

\begin{figure}
\begin{center}
 \includegraphics[width=0.5\textwidth]{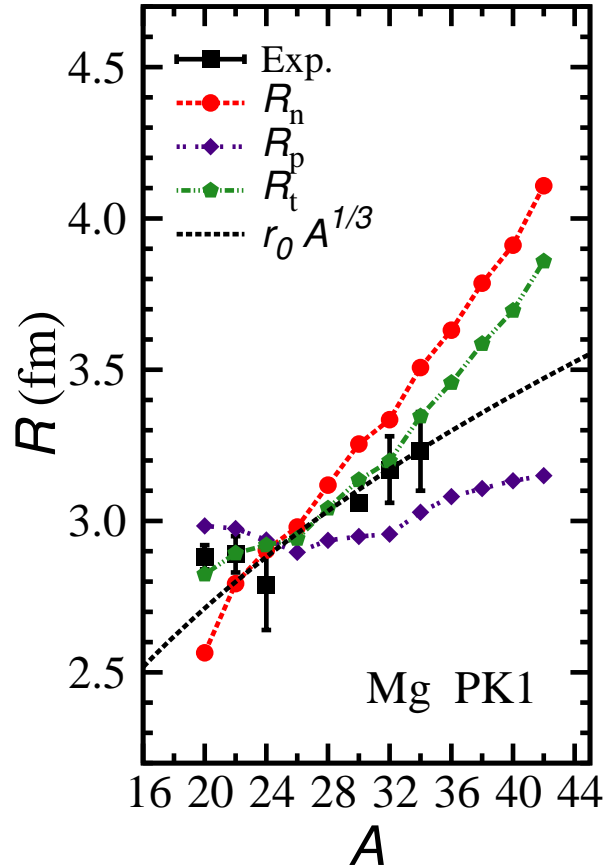}
\end{center}
\caption{(Color online) %
The root mean square (rms) radii for Mg isotopes calculated in DRHBc theory
plotted as functions of the neutron number.
The neutron radii $R_{\rm n}$, proton radii $R_{\rm p}$,
matter radii $R_{\rm t}$, the $r_0 A^{1/3}$ curve, and the available data for
$R_{\rm t}$~\cite{Suzuki1998_NPA630-661, Kanungo2011_PRC83-021302R} are given.
The neutron radius $R_{\rm n}$, proton radius $R_{\rm p}$ and total radius
$R_{\rm t}$ of Mg isotopes are calculated with PK1.
The $r_0 A^{1/3}$ curve is included to guide the eye.
The data for matter radii labeled by ``Exp.'' are taken from
Refs.~\cite{Suzuki1998_NPA630-661,Kanungo2011_PRC83-021302R}.
Taken from Ref.~\cite{Li2012_PRC85-024312}.
}
\label{fig:mg-radii}
\end{figure}

In Fig.~\ref{fig:mg-radii}, the rms radii for
Mg isotopes are plotted as functions of the neutron number,
including the neutron radii $R_{\rm n}$, proton radii $R_{\rm p}$,
matter radii $R_{\rm t}$, the $r_0 A^{1/3}$ curve, and the available data for
$R_{\rm t}$~\cite{Suzuki1998_NPA630-661, Kanungo2011_PRC83-021302R}.
The calculated matter radius follows roughly the $r_0 A^{1/3}$ curve up to $A=34$.
From $^{36}$Mg on, the matter radius lies much high above the $r_0 A^{1/3}$ curve.
This indicates some exotic structure in these nuclei.

\subsection{Density distributions and nuclear shape}

\begin{figure}
\begin{center}
 \includegraphics[width=0.5\textwidth]{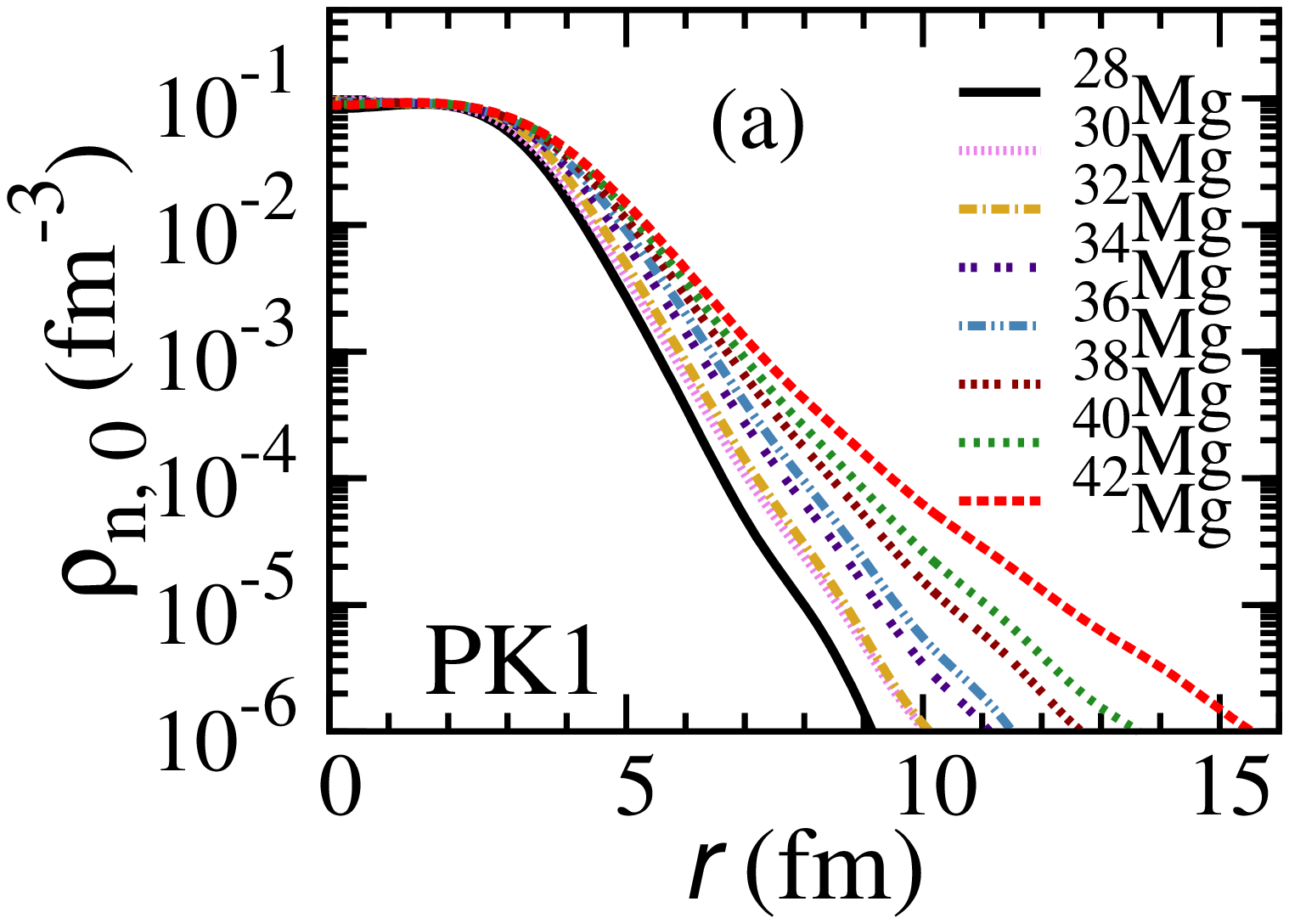}
~~
 \includegraphics[width=0.5\textwidth]{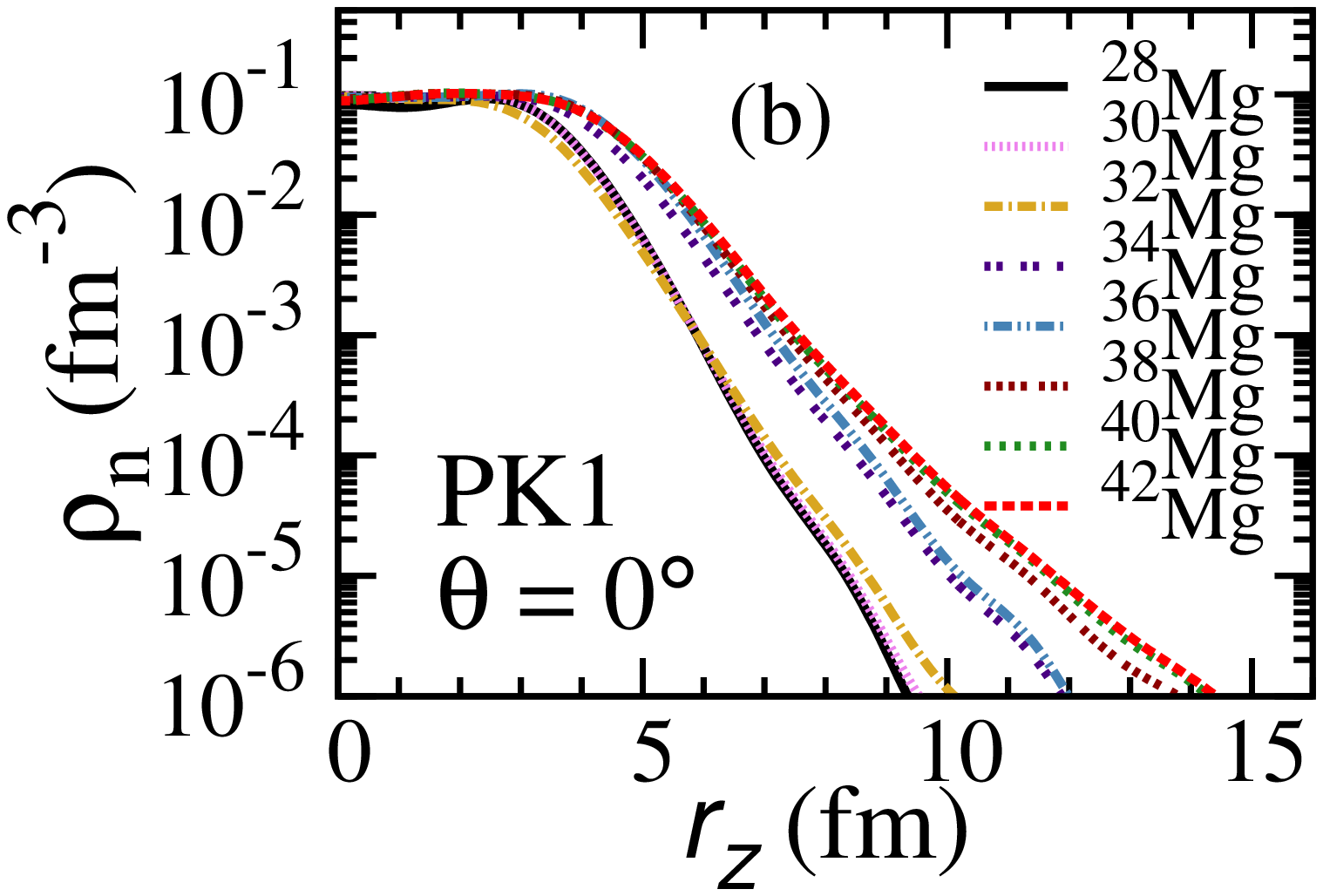}
~~
 \includegraphics[width=0.5\textwidth]{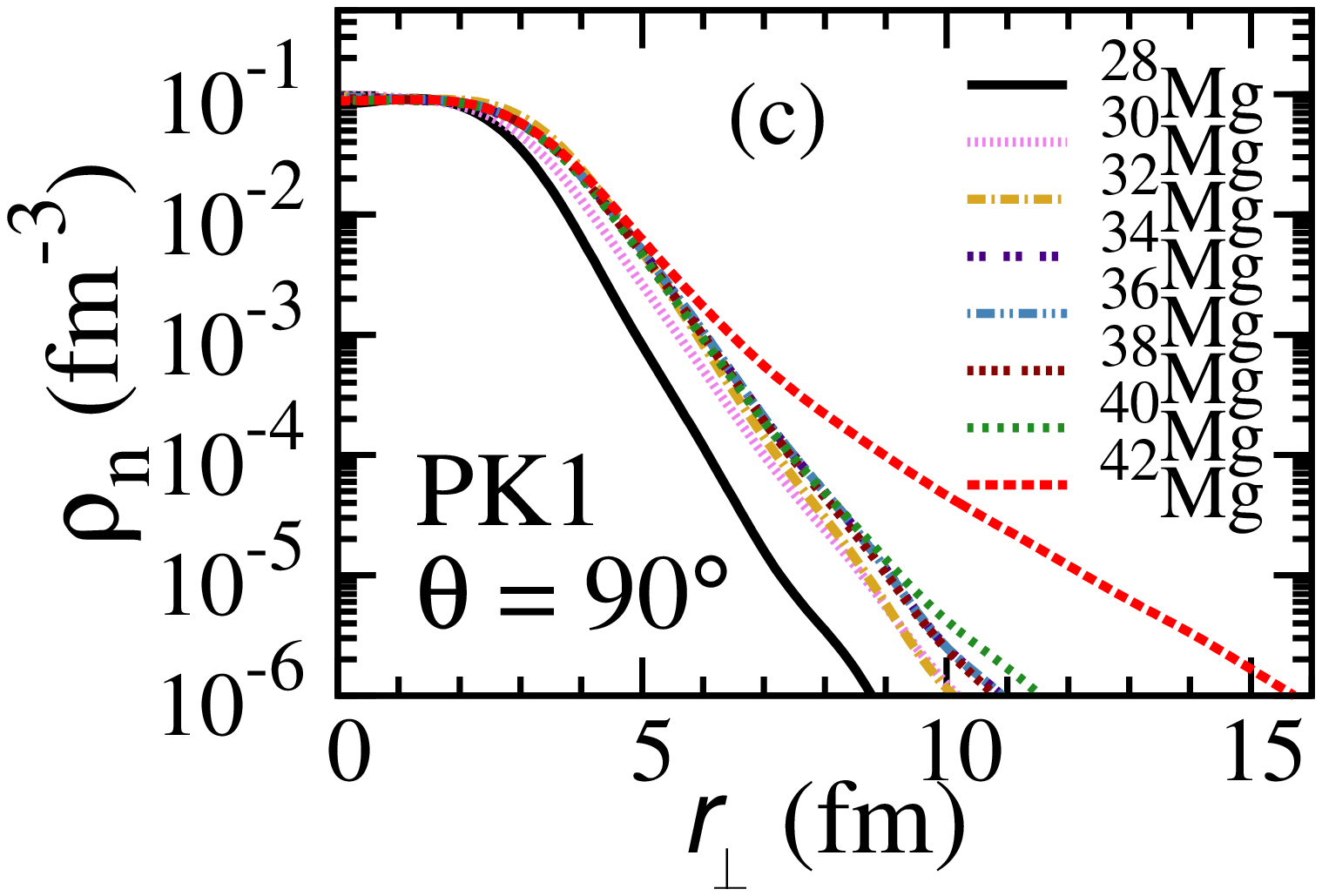}
\end{center}
\caption{(Color online) %
Neutron density profiles of even-even Mg isotopes with $A \ge 28$
calculated in DRHBc theory with the parameter set PK1.
The upper panel represents the spherical
component of the neutron density distribution $\rho_{\mathrm{n},\lambda=0}(r)$.
The middle panel refers to the density distribution $\rho_\mathrm{n}(z,r_\perp=0)$
along the symmetry axis $z$
( $\theta = 0^\circ$)
with $r_\perp = \sqrt{x^2+y^2}$ and the lower panel refers
to $\rho_\mathrm{n}(z=0,r_\perp)$  that perpendicular to
the symmetry axis $z$ ($\theta = 90^\circ$).
Taken from Ref.~\cite{Li2012_PRC85-024312}.
}
\label{fig:mg-dens-profile}
\end{figure}

Figure~\ref{fig:mg-dens-profile} shows
neutron density profiles of even-even Mg isotopes with $A \ge 28$
calculated in DRHBc theory with
the parameter set PK1~\cite{Li2012_PRC85-024312}.
The spherical component of the neutron
density distribution is represented by $\rho_{\mathrm{n},\lambda=0}(r)$.
The density distributions along the symmetry axis $z$ ($\theta = 0^\circ$) and
perpendicular to the symmetry axis $z$ ($\theta = 90^\circ$) are respectively represented by
$\rho_\mathrm{n}(z,r_\perp=0)$  and $\rho_\mathrm{n}(z=0,r_\perp)$
with $r_\perp = \sqrt{x^2+y^2}$.

With increasing $A$, the spherical component of the neutron density distribution
$\rho_{\mathrm{n},\lambda=0}(r)$ changes quickly at
$^{34}$Mg and abruptly at $^{42}$Mg.
The density distribution along the symmetry axis $\rho_\mathrm{n}(z,r_\perp=0)$
changes abruptly from $^{32}$Mg to $^{34}$Mg. This can be understood easily
from the shape change in these two nuclei: $^{32}$Mg is spherical but
$^{34}$Mg is prolate which means that it is elongated in the $z$ axis.
In the direction perpendicular to the symmetry axis, the neutron density
$\rho_\mathrm{n}(z=0,r_\perp)$ of $^{42}$Mg extends very far away from
the center of the nucleus and a long tail emerges, revealing the formation
of a halo.

By comparing $\rho_\mathrm{n}(z,r_\perp=0)$ and $\rho_\mathrm{n}(z=0,r_\perp)$
for $^{42}$Mg, it was found that in the tail part, the neutron density
distributes more along the direction perpendicular to the symmetry axis.
Since this nucleus as a whole is prolate, it indicates that the tail
part has a different shape from the nucleus itself.
This is the shape decoupling which was first predicted in Ref.~\cite{Zhou2010_PRC82-011301R}.

\begin{figure}
   \begin{center}
   \includegraphics[width=0.5\textwidth]{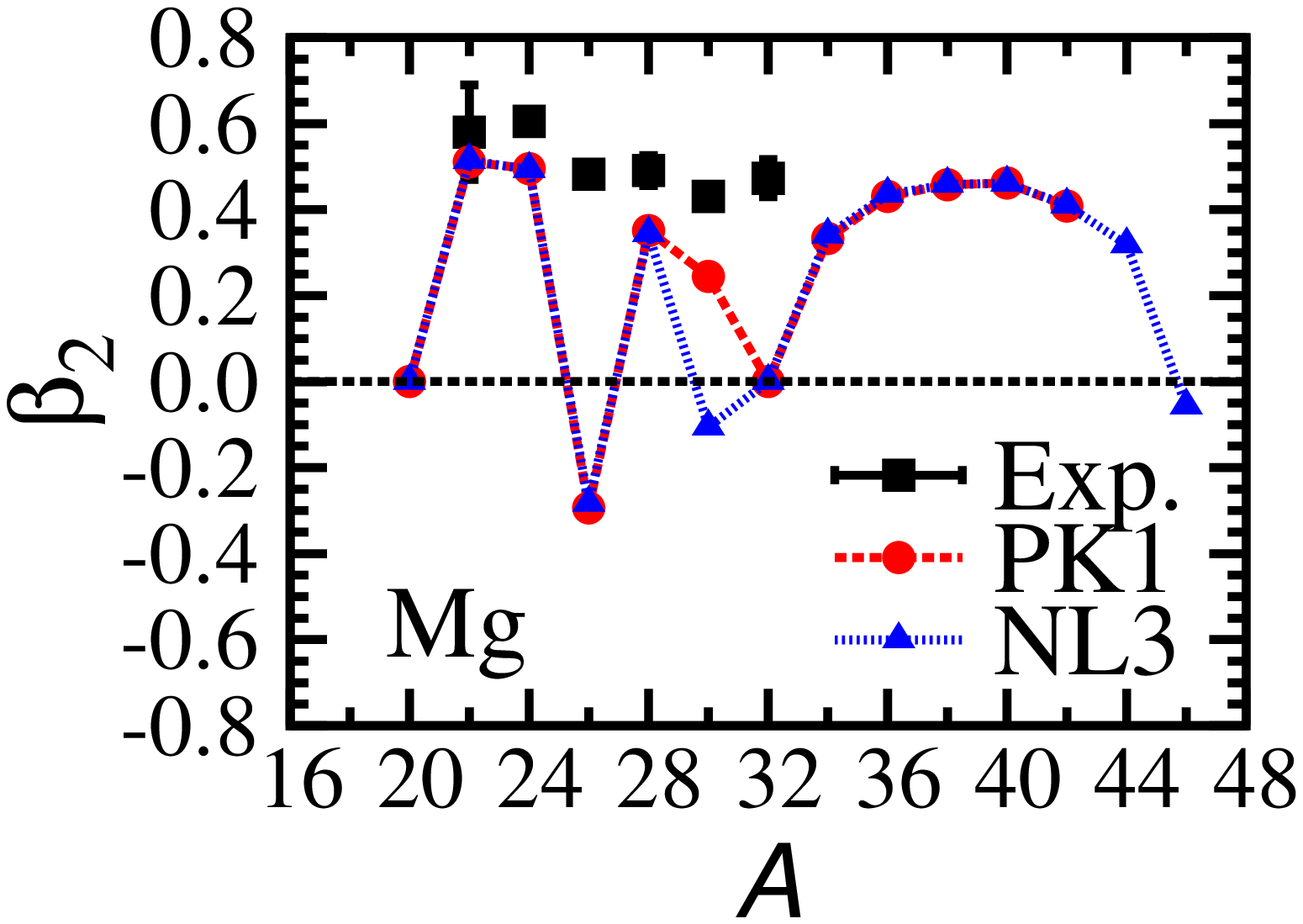}
   \end{center}
   \caption{(Color online) %
   The quadruple deformation parameter $\beta$ of Mg isotopes calculated with the parameter sets NL3 and PK1. The data labeled with ``Exp.'' are taken from~\cite{Raman2001_ADNDT78-1}.
   Taken from Ref.~\cite{Li2012_PRC85-024312}.
}
\label{fig:mg-beta}
\end{figure}

The comparison of quadrupole deformation $\beta$ between the deformed RHB calculation
and the experiment is given in Fig.~\ref{fig:mg-beta}.
The experimental value of $\beta$ is extracted from the measured
$B(E2:0^+_1 \rightarrow 2^+_1)$ and only the absolute value could be
obtained~\cite{Raman2001_ADNDT78-1}.
The results of NL3 are almost
the same as the results of PK1 except for $^{30}$Mg. It is slightly oblate
with NL3 but it has a prolate shape with PK1.
For $^{32}$Mg, the gap between the neutron levels $1d_{3/2}$ and $1f_{7/2}$
is almost $7$~MeV which results in a strong closed shell at $N = 20$.
Therefore the deformed RHB calculations with both parameter sets predict
spherical shapes for this nucleus.
This results in the large discrepancy in $S_\mathrm{2n}$ for $^{32}$Mg as
is seen in Fig.~\ref{fig:Fig18}.
Other mean field models also predict spherical or almost spherical shapes for
$^{32}$Mg~\cite{Patra1991_PLB273-13, Ren1996_PLB380-241, Terasaki1997_NPA621-706,
Yao2010_PRC81-044311, Yao2011_PRC83-014308}.
However, this nucleus is well deformed as is indicated by the large
$B(E2:0^+_1\rightarrow2^+_1)$ value~\cite{Church2005_PRC72-054320}.
The same is true for some neighboring nuclei with $N \sim 20$.
These nuclei form the so called ``island of inversion''~\cite{Sorlin2008_PPNP61-602,
Tripathi2008_PRL101-142504, Doornenbal2009_PRL103-032501, Wimmer2010_PRL105-252501}
which is related to the quenching of the $N=20$ shell closure.
In order to describe the shell quenching effect at $N = 20$, one has to
go beyond the mean field~\cite{Rodriguez-Guzman2000_PRC62-054319, Yao2011_PRC83-014308}.

\subsection{Deformed halo and shape decoupling}

The deformed relativistic Hartree-Bogoliubov theory in continuum with
the parameter set PK1 predicted the last bound neutron rich nucleus
for Mg isotopes as
$^{42}$Mg~\cite{Li2012_PRC85-024312}. There are both prolate and oblate
stable solutions for $^{42}$Mg. The ground state of $^{42}$Mg is prolate. With the
triaxially deformed relativistic mean field calculations~\cite{Lu2011_PRC84-014328},
it is shown that the obtained oblate minimum is not an isomeric state but a saddle
point in the potential energy surface.

It is found that
the ground state of $^{42}$Mg is well deformed with
$\beta \approx 0.41$ and two neutron separation energy
$S_{2n} \approx 0.22$ MeV~\cite{Li2012_PRC85-024312}.
The density distribution of such weakly bound nucleus has a very long tail
in the direction perpendicular to the symmetry axis, see Fig.~\ref{fig:mg-dens-profile},
which indicates the prolate $^{42}$Mg has an oblate halo.

\begin{figure}
\begin{center}
 \includegraphics[width=0.45\textwidth]{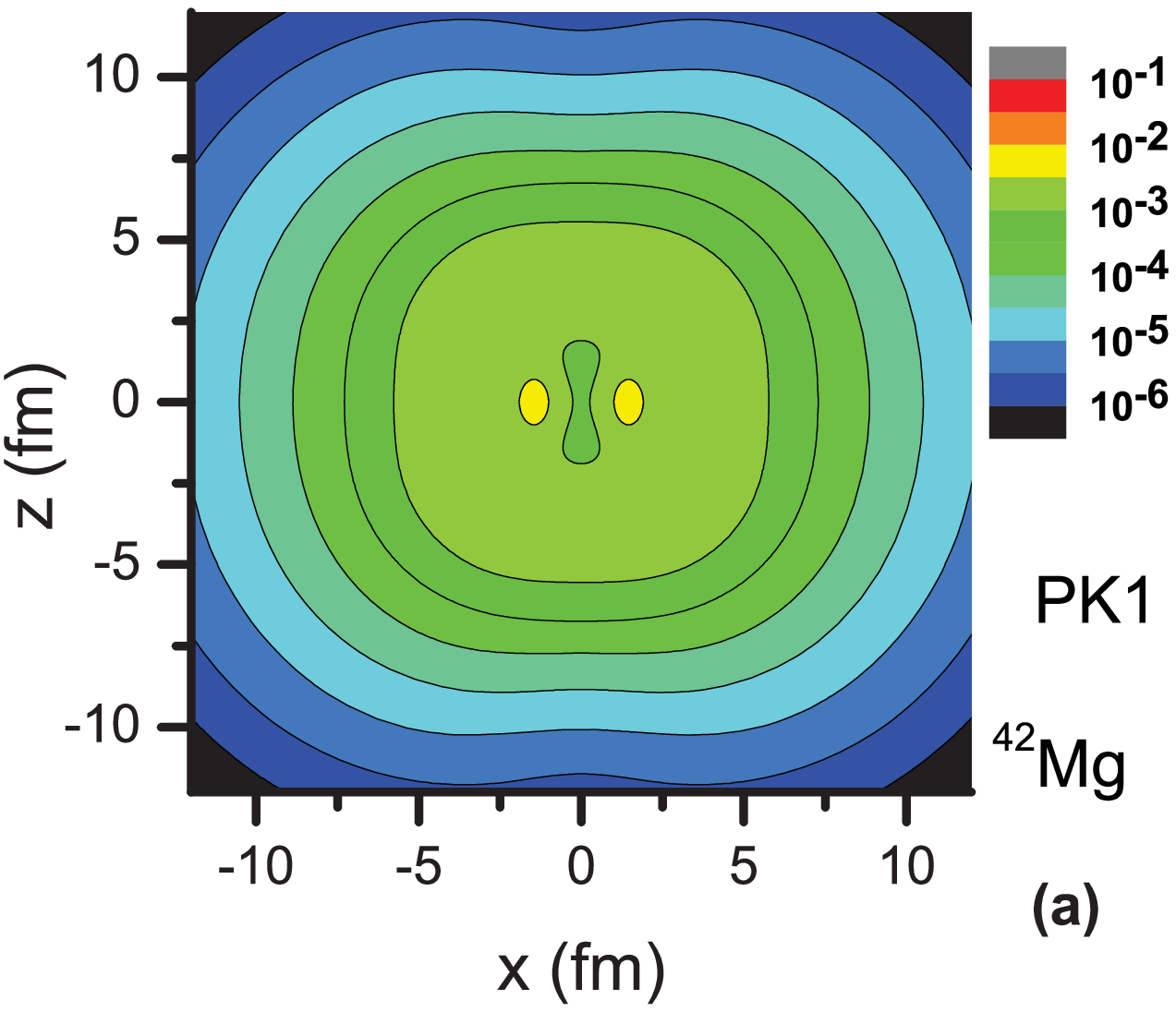}
~~~~~~~~~~
 \includegraphics[width=0.45\textwidth]{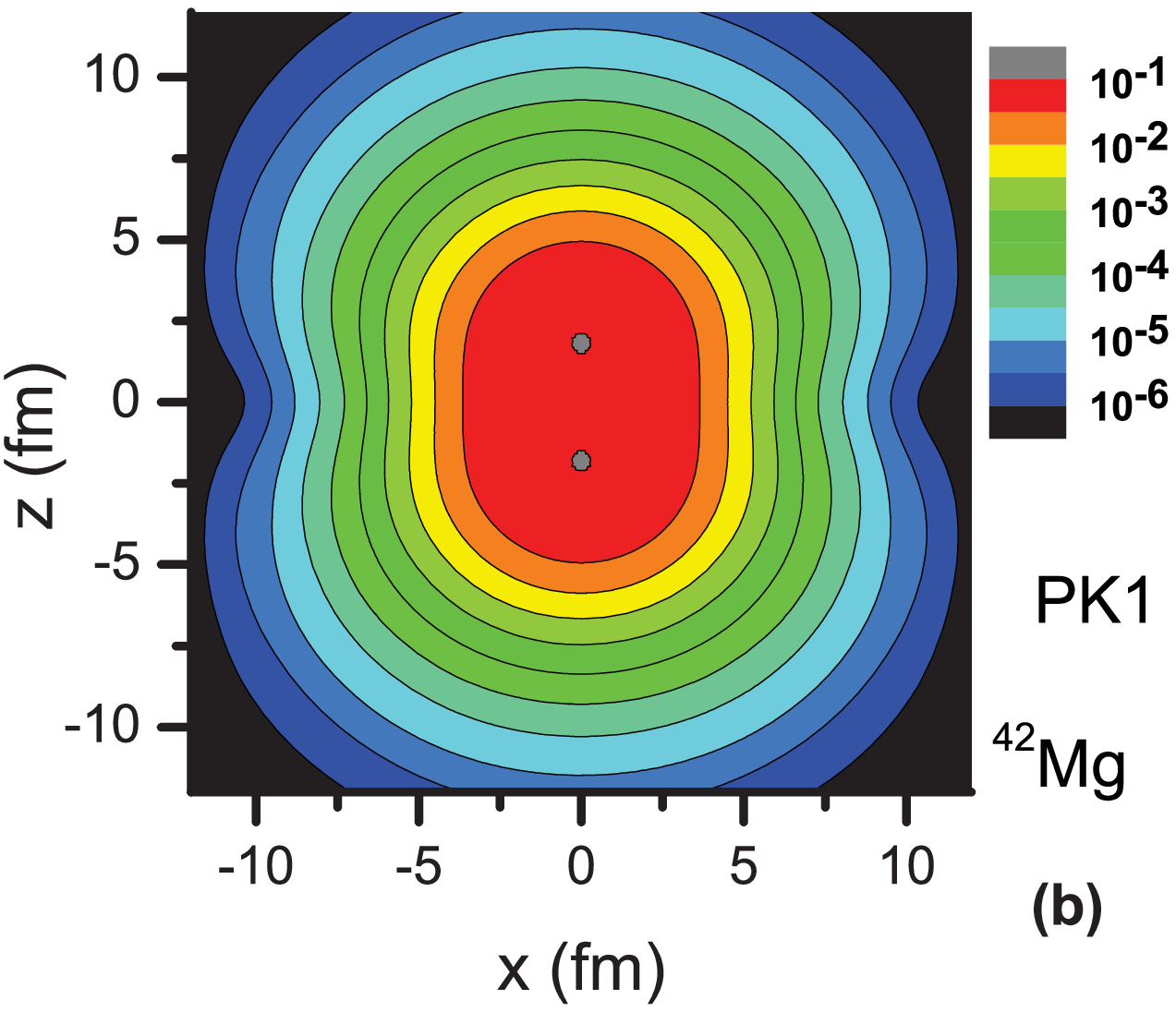}
\end{center}
\caption{(Color online)
\label{fig:Mg42_pro_2d}
Density distributions of the ground state of $^{42}$Mg with the
$z$ axis as the symmetry axis: (a) the neutron halo, and (b) the neutron core.
Taken from Ref.~\cite{Li2012_PRC85-024312}.
}
\end{figure}

The density distribution is divided into contributions of the oblate ``halo''
and of the prolate ``core'' in Fig.~\ref{fig:Mg42_pro_2d}.
The density distribution of this weakly bound nucleus has a very long tail
in the direction perpendicular to the symmetry axis which indicates
the prolate nucleus $^{42}$Mg has an oblate halo and
there is a decoupling between the shapes of the core and the halo.
It should be noted that in the calculations with NL3, $^{46}$Mg is
the last nucleus~\cite{Zhou2010_PRC82-011301R}.

\begin{figure}[tbh]
\begin{center}
\includegraphics[width=10cm]{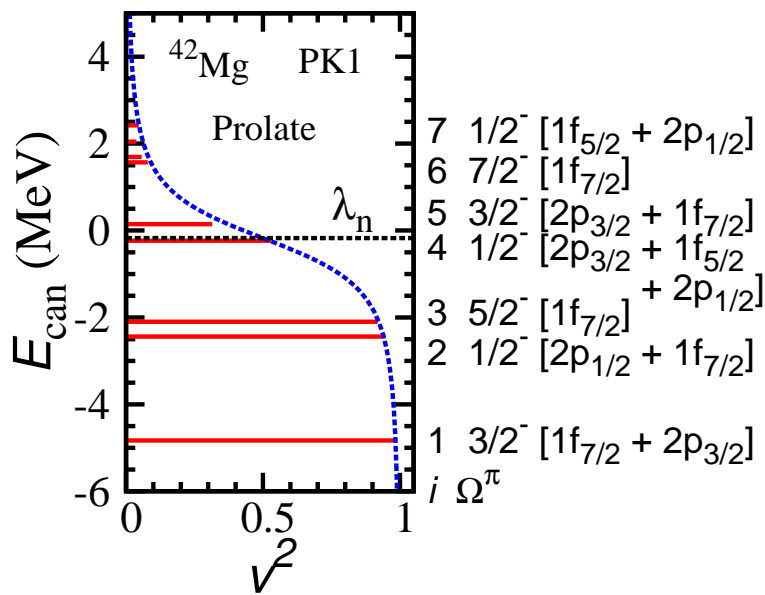}
\caption{(Color online)
Single neutron levels of ground state of $^{42}$Mg in the canonical basis
as a function of the occupation probability $v^2$.
The blue dashed line corresponds to the BCS-formula with an average pairing gap.
Taken from Ref.~\cite{Li2012_PRC85-024312}.
}
\label{fig:Mg42_pro_lev}
\end{center}
\end{figure}

The single particle spectrum around the Fermi level for the ground state of
$^{42}$Mg is shown in Fig.~\ref{fig:Mg42_pro_lev}~\cite{Li2012_PRC85-024312}.
The good quantum numbers of each single particle state are also shown.
The occupation probabilities $v^2$ in the canonical basis have BCS-form~\cite{Ring1980}
and are given by the length of the horizontal lines in Fig.~\ref{fig:Mg42_pro_lev}.
The levels close to the threshold are labeled by the number $i$ according
to their energies, and their conserved quantum number
$\Omega^\pi$ as well as the main spherical components are given at the right hand side.
The neutron Fermi level is within the $pf$ shell and most of the single
particle levels have negative parities.
Since the chemical potential $\lambda_n$ is close to the continuum, orbitals
above the threshold have noticeable occupations due to the pairing correlations.
The single neutron levels of $^{42}$Mg can be divided into two parts, the
deeply bound levels ($\varepsilon_{\rm can} < -2$ MeV) corresponding
to the ``core'', and the remaining weakly bound levels close to the
threshold ($\varepsilon_{\rm can} > -0.3$ MeV) and in the continuum
corresponding to the ``halo''.

As discussed in Refs.~\cite{Zhou2010_PRC82-011301R, Li2012_PRC85-024312, Zhou2011_JPCS312-092067},
the shape of the halo originates from the intrinsic structure of the weakly
bound or continuum orbitals.
By examining the neutron density distribution, it was found that
for the ground state of $^{42}$Mg, the halo is mainly
formed by level 4 and level 5.
A detailed analysis of the intrinsic properties revealed that in $^{42}$Mg the shape of the halo is oblate and decouples from the prolate core.

We note that this kind of shape decoupling was later also found
in non-relativistic HFB calculations~\cite{Pei2013_PRC87-051302R}.
An exotic ``egg''-like structure
consisting of a spherical core plus a prolate halo was predicted
in $^{38}$Ne, in which the near-threshold nonresonant continuum
plays an essential role.

\subsection{Recent progress}

For odd particle system, the halo could depend strongly on the interplay
among the odd-even effects, continuum and pairing effects, deformation effects, etc.
Combining the necessity for the improved descriptions of
the equation of state at high density, the asymmetric nuclear matter and
the isovector properties of nuclei far from stability with
the density-dependent meson-nucleon couplings
\cite{Brockmann1992_PRL68-3408, Fuchs1995_PRC52-3043,
Typel1999_NPA656-331, Niksic2002_PRC66-024306,
Long2004_PRC69-034319, Lalazissis2005_PRC71-024312},
it is necessary to extend the DRHBc theory with the density-dependent
meson-nucleon couplings.
Here we will briefly review these progress, i.e.,
the exotic nuclear structure in unstable odd-$A$ or odd-odd nuclei
\cite{Li2012_CPL29-042101} and the density-dependent deformed relativistic
Hartree-Bogoliubov (DDDRHB) theory in continuum
\cite{Chen2012_PRC85-067301}.

In order to describe the exotic nuclear structure in unstable odd-$A$ or odd-odd
nuclei, the blocking effect of one or several nucleons has to be taken into account.
In Ref.~\cite{Li2012_CPL29-042101}, the DRHBc theory
\cite{Zhou2008_ISPUN2007, Zhou2010_PRC82-011301R, Li2012_PRC85-024312}
is extended to incorporate the blocking effect due to odd particle system.
In such a way, pairing correlations, continuum, deformation, blocking effects,
and extended spatial density distributions in exotic odd-$A$ or odd-odd nuclei can be
taken into account microscopically and self-consistently.

For an odd nucleon system with the $i$-th level blocked, the time reversal symmetry is violated and there appear currents in the system. In principle, one has to diagonalize the RHB equation twice.
Using the equal filling approximation which is usually made~\cite{Schunck2010_PRC81-024316,
Perez-Martin2008_PRC78-014304--12}, the currents are neglected and the two configurations of a particle in the time-reversal partner space are averaged in a statistical manner~\cite{Li2012_CPL29-042101}.
The corresponding currents cancel each other and in this way, in each step of the iteration meson fields with the time reversal symmetry is obtained.
In practice, the density matrix $\rho$ and the abnormal
density $\kappa$ in two subspaces are averaged and replaced by
\begin{eqnarray}
  \rho' & = & \rho_{ _{M \times M} }
         + \frac{1}{2}
           \left( U_{k_b} U^{*T}_{k_b} -  V^*_{k_b} V^T_{k_b} \right) ,
    \label{eq:rhodd} \\
  \kappa' & = & \kappa_{ _{M \times M} }
         -  \frac{1}{2} \left( U_{k_b} V^{*T}_{k_b} + V^*_{k_b} U^T_{k_b} \right) ,
    \label{eq:kapodd}
\end{eqnarray}
where $V_{k_b}$ and $U_{k_b}$ are column vectors in the matrices $V$ and $U$
corresponding to the blocked level.

\begin{figure}
\begin{center}
\includegraphics[width=0.40\textwidth]{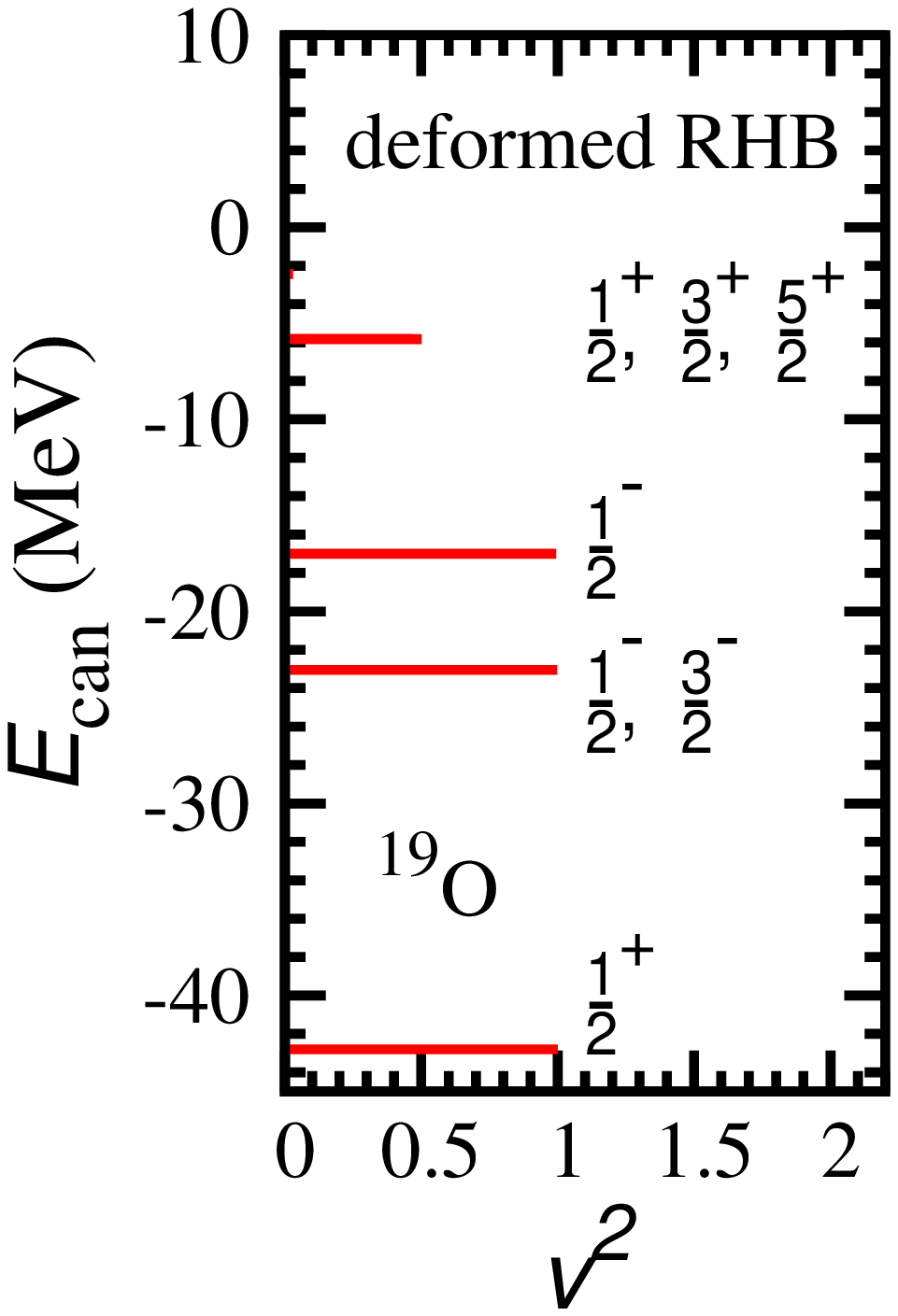}
~~~
\includegraphics[width=0.40\textwidth]{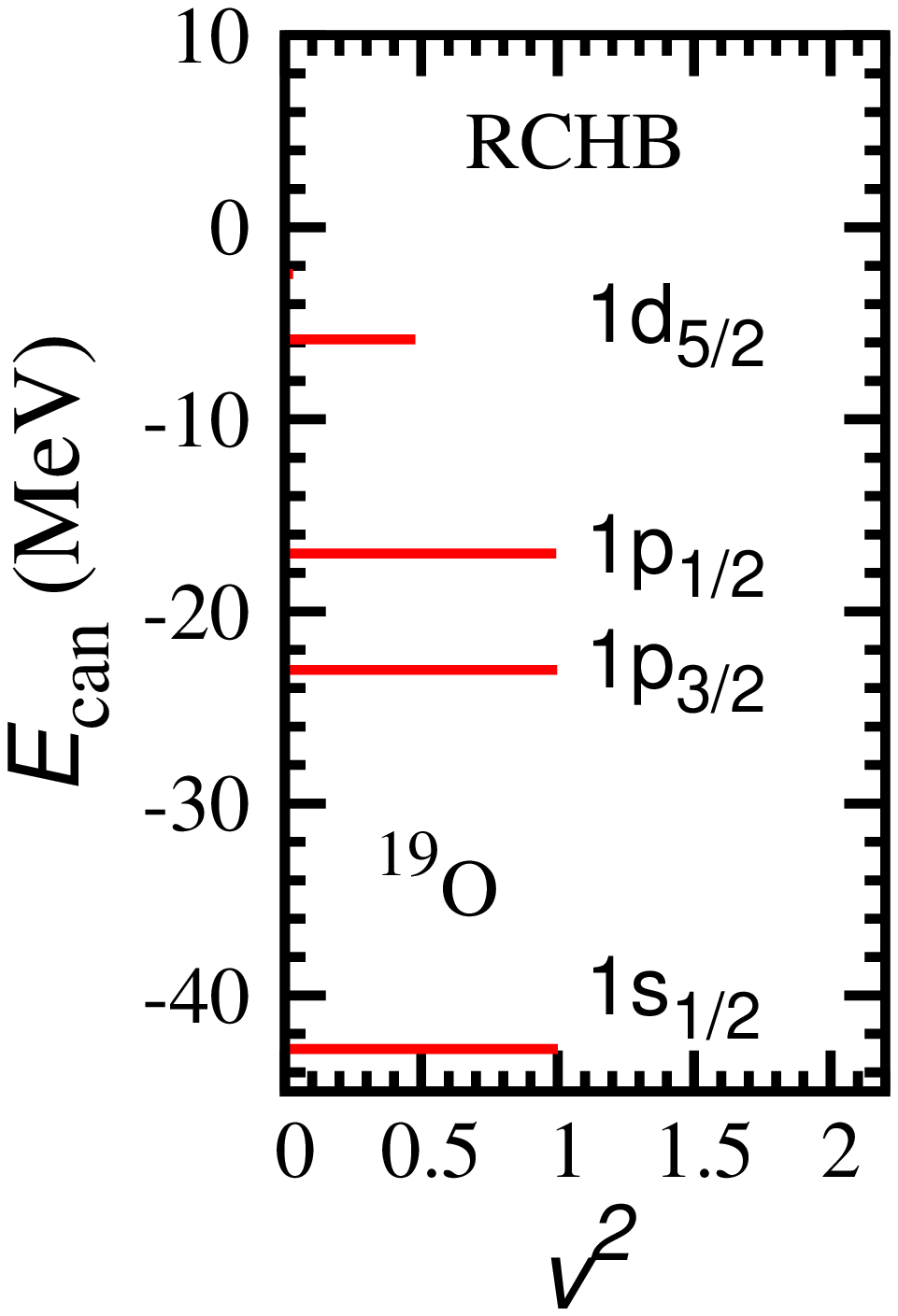}
\end{center}
\caption{(Color online)
\label{fig:O19_lev}
Neutron single particle levels in the canonical basis and their occupation
probability $v^2$ for $^{19}$O from the DRHBc and RCHB calculations. Taken from Ref.~\cite{Li2012_CPL29-042101}.
}
\end{figure}

Using the density functional PK1~\cite{Long2004_PRC69-034319}
and a zero range density dependent pairing force~\cite{Meng1998_NPA635-3, Li2012_PRC85-024312},
the deformed RHB equations are solved in a spherical Dirac Woods-Saxon
basis~\cite{Zhou2003_PRC68-034323} and the deformed potentials and densities are
expanded in terms of the Legendre polynomials $P_\lambda(\cos \theta)$
(for details see Refs.~\cite{Zhou2008_ISPUN2007, Zhou2010_PRC82-011301R, Li2012_PRC85-024312}).
In order to check the accuracy of the DRHBc code with the blocking effect,
the spherical nucleus $^{19}$O is investigated and the results of
the DRHBc code allowing only the spherical components of the fields,
i.e., $\lambda = 0$, are compared with those obtained in RCHB theory~\cite{Meng1998_NPA635-3}.
It is found that both calculations agree well with each other with high precision~\cite{Li2012_CPL29-042101}.

In Fig.~\ref{fig:O19_lev}, the neutron single particle levels
in the canonical basis for $^{19}$O is shown in comparison with the RCHB results in order to understand the details and validity of the DRHBc calculation.
The length of each level is proportional to the occupation probability.
In the RCHB calculation, the blocked orbital is 1$d_{5/2}$.
In the DRHBc code, a spherical solution is enforced by allowing only
$\lambda=0$ component in the Legendre expansion mentioned earlier
and therefore the three sublevels of the 1$d_{5/2}$ orbital with
$\Omega^{\pi} ={1}/{2}^+$, ${3}/{2}^+$, and ${5}/{2}^+$ are degenerate.
The present results in Fig.~\ref{fig:O19_lev} are obtained with the level
$\Omega^{\pi} = {1}/{2}^+$ blocked in the DRHBc code,
but blocking any of these three levels gives the same results.
From Fig.~\ref{fig:O19_lev} one finds nice agreement between the results
from the DRHBc and RCHB calculations.

\begin{figure}
\begin{center}
\includegraphics[width=0.50\textwidth]{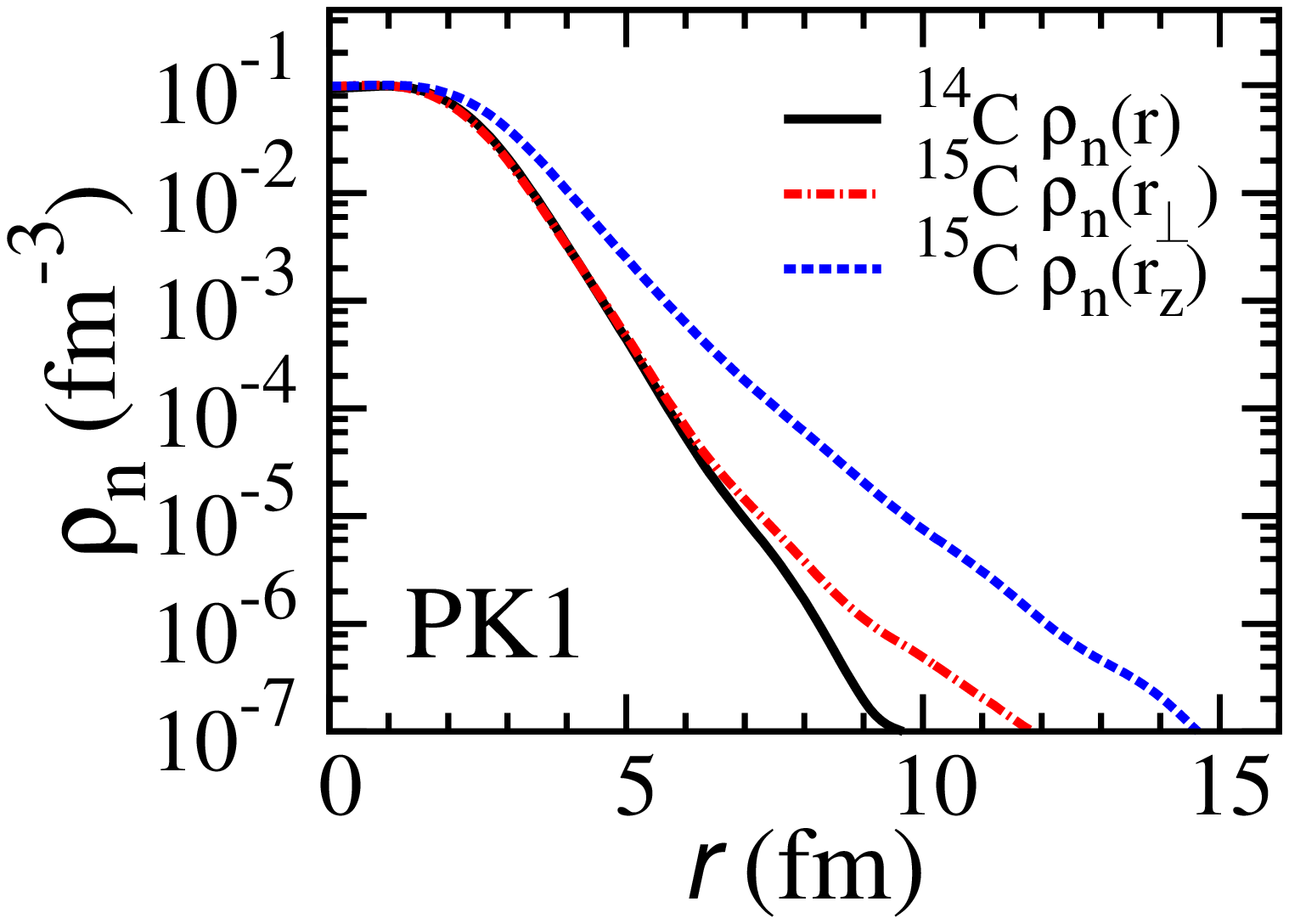}
~~~
\includegraphics[width=0.40\textwidth]{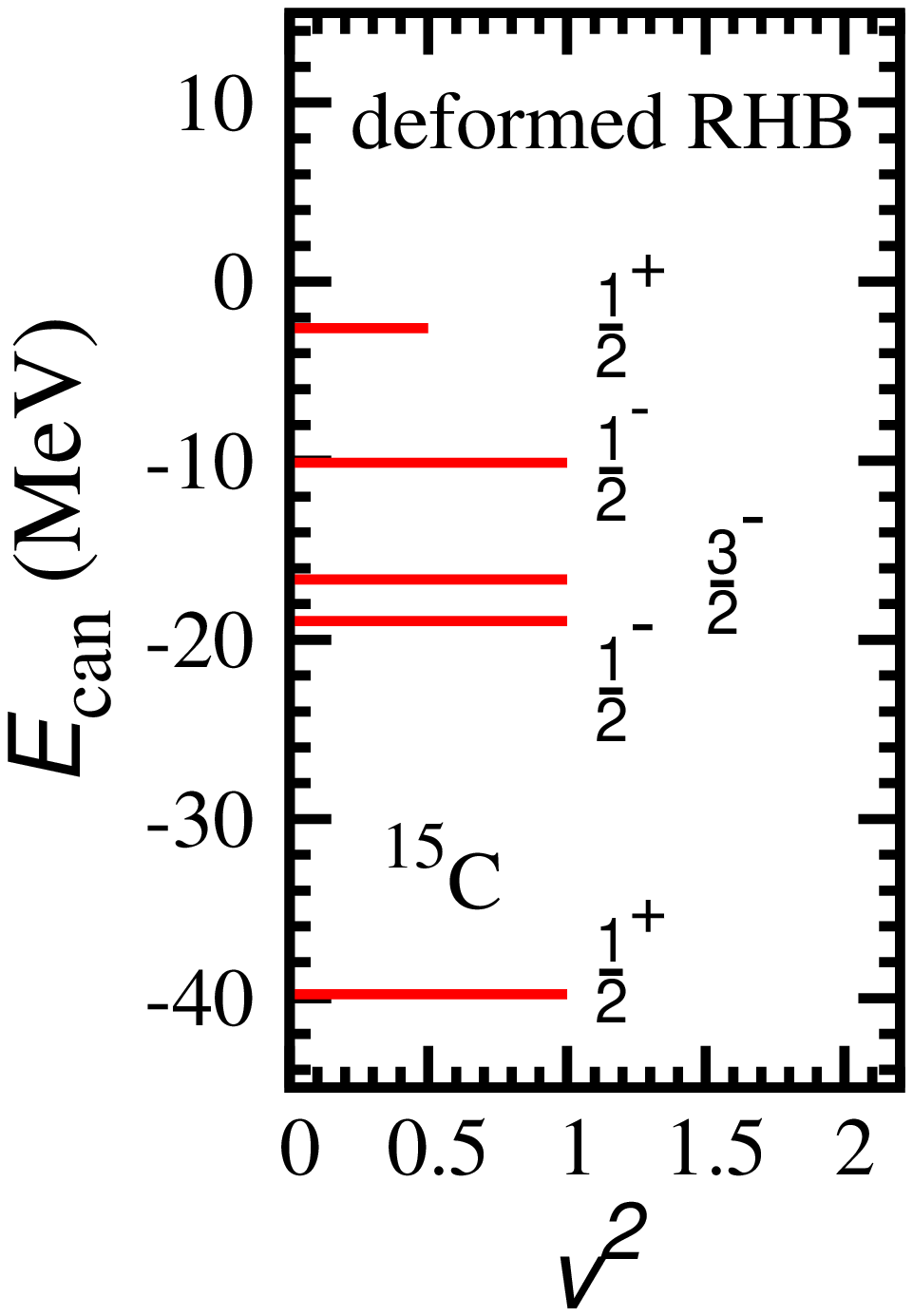}
\end{center}
\caption{(Color online) \label{fig:Fig25a}
(a) Neutron density profiles of $^{14}$C and $^{15}$C in DRHBc calculations
with the parameter set PK1.
\label{fig:Fig25b}
(b) Neutron single particle levels in the canonical basis and their occupation
probability $v^2$ for $^{15}$C in DRHBc calculation.
Taken from Ref.~\cite{Li2012_CPL29-042101}.
}
\end{figure}

An application of the DRHBc theory for an exotic and deformed nucleus with an odd number of particles is given in Ref.~\cite{Li2012_CPL29-042101} for the deformed halo candidate nucleus $^{15}$C~\cite{Fang2004_PRC69-034613} with the parameter set PK1~\cite{Long2004_PRC69-034319}.
In contrary to the spherical even-even nucleus $^{14}$C, the DRHBc theory predicts a deformation $\beta = 0.25$ for the odd-$A$ nucleus $^{15}$C.
The neutron root mean square radii of $^{14}$C and $^{15}$C are $2.56$~fm and $2.79$~fm, respectively.

The DRHBc calculations for $^{15}$C lead to a deformed density
distribution with an axial symmetry.
In Fig.~\ref{fig:Fig25a}, the neutron densities along the symmetry axis
$\rho_n(r_z, r_\perp = 0)$ and perpendicular to the symmetry axis
$\rho_n(r_z=0, r_\perp)$ are plotted as dashed and dashed-dotted lines.
The spherical density for $^{14}$C is also included as a reference by a solid line.
It is interesting to see that, in the direction perpendicular to
the symmetry axis, the neutron densities are almost the same for $^{14}$C
and $^{15}$C at least when $r_\perp < 6$ fm.
Along the symmetry axis the neutron density of $^{15}$C extends much further
than that of $^{14}$C.
This is partly due to that $^{15}$C is prolate and the weakly bound
${1}/{2}^+$ level is occupied.

The single particle levels of $^{15}$C in the canonical basis are plotted in Fig.~\ref{fig:Fig25b}.
Since it is a deformed nucleus, the $1p_{3/2}$ orbital is split into two levels with
$\Omega^{\pi} ={1}/{2}^-$ and $\Omega^{\pi} ={3}/{2}^-$, respectively.
There is one neutron occupying in $\Omega^{\pi} ={1}/{2}^+$ level near the threshold.
The occupation probability $v^2=0.5$, indicating that it is averaged over the two
configurations with $\Omega^{\pi}=\pm{1}/{2}^+$.
Because of the deformation, this level is a mixture of the spherical orbitals
$1d_{5/2}$ ($62\%$) and $2s_{1/2}$ ($36\%$).
The weakly-bound feature and the relatively large $s$-wave component of this level
results in that the neutron density of $^{15}$C extends further along the symmetry axis.

After the comparison between the DRHBc and the spherical RCHB results for $^{19}$O as well as the application for the deformed neutron-rich nucleus $^{15}$C, the DRHBc with blocking effect proved to be a powerful tool for the study of the odd-$A$ or odd-odd exotic nuclei.

In the development of the DRHBc with the density-dependent meson-nucleon couplings~\cite{Chen2012_PRC85-067301}, the challenge is the treatment of the density-dependent couplings, meson fields, and potentials in axially deformed system with partial wave method.

\begin{figure}[tbh]
\begin{center}
      \includegraphics[width=9cm]{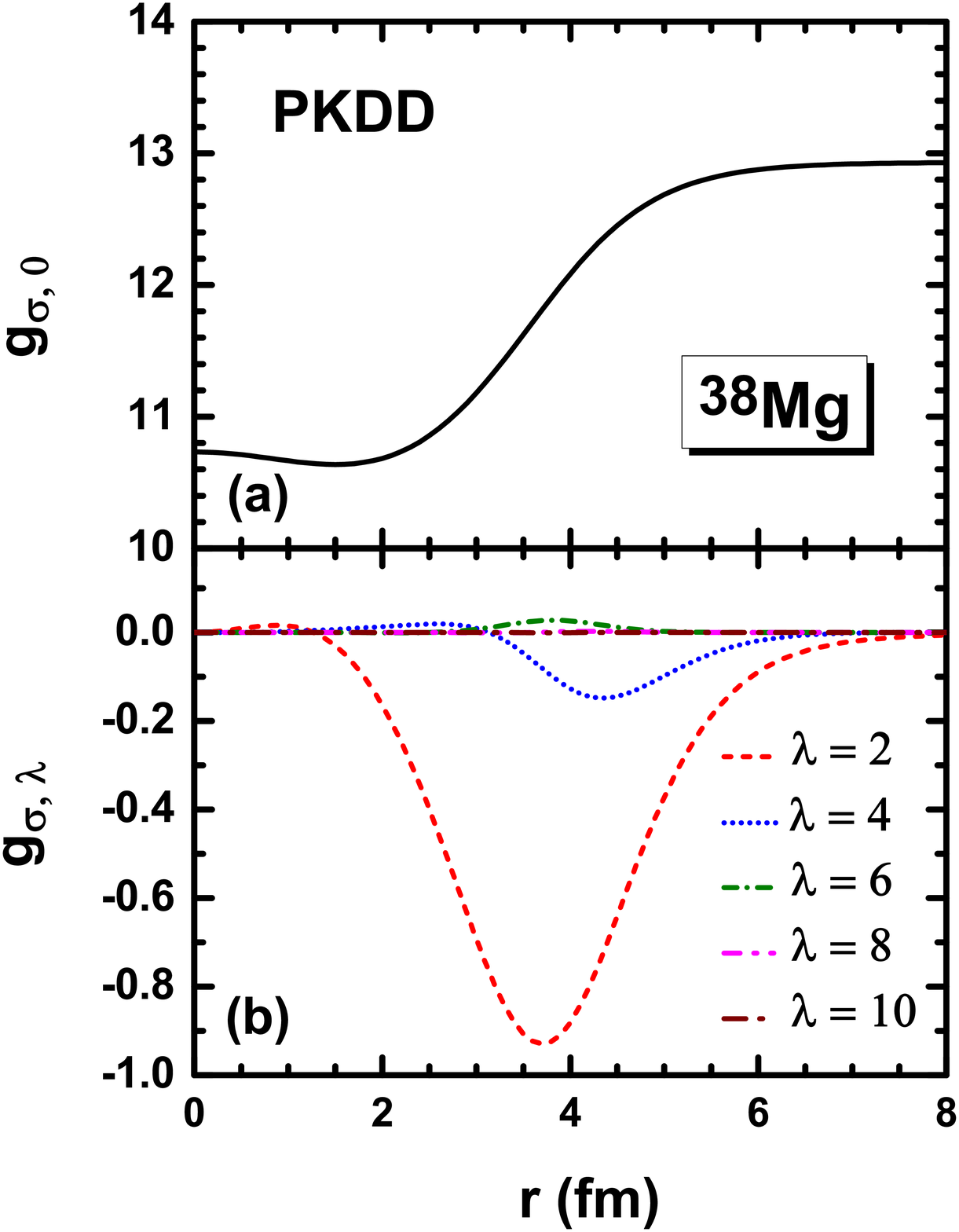}
\end{center}
      \caption{(Color online) Decompositions of coupling strength $g_{\sigma,\lambda}$ with $\lambda=0$ (a) and $\lambda=2,4,6,8,10$ (b) for $^{38}$Mg  calculated with density-dependent DRHBc theory. Taken from Ref.~\cite{Chen2012_PRC85-067301}.
      }
      \label{fig:Fig26}
\end{figure}

In Fig.~\ref{fig:Fig26}, by taking the $\sigma$ meson as an example, the partial waves of density-dependent coupling strengths $g_{\sigma,\lambda}(r)$ for $^{38}$Mg are shown with
$g_{\sigma} (\rho_v) = \sum_{\lambda}g_{\sigma,\lambda}(r)P_{\lambda}(\cos\theta)$.
It can be seen that the major component is that with $\lambda=0$, which is more than one order of magnitude larger than the others. The amplitudes of $g_{\sigma,\lambda}$ decrease quickly with increasing $\lambda$ and become negligible when $\lambda \geq 8$. This means that a cutoff $\lambda_{g,\rm{max}}=10$ is enough to provide accurate results.

By comparing the calculated bulk and single particle properties of neutron-rich nucleus $^{38}$Mg in spherical case with $\lambda_{\rho,\rm{max}}=0$ with those obtained by RCHB~\cite{Meng1998_NPA635-3},
the density-dependent DRHBc theory is verified~\cite{Chen2012_PRC85-067301}. It therefore opens the door to investigate the deformed exotic nuclei with effective density-dependent interactions.

\section{\label{sec:summary}Summary and perspectives}

In this Topical Review, we have reviewed the progresses on
the covariant density functional theory in continuum for
the neutron halo phenomenon in both spherical and deformed nuclei.

In the mean field study of nuclear halos, a proper description of
nuclear pairing and the continuum contribution is very important.
To include properly the contribution of continuum states,
the Bogoliubov transformation has been justified to be very useful.
The relativistic continuum Hartree-Bogoliubov (RCHB) theory has been
developed in which the coupled differential equations
are solved in coordinate ($r$) space and
the mean field effects of the coupling to the continuum
can be fully taken into account.

From the RCHB theory, a self-consistent description of
the neutron halo in $^{11}$Li has been achieved.
The calculated density distribution agrees well with
the experiment.
It has further been shown that the halo in $^{11}$Li is formed by Cooper-pairs
scattered in the two levels $1p_{1/2}$ and $2s_{1/2}$ with the
former below the Fermi surface and the latter in the continuum.
In the RCHB theory,
a halo appears if there are several single particle valence orbitals
with small orbital angular momenta and correspondingly
small centrifugal barriers close to the continuum limit.

For medium and heavy nuclei, the RCHB theory has treated the influence of pairing correlations
and many-body effects properly and predicted the existence of more neutrons in the halo.
Giant neutron halos, consisting of up to six halo neutrons,
have been predicted in Zr and Ca isotopes
close to the neutron drip line.
When approaching the neutron drip line,
the two-neutron separation energies for several nuclei in Ca and Zr isotopes
are extremely close to zero, which means that
the valence neutrons
gradually occupy the loosely bound states and the continuum.

The phenomena of halo and giant halo of neutron-rich even-Ca
isotopes were investigated and compared in the framework of
the RCHB and non-relativistic Skyrme HFB calculations.
It has been found that although halo phenomena exist for Ca isotopes
near neutron drip line in both calculations, the halo in the Skyrme HFB
calculations starts at a more neutron-rich nucleus than that in the RCHB calculations,
and the RCHB calculations have larger neutron root-mean-square radii
systematically in $N\ge 40$ than the Skyrme HFB calculations.
The former difference comes from the difference in shell structure.
The reasons for the latter can be partly explained by the neutron
$3s_{1/2}$ orbital, which causes more than $50\%$ of the difference
for neutron radii at $^{66}$Ca.

The relativistic Hartree-Fock-Bogoliubov theory in continuum (RHFBc theory) with
density-dependent meson-nucleon couplings has been developed.
With this theory, the influence of exchange terms on the formation
of neutron halos has been studied.
Giant halos as well as ordinary ones were found in Ce isotopes
close to the neutron drip line.
It was demostrated that the shell structure evolution, when approaching
the drip line, is important for the appearance of neutron halos.

The RCHB theory has also been extended to including hyperons and
even Ca isotopes ranging from the proton drip line to the neutron drip line
have been studied systematically.
Compared with ordinary nuclei, the additional one or two $\Lambda$-hyperons
lower the Fermi level and the last bound hypernuclei have two more neutrons
than the corresponding ordinary nuclei.
Giant halo phenomena due to the pairing correlation and the contribution
from the continuum have been predicted in Ca hypernuclei,
similar to those appearing in ordinary Ca isotopes.

A deformed relativistic Hartree-Bogoliubov theory in continuum (DRHBc theory)
has been developed aiming at a proper description of exotic nuclei,
particularly those with a large spatial extension.
In order to give an adequate consideration of both the contribution of
the continuum and the large spatial distribution in exotic nuclei,
the deformed RHB equations were solved in a Woods-Saxon (WS) basis in which
the radial wave functions have a proper asymptotic behavior at large distance
from the nuclear center.

Halo phenomena in deformed nuclei were investigated within
the DRHBc theory.
These weakly bound quantum systems present interesting examples for
the study of the interdependence between the deformation of the core
and the particles in the halo.
The Mg and Ne isotopes have been studied in details.
It was found that in $^{42,44}$Mg, there appears a decoupling of
the halo orbitals from the deformation of the core:
The core is prolate, but the halo has a slightly oblate shape.
The generic conditions for the occurrence of this decoupling effects
have been given: the existence and the deformation of a possible neutron halo depends
essentially on the quantum numbers of the main components of
the single particle orbitals in the vicinity of the Fermi surface.

Definitely there exist lots of open questions on the study of nuclear halos
with the covariant density functional theory.
We list several of them below.

In the RCHB, RHFBc and DRHBc theories, the box boundary condition
has been applied and the continuum is discretized.
Therefore one cannot get the width for resonances.
With the Green's function method, the correct asymptotic behavior
can be given on the wave function for the continuum states.
The Green's function method has been implemented in
the non-relativistic HFB theory \cite{Zhang2011_PRC83-054301}.
The effort to implement the Green's function approach in the RHB
theory is undertaken and promising preliminary results have been obtained
\cite{Sun2014_PRC90-054321}.

In the systematic study of giant halos in severl isotope chains, it
has been found that the appearance of giant halos and the contribution of continuum
are very important in determining the boundary of nuclear chart.
Furthermore, the deformation effects will also influence very much
the position of drip lines.
More systematic study is now carried out to reveal the global
impact of the continuum and deformation effects on the nuclear boundary.

If the core and the halo of a nucleus have different shapes,
many interesting questions arise: For example,
What will happen when the nucleus rotates?
What role could the shape fluctuation play?
To answer these questions, future works are welcome
to implement the angular momentum projection in the DRHBc theory and
to use the generator coordinate method (GCM) or
to solve Bohr Hamiltonian based on the potential energy surface obtained
from constraint DRHBc calculations.

The possible experimental signals of shape decoupling in deformed
halo nuclei are certainly interesting topics.
The halo feature is connected with relatively large cross sections
and narrow longitudinal momentum distributions in knockout reactions.
The decoupling between the deformations of the core and the halo
may manifest itself by some new experimental observables, e.g.,
the double-hump shape of longitudinal momentum distribution in
single particle removal reactions and new dipole modes, etc.
In particular, a combination of the experimental method proposed in
Ref.~\cite{Navin1998_PRL81-5089} and the theoretical approach developed
in Ref.~\cite{Sakharuk1999_PRC61-014609} would be useful in the study of
longitudinal momentum distribution in single particle removal reactions
with deformed halo nuclei as projectiles.
The shape decoupling effects may also has some influence on the
sub-barrier capture process in heavy ion
collisions~\cite{Sargsyan2011_PRC84-064614}.

\ack

We would like to thank
Y. Chen,
L. S. Geng, N. V. Giai,
Y. Kim,
L. L. Li, H. Z. Liang, M. Liu, W. H. Long, H. F. L\"u,
P. Ring,
H. Sagawa,
I. Tanihata,
J. Terasaki, H. Toki,
N. Wang,
X. Z. Wu,
X. W. Xia,
S. Yamaji,
J. Y. Zeng, S. Q. Zhang,
and
E. G. Zhao for valuable discussions and fruitful collaborations.
This work was supported in part by
the National Key Basic Research Program of China (Grant No. 2013CB834400),
the Natural Science Foundation of China (Grants
No. 10975008,
No. 11175002, No. 11105005, No. 11105010, No. 11035007, No. 11128510,
No. 11121403, No. 11120101005, No. 11211120152,
No. 11235002, No. 11275248, No. 11205004,
and No. 11335002)，
the Research Fund for the Doctoral Program of Higher Education
(Grant No. 20110001110087),
and
the Chinese Academy of Sciences (Grant No. KJCX2-EW-N01).

\section*{References}


\begin{thebibliography}{100}
\expandafter\ifx\csname url\endcsname\relax
  \def\url#1{{\tt #1}}\fi
\expandafter\ifx\csname urlprefix\endcsname\relax\def\urlprefix{URL }\fi
\providecommand{\eprint}[2][]{\url{#2}}

\bibitem{Thoennessen2013_RPP76-056301}
Thoennessen M 2013 {\em Rep. Prog. Phys.\/} {\bf 76} 056301--14 
  {see also \url{http://www.nscl.msu.edu/~thoennes/isotopes/}}
  \urlprefix\url{http://stacks.iop.org/0034-4885/76/i=5/a=056301}

\bibitem{Erler2012_Nature486-509}
Erler J, Birge N, Kortelainen M, Nazarewicz W, Olsen E, Perhac A~M and Stoitsov
  M 2012 {\em Nature\/} {\bf 486} 509--512 
  \urlprefix\url{http://dx.doi.org/10.1038/nature11188}

\bibitem{Afanasjev2013_PLB726-680}
Afanasjev A, Agbemava S, Ray D and Ring P 2013 {\em Phys. Lett. B\/} {\bf 726}
  680--684 
  \urlprefix\url{http://www.sciencedirect.com/science/article/pii/S037026931300734X}

\bibitem{Agbemava2014_PRC89-054320}
Agbemava S~E, Afanasjev A~V, Ray D and Ring P 2014 {\em Phys. Rev. C\/} {\bf
  89} 054320--37
  \urlprefix\url{http://dx.doi.org/10.1103/PhysRevC.89.054320}

\bibitem{Qu2013_SciChinaPMA56-2031}
Qu X, Chen Y, Zhang S, Zhao P, Shin I, Lim Y, Kim Y and Meng J 2013 {\em Sci.
  China-Phys. Mech. Astron.\/} {\bf 56} 2031--2036 
  \urlprefix\url{http://dx.doi.org/10.1007/s11433-013-5329-5}

\bibitem{Motobayashi2010_NPA834-707c}
Motobayashi T 2010 {\em Nucl. Phys. A\/} {\bf 834} 707c--712c 
  \urlprefix\url{http://dx.doi.org/10.1016/j.nuclphysa.2010.01.128}

\bibitem{Sturm2010_NPA834-682c}
Sturm C, Sharkov B and St\"ocker H 2010 {\em Nucl. Phys. A\/} {\bf 834}
  682c--687c 
  \urlprefix\url{http://dx.doi.org/10.1016/j.nuclphysa.2010.01.124}

\bibitem{Gales2010_NPA834-717c}
Gales S 2010 {\em Nucl. Phys. A\/} {\bf 834} 717c--723c 
  \urlprefix\url{http://dx.doi.org/10.1016/j.nuclphysa.2010.01.130}

\bibitem{Thoennessen2010_NPA834-688c}
Thoennessen M 2010 {\em Nucl. Phys. A\/} {\bf 834} 688c--693c 
  \urlprefix\url{http://dx.doi.org/10.1016/j.nuclphysa.2010.01.125}

\bibitem{Choi2010_ISNPA}
Choi S {KoRIA} project - {RI} accelerator in {Korea} invited talk given in
  International Symposium on Nuclear Physics in Asia, 14-15 October, 2010,
  Beihang University, Beijing
  \urlprefix\url{http://indico.ihep.ac.cn/conferenceTimeTable.py?confId=1714}

\bibitem{Xia2002_NIMA488-11}
Xia J~W, Zhan W~L, Wei B~W, Yuan Y~J, Song M~T, Zhang W~Z, Yang X~D, Yuan P,
  Gao D~Q, Zhao H~W, Yang X~T, Xiao G~Q, Man K~T, Dang J~R, Cai X~H, Wang Y~F,
  Tang J~Y, Qiao W~M, Rao Y~N, He Y, Mao L~Z and Zhou Z~Z 2002 {\em Nucl.
  Instrum. Methods Phys. Res. A\/} {\bf 488} 11--25 
  \urlprefix\url{http://dx.doi.org/10.1016/S0168-9002(02)00475-8}

\bibitem{Zhan2010_NPA834-694c}
Zhan W, Xu H, Xiao G, Xia J, Zhao H and Yuan Y 2010 {\em Nucl. Phys. A\/} {\bf
  834} 694c--700c 
  \urlprefix\url{http://dx.doi.org/10.1016/j.nuclphysa.2010.01.126}

\bibitem{Mueller1993_ARNPS43-529}
M\"uller A~C and Sherrill B~M 1993 {\em Annu. Rev. Nucl. Part. Sci.\/} {\bf 43}
  529--583
  \urlprefix\url{http://dx.doi.org/10.1146/annurev.ns.43.120193.002525}

\bibitem{Tanihata1995_PPNP35-505}
Tanihata I 1995 {\em Prog. Part. Nucl. Phys.\/} {\bf 35} 505--573
  \urlprefix\url{http://dx.doi.org/10.1016/0146-6410(95)00046-L}

\bibitem{Hansen1995_ARNPS45-591}
Hansen P~G, Jensen A~S and Jonson B 1995 {\em Annu. Rev. Nucl. Part. Sci.\/}
  {\bf 45} 591--634 
  \urlprefix\url{http://dx.doi.org/10.1146/annurev.ns.45.120195.003111}

\bibitem{Casten2000_PPNP45-S171}
Casten R~F and Sherrill B~M 2000 {\em Prog. Part. Nucl. Phys.\/} {\bf 45}
  S171--S223 \urlprefix\url{http://dx.doi.org/10.1016/S0146-6410(00)90013-9}

\bibitem{Bertulani2001_PhysRIB}
Bertulani C~A, Hussein M~S and M\"unzenberg G 2001 {\em Physics of Radioactive
  Beams\/} (Nova Science Publishers, Inc., Huntington, New York)
  \urlprefix\url{http://books.google.com/books?id=sxWMQAAACAAJ}

\bibitem{Johnson2004_PR389-1}
Johnson B 2004 {\em Phys. Rep.\/} {\bf 389} 1--59
  \urlprefix\url{http://dx.doi.org/10.1016/j.physrep.2003.07.004}

\bibitem{Jensen2004_RMP76-215}
Jensen A~S, Riisager K, Fedorov D~V and Garrido E 2004 {\em Rev. Mod. Phys.\/}
  {\bf 76} 215--261 \urlprefix\url{http://link.aps.org/abstract/RMP/v76/p215}

\bibitem{Meng2006_PPNP57-470}
Meng J, Toki H, Zhou S~G, Zhang S~Q, Long W~H and Geng L~S 2006 {\em Prog.
  Part. Nucl. Phys.\/} {\bf 57} 470--563
  \urlprefix\url{http://dx.doi.org/10.1016/j.ppnp.2005.06.001}

\bibitem{Ershov2010_JPG37-064026}
Ershov S~N, Grigorenko L~V, Vaagen J~S and Zhukov M~V 2010 {\em J. Phys. G:
  Nucl. Phys.\/} {\bf 37} 064026--15 
  \urlprefix\url{http://stacks.iop.org/0954-3899/37/i=6/a=064026}

\bibitem{Frederico2012_PPNP67-939}
Frederico T, Delfino A, Tomio L and Yamashita M~T 2012 {\em Prog. Part. Nucl.
  Phys.\/} {\bf 67} 939--994 
  \urlprefix\url{http://www.sciencedirect.com/science/article/pii/S0146641012000907}

\bibitem{Riisager2013_PST152-014001}
Riisager K 2013 {\em Phys. Scr.\/} {\bf T152} 014001--13
  \urlprefix\url{http://dx.doi.org/10.1088/0031-8949/2013/T152/014001}

\bibitem{Tanihata1985_PRL55-2676}
Tanihata I, Hamagaki H, Hashimoto O, Shida Y, Yoshikawa N, Sugimoto K, Yamakawa
  O, Kobayashi T and Takahashi N 1985 {\em Phys. Rev. Lett.\/} {\bf 55}
  2676--2679 \urlprefix\url{http://link.aps.org/abstract/PRL/v55/p2676}

\bibitem{Ozawa2000_PRL84-5493}
Ozawa A, Kobayashi T, Suzuki T, Yoshida K and Tanihata I 2000 {\em Phys. Rev.
  Lett.\/} {\bf 84} 5493--5495
  \urlprefix\url{http://link.aps.org/doi/10.1103/PhysRevLett.84.5493}

\bibitem{Warburton1990_PRC41-1147}
Warburton E~K, Becker J~A and Brown B~A 1990 {\em Phys. Rev. C\/} {\bf 41}
  1147--1166 \urlprefix\url{http://link.aps.org/doi/10.1103/PhysRevC.41.1147}

\bibitem{Adrich2005_PRL95-132501}
Adrich P, Klimkiewicz A, Fallot M, Boretzky K, Aumann T, Cortina-Gil D,
  Pramanik U~D, Elze T~W, Emling H, Geissel H, Hellstrom M, Jones K~L, Kratz
  J~V, Kulessa R, Leifels Y, Nociforo C, Palit R, Simon H, Surowka G, Summerer
  K and Walus W 2005 {\em Phys. Rev. Lett.\/} {\bf 95} 132501--4
  \urlprefix\url{http://link.aps.org/abstract/PRL/v95/e132501}

\bibitem{Meng1998_PRL80-460}
Meng J and Ring P 1998 {\em Phys. Rev. Lett.\/} {\bf 80} 460--463
  \urlprefix\url{http://link.aps.org/abstract/PRL/v80/p460}

\bibitem{Meng2002_PRC65-041302R}
Meng J, Toki H, Zeng J~Y, Zhang S~Q and Zhou S~G 2002 {\em Phys. Rev. C\/} {\bf
  65} 041302(R)--4 \urlprefix\url{http://link.aps.org/abstract/PRC/v65/e041302}

\bibitem{Zhou2010_PRC82-011301R}
Zhou S~G, Meng J, Ring P and Zhao E~G 2010 {\em Phys. Rev. C\/} {\bf 82}
  011301(R)--5
  \urlprefix\url{http://link.aps.org/doi/10.1103/PhysRevC.82.011301}

\bibitem{Li2012_PRC85-024312}
Li L, Meng J, Ring P, Zhao E~G and Zhou S~G 2012 {\em Phys. Rev. C\/} {\bf 85}
  024312--17 \urlprefix\url{http://link.aps.org/doi/10.1103/PhysRevC.85.024312}

\bibitem{Dobaczewski2007_PPNP59-432}
Dobaczewski J, Michel N, Nazarewicz W, Ploszajczak M and Rotureau J 2007 {\em
  Prog. Part. Nucl. Phys.\/} {\bf 59} 432--445 
  \urlprefix\url{http://dx.doi.org/10.1016/j.ppnp.2007.01.022}

\bibitem{Bertsch1991_APNP209-63}
Bertsch G~F and Esbensen H 1991 {\em Annals of Physics\/} {\bf 209} 327--363
  \urlprefix\url{http://dx.doi.org/10.1016/0003-4916(91)90033-5}

\bibitem{Meng1996_PRL77-3963}
Meng J and Ring P 1996 {\em Phys. Rev. Lett.\/} {\bf 77} 3963--3966
  \urlprefix\url{http://link.aps.org/abstract/PRL/v77/p3963}

\bibitem{Meng1998_PRC57-1229}
Meng J 1998 {\em Phys. Rev. C\/} {\bf 57} 1229--1232 
  \urlprefix\url{http://dx.doi.org/10.1103/PhysRevC.57.1229}

\bibitem{Bulgac1980_nucl-th9907088}
Bulgac A 1980 {Hartree}-{Fock}-{Bogoliubov} approximation for finite systems
  IPNE FT-194-1980, Bucharest (arXiv: nucl-th/9907088)
  \urlprefix\url{http://arxiv.org/abs/nucl-th/9907088}

\bibitem{Dobaczewski1984_NPA422-103}
Dobaczewski J, Flocard H and Treiner J 1984 {\em Nucl. Phys. A\/} {\bf 422}
  103--139 
  \urlprefix\url{http://dx.doi.org/10.1016/0375-9474(84)90433-0}

\bibitem{Poschl1997_PRL79-3841}
P\"{o}schl W, Vretenar D, Lalazissis G~A and Ring P 1997 {\em Phys. Rev.
  Lett.\/} {\bf 79} 3841--3844
  \urlprefix\url{http://link.aps.org/abstract/PRL/v79/p3841}

\bibitem{Meng1998_NPA635-3}
Meng J 1998 {\em Nucl. Phys. A\/} {\bf 635} 3--42 
  \urlprefix\url{http://dx.doi.org/10.1016/S0375-9474(98)00178-X}

\bibitem{Yang2001_CPL18-196}
Yang S~C, Meng J and Zhou S~G 2001 {\em Chin. Phys. Lett.\/} {\bf 18} 196--198
  \urlprefix\url{http://stacks.iop.org/0256-307X/18/196}

\bibitem{Cao2002_PRC66-024311}
Cao L~G and Ma Z~Y 2002 {\em Phys. Rev. C\/} {\bf 66} 024311--5
  \urlprefix\url{http://link.aps.org/doi/10.1103/PhysRevC.66.024311}

\bibitem{Zhang2004_PRC70-034308}
Zhang S~S, Meng J, Zhou S~G and Hillhouse G~C 2004 {\em Phys. Rev. C\/} {\bf
  70} 034308--8 (\textit{Preprint} \eprint{nucl-th/0403013})
  \urlprefix\url{http://link.aps.org/abstract/PRC/v70/e034308}

\bibitem{Zhang2008_PRC77-014312}
Zhang L, Zhou S~G, Meng J and Zhao E~G 2008 {\em Phys. Rev. C\/} {\bf 77}
  014312--6 (\textit{Preprint} \eprint{0712.3087})
  \urlprefix\url{http://link.aps.org/doi/10.1103/PhysRevC.77.014312}

\bibitem{Zhou2009_JPB42-245001}
Zhou S~G, Meng J and Zhao E~G 2009 {\em J. Phys. B: At. Mol. Phys.\/} {\bf 42}
  245001--5 (\textit{Preprint} \eprint{0911.1171})
  \urlprefix\url{http://dx.doi.org/10.1088/0953-4075/42/24/245001}

\bibitem{Li2010_PRC81-034311}
Li Z~P, Meng J, Zhang Y, Zhou S~G and Savushkin L~N 2010 {\em Phys. Rev. C\/}
  {\bf 81} 034311--6
  \urlprefix\url{http://link.aps.org/doi/10.1103/PhysRevC.81.034311}

\bibitem{Guo2010_PRC82-034318}
Guo J~Y, Fang X~Z, Jiao P, Wang J and Yao B~M 2010 {\em Phys. Rev. C\/} {\bf
  82} 034318--8
  \urlprefix\url{http://link.aps.org/doi/10.1103/PhysRevC.82.034318}

\bibitem{Sandulescu2000_PRC61-061301R}
Sandulescu N, Van~Giai N and Liotta R~J 2000 {\em Phys. Rev. C\/} {\bf 61}
  061301(R)--5 \urlprefix\url{http://link.aps.org/abstract/PRC/v61/e061301}

\bibitem{Sandulescu2003_PRC68-054323}
Sandulescu N, Geng L~S, Toki H and Hillhouse G~C 2003 {\em Phys. Rev. C\/} {\bf
  68} 054323--6
  \urlprefix\url{http://link.aps.org/doi/10.1103/PhysRevC.68.054323}

\bibitem{Zhang2013_EPJA49-77}
Zhang S~S, Zhao E~G and Zhou S~G 2013 {\em Euro. Phys. J. A\/} {\bf 49} 77 
  (\textit{Preprint} \eprint{1105.0504})
  \urlprefix\url{http://dx.doi.org/10.1140/epja/i2013-13077-8}

\bibitem{Zhang2014_PLB730-30}
Zhang S~S, Smith M~S, Kang Z~S and Zhao J 2014 {\em Phys. Lett. B\/} {\bf 730}
  30--35
  \urlprefix\url{http://www.sciencedirect.com/science/article/pii/S0370269314000355}

\bibitem{Belyaev1987_SovJNP45-783}
Belyaev S~T, Smirnov A~V, Tolokonnikov S~V and Fayans S~A 1987 {\em Sov. J. Nucl.
  Phys.\/} {\bf 45} 783

\bibitem{Shlomo1975_NPA243-18}
Shlomo S and Bertsch G 1975 {\em Nucl. Phys. A\/} {\bf 243} 507--518 
  \urlprefix\url{http://dx.doi.org/10.1016/0375-9474(75)90292-4}

\bibitem{Matsuo2001_NPA696-371}
Matsuo M 2001 {\em Nucl. Phys. A\/} {\bf 696} 371--395 
  \urlprefix\url{http://dx.doi.org/10.1016/S0375-9474(01)01133-2}

\bibitem{Oba2009_PRC80-024301}
Oba H and Matsuo M 2009 {\em Phys. Rev. C\/} {\bf 80} 024301--14
  \urlprefix\url{http://link.aps.org/doi/10.1103/PhysRevC.80.024301}

\bibitem{Zhang2011_PRC83-054301}
Zhang Y, Matsuo M and Meng J 2011 {\em Phys. Rev. C\/} {\bf 83} 054301--11
  \urlprefix\url{http://link.aps.org/doi/10.1103/PhysRevC.83.054301}

\bibitem{Zhang2012_PRC86-054318}
Zhang Y, Matsuo M and Meng J 2012 {\em Phys. Rev. C\/} {\bf 86} 054318--12
  \urlprefix\url{http://link.aps.org/doi/10.1103/PhysRevC.86.054318}

\bibitem{Zhang2014_PRC90-034313}
Zhang Y, Matsuo M and Meng J 2014 {\em Phys. Rev. C\/} {\bf 90} 034313--5 
  \urlprefix\url{http://dx.doi.org/10.1103/PhysRevC.90.034313}

\bibitem{Sun2014_PRC90-054321}
Sun T~T, Zhang S~Q, Zhang Y, Hu J~N and Meng J 2014 {\em Phys. Rev. C\/} {\bf
  90}(5) 054321
  \urlprefix\url{http://link.aps.org/doi/10.1103/PhysRevC.90.054321}

\bibitem{Vautherin1972_PRC5-626}
Vautherin D and Brink D~M 1972 {\em Phys. Rev. C\/} {\bf 5} 626--647
  \urlprefix\url{http://link.aps.org/abstract/PRC/v5/p626}

\bibitem{Decharge1980_PRC21-1568}
Decharge J and Gogny D 1980 {\em Phys. Rev. C\/} {\bf 21} 1568--1593
  \urlprefix\url{http://link.aps.org/doi/10.1103/PhysRevC.21.1568}

\bibitem{Gambhir1990_APNY198-132}
Gambhir Y~K, Ring P and Thimet A 1990 {\em Ann. Phys. (NY)\/} {\bf 198}
  132--179 \urlprefix\url{http://dx.doi.org/10.1016/0003-4916(90)90330-Q}

\bibitem{Zhou2000_CPL17-717}
Zhou S~G, Meng J, Yamaji S and Yang S~C 2000 {\em Chin. Phys. Lett.\/} {\bf 17}
  717--719 
  \urlprefix\url{http://dx.doi.org/10.1088/0256-307X/17/10/006}

\bibitem{Stoitsov2000_PRC61-034311}
Stoitsov M~V, Dobaczewski J, Ring P and Pittel S 2000 {\em Phys. Rev. C\/} {\bf
  61} 034311--14 \urlprefix\url{http://link.aps.org/abstract/PRC/v61/e034311}

\bibitem{Stoitsov2003_PRC68-054312}
Stoitsov M~V, Dobaczewski J, Nazarewicz W, Pittel S and Dean D~J 2003 {\em
  Phys. Rev. C\/} {\bf 68} 054312--11
  \urlprefix\url{http://link.aps.org/abstract/PRC/v68/e054312}

\bibitem{Zhou2003_PRC68-034323}
Zhou S~G, Meng J and Ring P 2003 {\em Phys. Rev. C\/} {\bf 68} 034323--12
  (\textit{Preprint} \eprint{nucl-th/0303031})
  \urlprefix\url{http://link.aps.org/abstract/PRC/v68/e034323}

\bibitem{Serot1986_ANP16-1}
Serot B~D and Walecka J~D 1986 {\em Adv. Nucl. Phys.\/} {\bf 16} 1--327

\bibitem{Ring1996_PPNP37-193}
Ring P 1996 {\em Prog. Part. Nucl. Phys.\/} {\bf 37} 193--263 
  \urlprefix\url{http://dx.doi.org/10.1016/0146-6410(96)00054-3}

\bibitem{Vretenar2005_PR409-101}
Vretenar D, Afanasjev A~V, Lalazissis G~A and Ring P 2005 {\em Phys. Rep.\/}
  {\bf 409} 101--259
  \urlprefix\url{http://dx.doi.org/10.1016/j.physrep.2004.10.001}

\bibitem{Niksic2011_PPNP66-519}
Nik\v{s}i\'{c} T, Vretenar D and Ring P 2011 {\em Prog. Part. Nucl. Phys.\/}
  {\bf 66} 519--548 
  \urlprefix\url{http://dx.doi.org/10.1016/j.ppnp.2011.01.055}

\bibitem{Meng2013FrontiersofPhysics55}
Meng J, Peng J, Zhang S~Q and Zhao P~W 2013 {\em Frontiers of Physics\/} {\bf
  8}(1) 55--79 

\bibitem{Brockmann1990_PRC42-1965}
Brockmann R and Machleidt R 1990 {\em Phys. Rev. C\/} {\bf 42} 1965--1980
  \urlprefix\url{http://link.aps.org/abstract/PRC/v42/p1965}

\bibitem{Sharma1993_PLB317-3}
Sharma M~M, Lalazissis G~A and Ring P 1993 {\em Phys. Lett. B\/} {\bf 317}
  9--13 
  \urlprefix\url{http://dx.doi.org/10.1016/0370-2693(93)91561-Z}

\bibitem{Arima1969_PLB30-517}
Arima A, Harvey M and Shimizu K 1969 {\em Phys. Lett. B\/} {\bf 30} 517--522
  \urlprefix\url{http://www.sciencedirect.com/science/article/B6TVN-470W8D8-1VW/1/3dec596cd3ba7d158a639c410a814db7}

\bibitem{Hecht1969_NPA137-129}
Hecht K~T and Adler A 1969 {\em Nucl. Phys. A\/} {\bf 137} 129--143
  \urlprefix\url{http://www.sciencedirect.com/science/article/B6TVB-47201GS-1D2/1/81a3cea8030569b83b7cce4f6d25d1ee}

\bibitem{Ginocchio1997_PRL78-436}
Ginocchio J~N 1997 {\em Phys. Rev. Lett.\/} {\bf 78} 436--439
  \urlprefix\url{http://link.aps.org/abstract/PRL/v78/p436}

\bibitem{Meng1998_PRC58-R628}
Meng J, Sugawara-Tanabe K, Yamaji S, Ring P and Arima A 1998 {\em Phys. Rev.
  C\/} {\bf 58} R628--631
  \urlprefix\url{http://link.aps.org/abstract/PRC/v58/pR628}

\bibitem{Meng1999_PRC59-154}
Meng J, Sugawara-Tanabe K, Yamaji S and Arima A 1999 {\em Phys. Rev. C\/} {\bf
  59} 154--163 \urlprefix\url{http://link.aps.org/abstract/PRC/v59/p154}

\bibitem{Long2006_PLB639-242}
Long W~H, Sagawa H, Meng J and Van~Giai N 2006 {\em Phys. Lett. B\/} {\bf 639}
  242--247 \urlprefix\url{http://dx.doi.org/10.1016/j.physletb.2006.05.065}

\bibitem{Liang2011_PRC83-041301R}
Liang H, Zhao P, Zhang Y, Meng J and Giai N~V 2011 {\em Phys. Rev. C\/} {\bf
  83} 041301(R)--5
  \urlprefix\url{http://link.aps.org/doi/10.1103/PhysRevC.83.041301}

\bibitem{Lu2012_PRL109-072501}
Lu B~N, Zhao E~G and Zhou S~G 2012 {\em Phys. Rev. Lett.\/} {\bf 109} 072501--5
  (\textit{Preprint} \eprint{1204.5069})
  \urlprefix\url{http://link.aps.org/doi/10.1103/PhysRevLett.109.072501}

\bibitem{Lu2013_PRC88-024323}
Lu B~N, Zhao E~G and Zhou S~G 2013 {\em Phys. Rev. C\/} {\bf 88} 024323--11
  \urlprefix\url{http://link.aps.org/doi/10.1103/PhysRevC.88.024323}

\bibitem{Liang2015PhysicsReports1}
Liang H, Meng J and Zhou S~G 2015 {\em Physics Reports\/} {\bf 570} 1 -- 84
  \urlprefix\url{http://www.sciencedirect.com/science/article/pii/S0370157315000502}

\bibitem{Zhou2003_PRL91-262501}
Zhou S~G, Meng J and Ring P 2003 {\em Phys. Rev. Lett.\/} {\bf 91} 262501--4
  (\textit{Preprint} \eprint{nucl-th/0304067})
  \urlprefix\url{http://link.aps.org/abstract/PRL/v91/e262501}

\bibitem{Liang2010_EPJA44-119}
Liang H, Hui~Long W, Meng J and Van~Giai N 2010 {\em Eur. Phys. J. A\/} {\bf
  44} 119--124 \urlprefix\url{http://dx.doi.org/10.1140/epja/i2010-10938-6}

\bibitem{Koepf1989_NPA493-61}
Koepf W and Ring P 1989 {\em Nucl. Phys. A\/} {\bf 493} 61--82 
  \urlprefix\url{http://dx.doi.org/10.1016/0375-9474(89)90532-0}

\bibitem{Yao2006_PRC74-024307}
Yao J~M, Chen H and Meng J 2006 {\em Phys. Rev. C\/} {\bf 74} 024307--11
  \urlprefix\url{http://link.aps.org/doi/10.1103/PhysRevC.74.024307}

\bibitem{Arima2011_SciChinaPMA54-93}
Arima A 2011 {\em Sci. China-Phys. Mech. Astron.\/} {\bf 54} 188--193 
  \urlprefix\url{http://dx.doi.org/10.1007/s11433-010-4224-6}

\bibitem{Li2011_PTP125-1185}
Li J, Yao J~M, Meng J and Arima A 2011 {\em Prog. Theor. Phys.\/} {\bf 125}
  1185--1192 \urlprefix\url{http://ptp.ipap.jp/link?PTP/125/1185/}

\bibitem{Wei2012_PTPSuppl196-400}
Wei J, Li J and Meng J 2012 {\em Prog. Theor. Phys. Suppl.\/} {\bf 196}
  400--406 \urlprefix\url{http://ptp.ipap.jp/link?PTPS/196/400/}

\bibitem{Afanasjev2000_NPA676-196}
Afanasjev A~V, Ring P and K\"onig J 2000 {\em Nucl. Phys. A\/} {\bf 676}
  196--244 
  \urlprefix\url{http://www.sciencedirect.com/science/article/pii/S0375947400001871}

\bibitem{Zhao2012_PRC85-054310}
Zhao P~W, Peng J, Liang H~Z, Ring P and Meng J 2012 {\em Phys. Rev. C\/} {\bf
  85} 054310--14
  \urlprefix\url{http://link.aps.org/doi/10.1103/PhysRevC.85.054310}

\bibitem{Zhao2011_PRL107-122501}
Zhao P~W, Peng J, Liang H~Z, Ring P and Meng J 2011 {\em Phys. Rev. Lett.\/}
  {\bf 107} 122501--5
  \urlprefix\url{http://link.aps.org/doi/10.1103/PhysRevLett.107.122501}

\bibitem{Zhao2011_PLB699-181}
Zhao P~W, Zhang S~Q, Peng J, Liang H~Z, Ring P and Meng J 2011 {\em Phys. Lett.
  B\/} {\bf 699} 181--186 
  \urlprefix\url{http://dx.doi.org/10.1016/j.physletb.2011.03.068}

\bibitem{Koepf1991_ZPA340-119}
Koepf W, Gambhir Y~K, Ring P and Sharma M~M 1991 {\em Z. Phys. A\/} {\bf 340}
  119--124 \urlprefix\url{http://dx.doi.org/10.1007/BF01303823}

\bibitem{Sharma1994_PRL72-1431}
Sharma M~M, Lalazissis G~A, Hillebrandt W and Ring P 1994 {\em Phys. Rev.
  Lett.\/} {\bf 72} 1431--1434
  \urlprefix\url{http://link.aps.org/doi/10.1103/PhysRevLett.72.1431}

\bibitem{Zhu1994_PLB328-1}
Zhu Z~Y, Shen W~Q, Cai Y~H and Ma Y~G 1994 {\em Phys. Lett. B\/} {\bf 328} 1--4
  \urlprefix\url{http://dx.doi.org/10.1016/0370-2693(94)90418-9}

\bibitem{Ren1995_PRC52-R20}
Ren Z, Mittig W, Chen B and Ma Z 1995 {\em Phys. Rev. C\/} {\bf 52} R20--R22
  \urlprefix\url{http://link.aps.org/doi/10.1103/PhysRevC.52.R20}

\bibitem{Ren1995_PLB351-11}
Ren Z, Xu G, Chen B, Ma Z and Mittig W 1995 {\em Phys. Lett. B\/} {\bf 351}
  11--17 \urlprefix\url{http://dx.doi.org/10.1016/0370-2693(95)00364-Q}

\bibitem{Ren1996_PRC53-R572}
Ren Z, Chen B, Ma Z and Xu G 1996 {\em Phys. Rev. C\/} {\bf 53} R572--575
  \urlprefix\url{http://link.aps.org/abstract/PRC/v53/pR572}

\bibitem{Ren1998_PRC57-2752}
Ren Z~Z, Faessler A and Bobyk A 1998 {\em Phys. Rev. C\/} {\bf 57} 2752--2755
  \urlprefix\url{http://link.aps.org/doi/10.1103/PhysRevC.57.2752}

\bibitem{Meng1998_PLB419-1}
Meng J, Tanihata I and Yamaji S 1998 {\em Phys. Lett. B\/} {\bf 419} 1--6
  \urlprefix\url{http://www.sciencedirect.com/science/article/B6TVN-3VGPWM4-44/1/f83ff8aa3662905a3a6f21e8b7464573}

\bibitem{Vretenar1998_PRC57-R1060}
Vretenar D, Poschl W, Lalazissis G~A and Ring P 1998 {\em Phys. Rev. C\/} {\bf
  57} R1060--R1063
  \urlprefix\url{http://link.aps.org/doi/10.1103/PhysRevC.57.R1060}

\bibitem{Lu2002_CPL19-1775}
L\"u H~F and Meng J 2002 {\em Chin. Phys. Lett.\/} {\bf 19} 1775--1778 
  \urlprefix\url{http://stacks.iop.org/0256-307X/19/i=12/a=310}

\bibitem{Lu2003_EPJA17-19}
L\"{u} H~F, Meng J, Zhang S~Q and Zhou S~G 2003 {\em Eur. Phys. J. A\/} {\bf
  17} 19--24 \urlprefix\url{http://dx.doi.org/10.1140/epja/i2002-10136-3}

\bibitem{Lu2003_HEPNP27-411}
Lu H~F, Meng J and Zhang S~Q 2003 {\em High Ener. Phys. Nucl. Phys.\/} {\bf 27}
  411--415 (in Chinese)

\bibitem{Lu2008_CPL25-3613}
L\"u H~F 2008 {\em Chin. Phys. Lett.\/} {\bf 25} 3613--3616 
  \urlprefix\url{http://stacks.iop.org/0256-307X/25/i=10/a=025}

\bibitem{Long2010_PRC81-031302R}
Long W~H, Ring P, Meng J, Van~Giai N and Bertulani C~A 2010 {\em Phys. Rev.
  C\/} {\bf 81} 031302(R)--5
  \urlprefix\url{http://dx.doi.org/10.1103/PhysRevC.81.031302}

\bibitem{Zhou2008_ISPUN2007}
Zhou S~G, Meng J and Ring P 2008 Deformed relativistic {Hartree-Bogoliubov}
  model for exotic nuclei {\em Physics of Unstable Nuclei\/} ed Khoa D~T,
  Egelhof P, Gales S, Van~Giai N and Motobayashi T (World Scientific) pp
  402--408 {Proceedings of the International Symposium on Physics of Unstable
  Nuclei, July 3-7, 2007, Hoi An, Vietnam} (\textit{Preprint}
  \eprint{0803.1376})
  \urlprefix\url{http://dx.doi.org/10.1142/9789812776150_0059}

\bibitem{Chen2012_PRC85-067301}
Chen Y, Li L, Liang H and Meng J 2012 {\em Phys. Rev. C\/} {\bf 85} 067301--5
  \urlprefix\url{http://link.aps.org/doi/10.1103/PhysRevC.85.067301}

\bibitem{Afanasjev1999_PR322-1}
Afanasjev A, Fossan D, Lane G and Ragnarsson I 1999 {\em Phys. Rep.\/} {\bf
  322} 1--124 
  \urlprefix\url{http://dx.doi.org/10.1016/S0370-1573(99)00035-6}

\bibitem{Ginocchio2005_PR414-165}
Ginocchio J~N 2005 {\em Phys. Rep.\/} {\bf 414} 165--261 
  \urlprefix\url{http://dx.doi.org/10.1016/j.physrep.2005.04.003}

\bibitem{Meng2013_FPC8-55}
Meng J, Peng J, Zhang S~Q and Zhao P~W 2013 {\em Front. Phys.\/} {\bf 8} 55--79
  \urlprefix\url{http://dx.doi.org/10.1007/s11467-013-0287-y}

\bibitem{Long2007_PRC76-034314}
Long W, Sagawa H, Giai N~V and Meng J 2007 {\em Phys. Rev. C\/} {\bf 76}
  034314--11 \urlprefix\url{http://link.aps.org/doi/10.1103/PhysRevC.76.034314}

\bibitem{Long2010_PRC81-024308}
Long W~H, Ring P, Giai N~V and Meng J 2010 {\em Phys. Rev. C\/} {\bf 81}
  024308--10 \urlprefix\url{http://link.aps.org/doi/10.1103/PhysRevC.81.024308}

\bibitem{Nikolaus1992_PRC46-1757}
Nikolaus B~A, Hoch T and Madland D~G 1992 {\em Phys. Rev. C\/} {\bf 46}
  1757--1781 \urlprefix\url{http://link.aps.org/doi/10.1103/PhysRevC.46.1757}

\bibitem{Burvenich2002_PRC65-044308}
Burvenich T, Madland D~G, Maruhn J~A and Reinhard P~G 2002 {\em Phys. Rev. C\/}
  {\bf 65} 044308--23
  \urlprefix\url{http://link.aps.org/doi/10.1103/PhysRevC.65.044308}

\bibitem{Niksic2008_PRC78-034318}
Nik\v{s}i\'{c} T, Vretenar D and Ring P 2008 {\em Phys. Rev. C\/} {\bf 78}
  034318--19 \urlprefix\url{http://link.aps.org/doi/10.1103/PhysRevC.78.034318}

\bibitem{Zhao2010_PRC82-054319}
Zhao P~W, Li Z~P, Yao J~M and Meng J 2010 {\em Phys. Rev. C\/} {\bf 82}
  054319--14 \urlprefix\url{http://link.aps.org/doi/10.1103/PhysRevC.82.054319}

\bibitem{Friar1996_PRC53-87}
Friar J, Madland D and Lynn B 1996 {\em Phys. Rev. C\/} {\bf 53} 3085--3087
  \urlprefix\url{http://dx.doi.org/10.1103/PhysRevC.53.3085}

\bibitem{Manohar1984_NPB234-12}
Manohar A and Georgi H 1984 {\em Nucl. Phys. B\/} {\bf 234} 189--212 
  \urlprefix\url{http://dx.doi.org/10.1016/0550-3213(84)90231-1}

\bibitem{Liang2012_PRC86-021302R}
Liang H, Zhao P, Ring P, Roca-Maza X and Meng J 2012 {\em Phys. Rev. C\/} {\bf
  86} 021302--5
  \urlprefix\url{http://link.aps.org/doi/10.1103/PhysRevC.86.021302}

\bibitem{Sulaksono2003_APNP308-70}
Sulaksono A, B\"urvenich T, Maruhn J~A, Reinhard P~G and Greiner W 2003 {\em
  Annals of Physics\/} {\bf 308} 354--370 
  \urlprefix\url{http://dx.doi.org/10.1016/S0003-4916(03)00146-5}

\bibitem{Kucharek1991_ZPA339-23}
Kucharek H and Ring P 1991 {\em Z. Phys. A\/} {\bf 339} 23--35
  \urlprefix\url{http://dx.doi.org/10.1007/BF01282930}

\bibitem{Berger1984_NPA428-23}
Berger J~F, Girod M and Gogny D 1984 {\em Nucl. Phys. A\/} {\bf 428} 23--36
  \urlprefix\url{http://dx.doi.org/10.1016/0375-9474(84)90240-9}

\bibitem{Meng1998Nucl.Phys.A3}
Meng J 1998 {\em Nucl. Phys. A\/} {\bf 635} 3--42

\bibitem{Bouyssy1987_PRC36-380}
Bouyssy A, Mathiot J~F, Van~Giai N and Marcos S 1987 {\em Phys. Rev. C\/} {\bf
  36} 380--401 \urlprefix\url{http://link.aps.org/doi/10.1103/PhysRevC.36.380}

\bibitem{Long2006_PLB640-150}
Long W~H, Van~Giai N and Meng J 2006 {\em Phys. Lett. B\/} {\bf 640} 150--154
  \urlprefix\url{http://dx.doi.org/10.1016/j.physletb.2006.07.064}

\bibitem{Koepf1991_ZPA339-81}
Koepf W and Ring P 1991 {\em Z. Phys. A\/} {\bf 339} 81--90
  \urlprefix\url{http://dx.doi.org/10.1007/BF01282936}

\bibitem{Terasaki1996_NPA600-371}
Terasaki J, Heenen P~H, Flocard H and Bonche P 1996 {\em Nucl. Phys. A\/} {\bf
  600} 371--386 
  \urlprefix\url{http://dx.doi.org/10.1016/0375-9474(96)00036-X}

\bibitem{Teran2003_PRC67-064314}
Teran E, Oberacker V~E and Umar A~S 2003 {\em Phys. Rev. C\/} {\bf 67}
  064314--13 \urlprefix\url{http://link.aps.org/abstract/PRC/v67/e064314}

\bibitem{Tajima2004_PRC69-034305}
Tajima N 2004 {\em Phys. Rev. C\/} {\bf 69} 034305--22
  \urlprefix\url{http://link.aps.org/abstract/PRC/v69/e034305}

\bibitem{Stoitsov2008_PRC77-054301}
Stoitsov M, Michel N and Matsuyanagi K 2008 {\em Phys. Rev. C\/} {\bf 77}
  054301--12 \urlprefix\url{http://link.aps.org/abstract/PRC/v77/e054301}

\bibitem{Nakada2008_NPA808-47}
Nakada H 2008 {\em Nucl. Phys. A\/} {\bf 808} 47--59 
  \urlprefix\url{http://www.sciencedirect.com/science/article/B6TVB-4SMNXRD-1/2/5b3d912dab1e2b213bc161417b54afd3}

\bibitem{Schunck2008_PRC78-064305}
Schunck N and Egido J~L 2008 {\em Phys. Rev. C\/} {\bf 78} 064305--14
  \urlprefix\url{http://link.aps.org/abstract/PRC/v78/e064305}

\bibitem{Zhou2006_AIPCP865-90}
Zhou S~G, Meng J and Ring P 2006 {\em AIP Conf. Proc.\/} {\bf 865} 90--95
  \urlprefix\url{http://link.aip.org/link/?APC/865/90/1}

\bibitem{Price1987_PRC36-354}
Price C~E and Walker G~E 1987 {\em Phys. Rev. C\/} {\bf 36} 354--364
  \urlprefix\url{http://link.aps.org/abstract/PRC/v36/p354}

\bibitem{Bonche1985_NPA443-39}
Bonche P, Flocard H, Heenen P~H, Krieger S~J and Weiss M~S 1985 {\em Nucl.
  Phys. A\/} {\bf 443} 39--63 
  \urlprefix\url{http://dx.doi.org/10.1016/0375-9474(85)90320-3}

\bibitem{Long2004_PRC69-034319}
Long W, Meng J, Giai N~V and Zhou S~G 2004 {\em Phys. Rev. C\/} {\bf 69}
  034319--15 (\textit{Preprint} \eprint{nucl-th/0311031})
  \urlprefix\url{http://link.aps.org/abstract/PRC/v69/e034319}

\bibitem{Zhao2009_CPL26-112102}
Zhao P~W, Sun B~Y and Meng J 2009 {\em Chin. Phys. Lett.\/} {\bf 26} 112102--3
  \urlprefix\url{http://stacks.iop.org/0256-307X/26/i=11/a=112102}

\bibitem{Meng2003_NPA722-C366}
Meng J, L\"{u} H~F, Zhang S~Q and Zhou S~G 2003 {\em Nucl. Phys. A\/} {\bf 722}
  366c--371c 
  \urlprefix\url{http://dx.doi.org/10.1016/S0375-9474(03)01391-5}

\bibitem{Meng1997_ZPA358-123}
Meng J, P\"oschl W and Ring P 1997 {\em Z. Phys. A\/} {\bf 358} 123--124 
  0939-7922 \urlprefix\url{http://dx.doi.org/10.1007/s002180050285}

\bibitem{Meng1999_NPA650-176}
Meng J and Tanihata I 1999 {\em Nucl. Phys. A\/} {\bf 650} 176--196
  \urlprefix\url{http://www.sciencedirect.com/science/article/B6TVB-3WSV1PV-3/1/a568ac0e55bc594c2a148681ea97c9e4}

\bibitem{Meng2002_PLB532-209}
Meng J, Zhou S~G and Tanihata I 2002 {\em Phys. Lett. B\/} {\bf 532} 209--214
  (\textit{Preprint} \eprint{nucl-th/0107039})
  \urlprefix\url{http://dx.doi.org/10.1016/S0370-2693(02)01574-5}

\bibitem{Zhang2002_CPL19-312}
Zhang S~Q, Meng J, Zhou S~G and Zeng J~Y 2002 {\em Chin. Phys. Lett.\/} {\bf
  19} 312--314 
  \urlprefix\url{http://dx.doi.org/10.1088/0256-307X/19/3/308}

\bibitem{Zhang2003_SciChinaG46-632}
Zhang S~Q, Meng J and Zhou S~G 2003 {\em Sci. China G\/} {\bf 46} 632--658
  (\textit{Preprint} \eprint{nucl-th/0302032})
  \urlprefix\url{http://www.scichina.com:8083/sciGe/EN/abstract/abstract408047.shtml}

\bibitem{Bertsch1989_PRC39-1154}
Bertsch G~F, Brown B~A and Sagawa H 1989 {\em Phys. Rev. C\/} {\bf 39}
  1154--1157 \urlprefix\url{http://link.aps.org/doi/10.1103/PhysRevC.39.1154}

\bibitem{Sagawa1992_PLB286-7}
Sagawa H 1992 {\em Physics Letters B\/} {\bf 286} 7--12 
  \urlprefix\url{http://dx.doi.org/10.1016/0370-2693(92)90150-3}

\bibitem{Rotival2009_PRC79-054308}
Rotival V and Duguet T 2009 {\em Phys. Rev. C\/} {\bf 79} 054308--24
  \urlprefix\url{http://link.aps.org/abstract/PRC/v79/e054308}

\bibitem{Rotival2009_PRC79-054309}
Rotival V, Bennaceur K and Duguet T 2009 {\em Phys. Rev. C\/} {\bf 79}
  054309--21 \urlprefix\url{http://link.aps.org/abstract/PRC/v79/e054309}

\bibitem{Meng1998Phys.Lett.B1}
Meng J, Tanihata I and Yamaji S 1998 {\em Phys. Lett. B\/} {\bf 419} 1--6

\bibitem{Im2000_PRC61-047302}
Im S and Meng J 2000 {\em Phys. Rev. C\/} {\bf 61} 047302--4
  \urlprefix\url{http://link.aps.org/doi/10.1103/PhysRevC.61.047302}

\bibitem{Mizutori2000_PRC61-044326}
Mizutori S, Dobaczewski J, Lalazissis G~A, Nazarewicz W and Reinhard P~G 2000
  {\em Phys. Rev. C\/} {\bf 61} 044326--14
  \urlprefix\url{http://link.aps.org/abstract/PRC/v61/e044326}

\bibitem{Sharma1993_PLB312-377}
Sharma M~M, Nagarajan M~A and Ring P 1993 {\em Phys. Lett. B\/} {\bf 312}
  377--381 
  \urlprefix\url{http://www.sciencedirect.com/science/article/B6TVN-470G13S-69/2/b3e2b625f3a9be50fc271b3986c87fda}

\bibitem{Zhao2012_PRC86-064324}
Zhao P~W, Song L~S, Sun B, Geissel H and Meng J 2012 {\em Phys. Rev. C\/} {\bf
  86} 064324--6
  \urlprefix\url{http://link.aps.org/doi/10.1103/PhysRevC.86.064324}

\bibitem{Sun2011_SciChinaPMA54-214}
Sun B, Zhao P and Meng J 2011 {\em Sci. China-Phys. Mech. Astron.\/} {\bf 54}
  210--214 
  \urlprefix\url{http://dx.doi.org/10.1007/s11433-010-4222-8}

\bibitem{Lu2012_PRC85-011301R}
Lu B~N, Zhao E~G and Zhou S~G 2012 {\em Phys. Rev. C\/} {\bf 85} 011301(R)--5
  \urlprefix\url{http://link.aps.org/doi/10.1103/PhysRevC.85.011301}

\bibitem{Lu2014_PRC89-014323}
Lu B~N, Zhao J, Zhao E~G and Zhou S~G 2014 {\em Phys. Rev. C\/} {\bf 89}
  014323--15 \urlprefix\url{http://link.aps.org/doi/10.1103/PhysRevC.89.014323}

\bibitem{Lalazissis1998_NPA632-363}
Lalazissis G~A, Vretenar D, P?schl W and Ring P 1998 {\em Nucl. Phys. A\/} {\bf
  632} 363--382 
  \urlprefix\url{http://www.sciencedirect.com/science/article/B6TVB-3T0XR5K-3/2/719c3043312d1ae0173bf408684416a5}

\bibitem{Bennaceur2000_PLB496-154}
Bennaceur K, Dobaczewski J and Ploszajczak M 2000 {\em Phys. Lett. B\/} {\bf
  496} 154--160 
  \urlprefix\url{http://www.sciencedirect.com/science/article/B6TVN-41XM9PX-4/2/d261a458d12e29f0df079cabe8081520}

\bibitem{Dobaczewski2001_NPA693-361}
Dobaczewski J, Nazarewicz W and Reinhard P~G 2001 {\em Nucl. Phys. A\/} {\bf
  693} 361--373 
  \urlprefix\url{http://dx.doi.org/10.1016/S0375-9474(01)00993-9}

\bibitem{Terasaki2006_PRC74-054318}
Terasaki J, Zhang S~Q, Zhou S~G and Meng J 2006 {\em Phys. Rev. C\/} {\bf 74}
  054318--6 (\textit{Preprint} \eprint{nucl-th/0603005})
  \urlprefix\url{http://link.aps.org/abstract/PRC/v74/e054318}

\bibitem{Terasaki2006_IJMPE15-1833}
Terasaki J, Zhang S~Q, Zhou S~G and Meng J 2006 {\em Int. J. Mod. Phys. E\/}
  {\bf 15} 1833--1841
  \urlprefix\url{http://www.worldscinet.com/cgi-bin/details.cgi?id=pii:S0218301306005010&type=html}

\bibitem{Grasso2006_PRC74-064317}
Grasso M, Yoshida S, Sandulescu N and Van~Giai N 2006 {\em Phys. Rev. C\/} {\bf
  74} 064317--7 \urlprefix\url{http://link.aps.org/abstract/PRC/v74/e064317}

\bibitem{Grasso2001_PRC64-064321}
Grasso M, Sandulescu N, Van~Giai N and Liotta R~J 2001 {\em Phys. Rev. C\/}
  {\bf 64} 064321--9
  \urlprefix\url{http://link.aps.org/abstract/PRC/v64/e064321}

\bibitem{Boguta1977Nucl.Phys.A413}
Boguta J and Bodmer A~R 1977 {\em Nucl. Phys. A\/} {\bf 292} 413--428

\bibitem{Brockmann1977_PLB69-167}
Brockmann R and Weise W 1977 {\em Phys. Lett. B\/} {\bf 69} 167--169 
  \urlprefix\url{http://dx.doi.org/10.1016/0370-2693(77)90635-9}

\bibitem{Horowitz1984_PLB140-86}
Horowitz C~J and Serot B~D 1984 {\em Phys. Lett. B\/} {\bf 140} 181--186 
  \urlprefix\url{http://dx.doi.org/10.1016/0370-2693(84)90916-X}

\bibitem{Bielajew1984_APNY156-215}
Bielajew A~F and Serot B~D 1984 {\em Annals of Physics\/} {\bf 156} 215--264
  \urlprefix\url{http://dx.doi.org/10.1016/0003-4916(84)90034-4}

\bibitem{Bouyssy1985_PRL55-733}
Bouyssy A, Marcos S, Mathiot J and Van~Giai N 1985 {\em Physical Review
  Letters\/} {\bf 55} 1731--1733 
  \urlprefix\url{http://dx.doi.org/10.1103/PhysRevLett.55.1731}

\bibitem{Blunden1987_PLB196-295}
Blunden P~G and Iqbal M~J 1987 {\em Phys. Lett. B\/} {\bf 196} 295--298 
  \urlprefix\url{http://dx.doi.org/10.1016/0370-2693(87)90734-9}

\bibitem{Bernardos1993_PRC48-2665}
Bernardos P, Fomenko V, Giai N, Quelle M, Marcos S, Niembro R and Savushkin L
  1993 {\em Phys. Rev. C\/} {\bf 48} 2665--2672 
  \urlprefix\url{http://dx.doi.org/10.1103/PhysRevC.48.2665}

\bibitem{Marcos2004_JPG30-703}
Marcos S, Savushkin L~N, Fomenko V~N, L\"opez-Quelle M and Niembro R 2004 {\em
  J. Phys. G\/} {\bf 30} 703--721 
  \urlprefix\url{http://dx.doi.org/10.1088/0954-3899/30/6/002}

\bibitem{Long2008_EPL82-12001}
Long W~H, Sagawa H, Meng J and Van~Giai N 2008 {\em Europhys. Lett.\/} {\bf 82}
  12001 \urlprefix\url{http://dx.doi.org/doi/10.1209/0295-5075/82/12001}

\bibitem{Liang2008_PRL101-122502}
Liang H, Van~Giai N and Meng J 2008 {\em Phys. Rev. Lett.\/} {\bf 101}
  122502--4
  \urlprefix\url{http://link.aps.org/doi/10.1103/PhysRevLett.101.122502}

\bibitem{Sun2008_PRC78-065805}
Sun B~Y, Long W~H, Meng J and Lombardo U 2008 {\em Phys. Rev. C\/} {\bf 78}
  065805--14 \urlprefix\url{http://link.aps.org/doi/10.1103/PhysRevC.78.065805}

\bibitem{Liang2009_PRC79-064316}
Liang H, Giai N~V and Meng J 2009 {\em Phys. Rev. C\/} {\bf 79} 064316--7
  \urlprefix\url{http://link.aps.org/doi/10.1103/PhysRevC.79.064316}

\bibitem{Liang2012_PRC85-064302}
Liang H, Zhao P and Meng J 2012 {\em Phys. Rev. C\/} {\bf 85}(6) 064302
  \urlprefix\url{http://link.aps.org/doi/10.1103/PhysRevC.85.064302}

\bibitem{Jaminon1989_PRC40-67}
Jaminon M and Mahaux C 1989 {\em Phys. Rev. C\/} {\bf 40} 354--367 \urlprefix\url{http://dx.doi.org/10.1103/PhysRevC.40.354}

\bibitem{Tarpanov2008_PRC77-054316}
Tarpanov D, Liang H, Van~Giai N and Stoyanov C 2008 {\em Phys. Rev. C\/} {\bf
  77} 054316 (pages~5)
  \urlprefix\url{http://link.aps.org/abstract/PRC/v77/e054316}

\bibitem{Long2009_PLB680-428}
Long W~H, Nakatsukasa T, Sagawa H, Meng J, Nakada H and Zhang Y 2009 {\em Phys.
  Lett. B\/} {\bf 680} 428--431 
  \urlprefix\url{http://dx.doi.org/10.1016/j.physletb.2009.09.034}

\bibitem{Moreno-Torres2010_PRC81-064327}
Moreno-Torres M, Grasso M, Liang H, De~Donno V, Anguiano M and Van~Giai N 2010
  {\em Phys. Rev. C\/} {\bf 81} 064327--13
  \urlprefix\url{http://dx.doi.org/10.1103/PhysRevC.81.064327}

\bibitem{Lalazissis2005_PRC71-024312}
Lalazissis G~A, Nik\v{s}i\'{c} T, Vretenar D and Ring P 2005 {\em Phys. Rev.
  C\/} {\bf 71} 024312--10
  \urlprefix\url{http://link.aps.org/abstract/PRC/v71/e024312}

\bibitem{Audi2003_NPA729-337}
Audi G, Wapstra A~H and Thibault C 2003 {\em Nucl. Phys. A\/} {\bf 729}
  337--676 \urlprefix\url{http://dx.doi.org/10.1016/j.nuclphysa.2003.11.003}

\bibitem{Rufa1990_PRC42-2469}
Rufa M, Schaffner J, Maruhn J, Stoecker H, Greiner W and Reinhard P~G 1990 {\em
  Phys. Rev. C\/} {\bf 42} 2469--2478
  \urlprefix\url{http://link.aps.org/doi/10.1103/PhysRevC.42.2469}

\bibitem{Rufa1991_PRC43-2020}
Rufa M, Schaffner J, Maruhn J, Stoecker H, Greiner W and Reinhard P~G 1991 {\em
  Phys. Rev. C\/} {\bf 43} 2020--2020 (Erratum)
  \urlprefix\url{http://link.aps.org/doi/10.1103/PhysRevC.43.2020}

\bibitem{Liang2015_PR570-1}
Liang H, Meng J and Zhou S~G 2015 {\em Phys. Rep.\/} {\bf 570} 1--84 
  \urlprefix\url{http://www.sciencedirect.com/science/article/pii/S0370157315000502}

\bibitem{Wang2014_PLB734-215}
Wang N, Liu M, Wu X and Meng J 2014 {\em Phys. Lett. B\/} {\bf 734} 215--219
  \urlprefix\url{http://www.sciencedirect.com/science/article/pii/S037026931400358X}

\bibitem{Ring1980}
Ring P and Schuck P 1980 {\em The nuclear many-body problem\/}
  (Springer-Verlag, New York)

\bibitem{Audi1995_NPA595-409}
Audi G and Wapstra A~H 1995 {\em Nucl. Phys. A\/} {\bf 595} 409--480
  \urlprefix\url{http://www.sciencedirect.com/science/article/B6TVB-3YYTR61-1/1/4d118297fb99a6b35c8527554b90e57d}

\bibitem{Wang2010_PRC81-044322}
Wang N, Liu M and Wu X 2010 {\em Phys. Rev. C\/} {\bf 81} 044322--
  \urlprefix\url{http://link.aps.org/doi/10.1103/PhysRevC.81.044322}

\bibitem{Wang2010_PRC82-044304}
Wang N, Liang Z, Liu M and Wu X 2010 {\em Phys. Rev. C\/} {\bf 82} 044304--
  \urlprefix\url{http://link.aps.org/doi/10.1103/PhysRevC.82.044304}

\bibitem{Liu2011_PRC84-014333}
Liu M, Wang N, Deng Y and Wu X 2011 {\em Phys. Rev. C\/} {\bf 84} 014333--8
  \urlprefix\url{http://link.aps.org/doi/10.1103/PhysRevC.84.014333}

\bibitem{Sobiczewski2014_PRC89-024311}
Sobiczewski A and Litvinov Y~A 2014 {\em Phys. Rev. C\/} {\bf 89}(2) 024311
  \urlprefix\url{http://link.aps.org/doi/10.1103/PhysRevC.89.024311}

\bibitem{Sobiczewski2014_PRC90-017302}
Sobiczewski A and Litvinov Y~A 2014 {\em Phys. Rev. C\/} {\bf 90} 017302--
  \urlprefix\url{http://link.aps.org/doi/10.1103/PhysRevC.90.017302}

\bibitem{LiZhu2012ActaPhysicaSinica72601}
Li~Zhu Niu Zhong-Ming S~B~H~W~N~M~J 2012 {\em Acta Physica Sinica\/} {\bf 61}
  72601 (pages~0)
  \urlprefix\url{http://wulixb.iphy.ac.cn/EN/abstract/article_47169.shtml}

\bibitem{Mendoza-Temis2014_arXiv1409.6135}
Mendoza-Temis J~d~J, Martinez-Pinedo G, Langanke K, Bauswein A and Janka H~T
  2014 On the robustness of the r-process in neutron-star mergers
  arXiv:1409.6135 [astro-ph.HE] (\textit{Preprint} \eprint{1409.6135})
  \urlprefix\url{http://arxiv.org/abs/1409.6135}

\bibitem{Audi2012_ChinPhysC36-1287}
Audi G, Wang M, Wapstra A~H, Kondev F~G, MacCormick M, X X and Pfeiffer B 2012
  {\em Chin. Phys. C\/} {\bf 36} 1287--1602
  \urlprefix\url{http://amdc.impcas.ac.cn/}

\bibitem{Chen2014_PRC89-014312}
Chen Y, Ring P and Meng J 2014 {\em Phys. Rev. C\/} {\bf 89} 014312--9
  \urlprefix\url{http://link.aps.org/doi/10.1103/PhysRevC.89.014312}

\bibitem{Misu1997_NPA614-44}
Misu T, Nazarewicz W and {\AA}berg S 1997 {\em Nucl. Phys. A\/} {\bf 614}
  44--70 
  \urlprefix\url{http://dx.doi.org/10.1016/S0375-9474(96)00458-7}

\bibitem{Li1996_PRC54-1617}
Li X and Heenen P~H 1996 {\em Phys. Rev. C\/} {\bf 54} 1617--1621
  \urlprefix\url{http://link.aps.org/abstract/PRC/v54/p1617}

\bibitem{Pei2006_NPA765-29}
Pei J~C, Xu F~R and Stevenson P~D 2006 {\em Nucl. Phys. A\/} {\bf 765} 29--38
  \urlprefix\url{http://dx.doi.org/10.1016/j.nuclphysa.2005.10.004}

\bibitem{Hamamoto2004_PRC69-041306R}
Hamamoto I 2004 {\em Phys. Rev. C\/} {\bf 69} 041306R--4
  \urlprefix\url{http://dx.doi.org/10.1103/PhysRevC.69.041306}

\bibitem{Nunes2005_NPA757-349}
Nunes F~M 2005 {\em Nucl. Phys. A\/} {\bf 757} 349--359 
  \urlprefix\url{http://dx.doi.org/10.1016/j.nuclphysa.2005.04.005}

\bibitem{Guo2003_CTP40-573}
Guo L, Zhao E~G and Sakata F 2003 {\em Commun. Theor. Phys.\/} {\bf 40}
  573--576 \urlprefix\url{http://ctp.itp.cn/qikan/Epaper/zhaiyao.asp?bsid=1081}

\bibitem{Li2012_CPL29-042101}
Li L, Meng J, Ring P, Zhao E~G and Zhou S~G 2012 {\em Chin. Phys. Lett.\/} {\bf
  29} 042101--4 (\textit{Preprint} \eprint{1203.1363})
  \urlprefix\url{http://dx.doi.org/10.1088/0256-307X/29/4/042101}

\bibitem{Baumann2007_Nature449-1022}
Baumann T, Amthor A~M, Bazin D, Brown B~A, III C~M~F, Gade A, Ginter T~N,
  Hausmann M, Matos M, Morrissey D~J, Portillo M, Schiller A, Sherrill B~M,
  Stolz A, Tarasov O~B and Thoennessen M 2007 {\em Nature\/} {\bf 449}
  1022--1024 
  \urlprefix\url{http://www.nature.com/nature/journal/v449/n7165/abs/nature06213.html}

\bibitem{Moller1995_ADNDT59-185}
M\"oller P, Nix J~R, Myers W~D and Swiatecki W~J 1995 {\em At. Data Nucl. Data
  Tables\/} {\bf 59} 185--381
  \urlprefix\url{http://dx.doi.org/10.1006/adnd.1995.1002}

\bibitem{Zhi2006_PLB638-166}
Zhi Q and Ren Z 2006 {\em Phys. Lett. B\/} {\bf 638} 166--170 
  \urlprefix\url{http://dx.doi.org/10.1016/j.physletb.2006.05.057}

\bibitem{Ren1996_PLB380-241}
Ren Z, Zhu Z~Y, Cai Y~H and Xu G 1996 {\em Phys. Lett. B\/} {\bf 380} 241--246
  \urlprefix\url{http://dx.doi.org/10.1016/0370-2693(96)00462-5}

\bibitem{Terasaki1997_NPA621-706}
Terasaki J, Flocard H, Heenen P~H and Bonche P 1997 {\em Nucl. Phys. A\/} {\bf
  621} 706--718 
  \urlprefix\url{http://dx.doi.org/10.1016/S0375-9474(97)00183-8}

\bibitem{Goriely2010_PRC82-035804}
Goriely S, Chamel N and Pearson J~M 2010 {\em Phys. Rev. C\/} {\bf 82}
  035804--18 and \href{http://www.astro.ulb.ac.be/pmwiki/Brusslib/Hfb17}{the
  HFB21 mass table}
  \urlprefix\url{http://link.aps.org/doi/10.1103/PhysRevC.82.035804}

\bibitem{Suzuki1998_NPA630-661}
Suzuki T, Geissel H, Bochkarev O, Chulkov L, Golovkov M, Fukunishi N, Hirata D,
  Irnich H, Janas Z, Keller H, Kobayashi T, Kraus G, M\"{u}zenberg G, Neumaier
  S, Nickel F, Ozawa A, Piechaczeck A, Roeckl E, Schwab W, S\"{u}merer K,
  Yoshida K and Tanihata I 1998 {\em Nucl. Phys. A\/} {\bf 630} 661 \urlprefix\url{http://dx.doi.org/10.1016/S0375-9474(98)00799-4}

\bibitem{Kanungo2011_PRC83-021302R}
Kanungo R, Prochazka A, Horiuchi W, Nociforo C, Aumann T, Boutin D, Cortina-Gil
  D, Davids B, Diakaki M, Farinon F, Geissel H, Gernhaeuser R, Gerl J, Janik R,
  Jonson B, Kindler B, Knoebel R, Kruecken R, Lantz M, Lenske H, Litvinov Y,
  Lommel B, Mahata K, Maierbeck P, Musumarra A, Nilsson T, Perro C,
  Scheidenberger C, Sitar B, Strmen P, Sun B, Suzuki Y, Szarka I, Tanihata I,
  Utsuno Y, Weick H and Winkler M 2011 {\em Phys. Rev. C\/} {\bf 83}
  021302(R)--4
  \urlprefix\url{http://link.aps.org/doi/10.1103/PhysRevC.83.021302}

\bibitem{Raman2001_ADNDT78-1}
Raman S, Nestor C~W and Tikkanen P 2001 {\em At. Data Nucl. Data Tables\/} {\bf
  78} 1--128

\bibitem{Patra1991_PLB273-13}
Patra S~K and Praharaj C~R 1991 {\em Phys. Lett. B\/} {\bf 273} 13--19 \urlprefix\url{http://dx.doi.org/10.1016/0370-2693(91)90545-2}

\bibitem{Yao2010_PRC81-044311}
Yao J~M, Meng J, Ring P and Vretenar D 2010 {\em Phys. Rev. C\/} {\bf 81}
  044311--13 \urlprefix\url{http://link.aps.org/doi/10.1103/PhysRevC.81.044311}

\bibitem{Yao2011_PRC83-014308}
Yao J~M, Mei H, Chen H, Meng J, Ring P and Vretenar D 2011 {\em Phys. Rev. C\/}
  {\bf 83} 014308--11
  \urlprefix\url{http://link.aps.org/doi/10.1103/PhysRevC.83.014308}

\bibitem{Church2005_PRC72-054320}
Church J~A, Campbell C~M, Dinca D~C, Enders J, Gade A, Glasmacher T, Hu Z,
  Janssens R~V~F, Mueller W~F, Olliver H, Perry B~C, Riley L~A and Yurkewicz
  K~L 2005 {\em Phys. Rev. C\/} {\bf 72} 054320--6
  \urlprefix\url{http://link.aps.org/abstract/PRC/v72/e054320}

\bibitem{Sorlin2008_PPNP61-602}
Sorlin O and Porquet M~G 2008 {\em Prog. Part. Nucl. Phys.\/} {\bf 61} 602--673
  \urlprefix\url{http://dx.doi.org/10.1016/j.ppnp.2008.05.001}

\bibitem{Tripathi2008_PRL101-142504}
Tripathi V, Tabor S~L, Mantica P~F, Utsuno Y, Bender P, Cook J, Hoffman C~R,
  Lee S, Otsuka T, Pereira J, Perry M, Pepper K, Pinter J~S, Stoker J, Volya A
  and Weisshaar D 2008 {\em Phys. Rev. Lett.\/} {\bf 101} 142504--4
  \urlprefix\url{http://link.aps.org/abstract/PRL/v101/e142504}

\bibitem{Doornenbal2009_PRL103-032501}
Doornenbal P, Scheit H, Aoi N, Takeuchi S, Li K, Takeshita E, Wang H, Baba H,
  Deguchi S, Fukuda N, Geissel H, Gernhauser R, Gibelin J, Hachiuma I, Hara Y,
  Hinke C, Inabe N, Itahashi K, Itoh S, Kameda D, Kanno S, Kawada Y, Kobayashi
  N, Kondo Y, Krucken R, Kubo T, Kuboki T, Kusaka K, Lantz M, Michimasa S,
  Motobayashi T, Nakamura T, Nakao T, Namihira K, Nishimura S, Ohnishi T,
  Ohtake M, Orr N~A, Otsu H, Ozeki K, Satou Y, Shimoura S, Sumikama T, Takechi
  M, Takeda H, Tanaka K~N, Tanaka K, Togano Y, Winkler M, Yanagisawa Y, Yoneda
  K, Yoshida A, Yoshida K and Sakurai H 2009 {\em Phys. Rev. Lett.\/} {\bf 103}
  032501--4 \urlprefix\url{http://link.aps.org/abstract/PRL/v103/e032501}

\bibitem{Wimmer2010_PRL105-252501}
Wimmer K, Kroell T, Kruecken R, Bildstein V, Gernhaeuser R, Bastin B, Bree N,
  Diriken J, Van~Duppen P, Huyse M, Patronis N, Vermaelen P, Voulot D, Van~de
  Walle J, Wenander F, Fraile L~M, Chapman R, Hadinia B, Orlandi R, Smith J~F,
  Lutter R, Thirolf P~G, Labiche M, Blazhev A, Kalkuehler M, Reiter P, Seidlitz
  M, Warr N, Macchiavelli A~O, Jeppesen H~B, Fiori E, Georgiev G, Schrieder G,
  Das~Gupta S, Lo~Bianco G, Nardelli S, Butterworth J, Johansen J and Riisager
  K 2010 {\em Phys. Rev. Lett.\/} {\bf 105} 252501--4
  \urlprefix\url{http://link.aps.org/doi/10.1103/PhysRevLett.105.252501}

\bibitem{Rodriguez-Guzman2000_PRC62-054319}
Rodriguez-Guzman R~R, Egido J~L and Robledo L~M 2000 {\em Phys. Rev. C\/} {\bf
  62} 054319--8
  \urlprefix\url{http://link.aps.org/doi/10.1103/PhysRevC.62.054319}

\bibitem{Lu2011_PRC84-014328}
Lu B~N, Zhao E~G and Zhou S~G 2011 {\em Phys. Rev. C\/} {\bf 84} 014328--10
  (\textit{Preprint} \eprint{1104.4638})
  \urlprefix\url{http://link.aps.org/doi/10.1103/PhysRevC.84.014328}

\bibitem{Zhou2011_JPCS312-092067}
Zhou S~G, Meng J, Ring P and Zhao E~G 2011 {\em J. Phys: Conf. Ser.\/} {\bf
  312} 092067--7 {arXiv:1101.3158v1 [nucl-th]}
  (\textit{Preprint} \eprint{1101.3158})
  \urlprefix\url{http://stacks.iop.org/1742-6596/312/i=9/a=092067}

\bibitem{Pei2013_PRC87-051302R}
Pei J~C, Zhang Y~N and Xu F~R 2013 {\em Phys. Rev. C\/} {\bf 87} 051302(R)--5
  \urlprefix\url{http://link.aps.org/doi/10.1103/PhysRevC.87.051302}

\bibitem{Brockmann1992_PRL68-3408}
Brockmann R and Toki H 1992 {\em Phys. Rev. Lett.\/} {\bf 68} 3408--3411
  \urlprefix\url{http://link.aps.org/doi/10.1103/PhysRevLett.68.3408}

\bibitem{Fuchs1995_PRC52-3043}
Fuchs C, Lenske H and Wolter H~H 1995 {\em Phys. Rev. C\/} {\bf 52} 3043--3060
  \urlprefix\url{http://link.aps.org/doi/10.1103/PhysRevC.52.3043}

\bibitem{Typel1999_NPA656-331}
Typel S and Wolter H~H 1999 {\em Nucl. Phys. A\/} {\bf 656} 331--364 \urlprefix\url{http://dx.doi.org/10.1016/S0375-9474(99)00310-3}

\bibitem{Niksic2002_PRC66-024306}
Nik\v{s}i\'{c} T, Vretenar D, Finelli P and Ring P 2002 {\em Phys. Rev. C\/}
  {\bf 66} 024306--15
  \urlprefix\url{http://link.aps.org/doi/10.1103/PhysRevC.66.024306}

\bibitem{Schunck2010_PRC81-024316}
Schunck N, Dobaczewski J, McDonnell J, More J, Nazarewicz W, Sarich J and
  Stoitsov M~V 2010 {\em Phys. Rev. C\/} {\bf 81} 024316--15
  \urlprefix\url{http://link.aps.org/doi/10.1103/PhysRevC.81.024316}

\bibitem{Perez-Martin2008_PRC78-014304--12}
Perez-Martin S and Robledo L~M 2008 {\em Phys. Rev. C\/} {\bf 78} 014304--12
  \urlprefix\url{http://link.aps.org/abstract/PRC/v78/e014304}

\bibitem{Fang2004_PRC69-034613}
Fang D~Q, Yamaguchi T, Zheng T, Ozawa A, Chiba M, Kanungo R, Kato T, Morimoto
  K, Ohnishi T, Suda T, Yamaguchi Y, Yoshida A, Yoshida K and Tanihata I 2004
  {\em Phys. Rev. C\/} {\bf 69} 034613--6
  \urlprefix\url{http://link.aps.org/doi/10.1103/PhysRevC.69.034613}

\bibitem{Navin1998_PRL81-5089}
Navin A, Bazin D, Brown B~A, Davids B, Gervais G, Glasmacher T, Govaert K,
  Hansen P~G, Hellstrom M, Ibbotson R~W, Maddalena V, Pritychenko B, Scheit H,
  Sherrill B~M, Steiner M, Tostevin J~A and Yurkon J 1998 {\em Phys. Rev.
  Lett.\/} {\bf 81} 5089--5092
  \urlprefix\url{http://link.aps.org/doi/10.1103/PhysRevLett.81.5089}

\bibitem{Sakharuk1999_PRC61-014609}
Sakharuk A and Zelevinsky V 1999 {\em Phys. Rev. C\/} {\bf 61} 014609--12
  \urlprefix\url{http://link.aps.org/abstract/PRC/v61/e014609}

\bibitem{Sargsyan2011_PRC84-064614}
Sargsyan V~V, Adamian G~G, Antonenko N~V, Scheid W and Zhang H~Q 2011 {\em
  Phys. Rev. C\/} {\bf 84} 064614--
  \urlprefix\url{http://link.aps.org/doi/10.1103/PhysRevC.84.064614}

\end{thebibliography}
\providecommand{\newblock}{}

\end{document}